\documentclass[twoside,12pt]{report}

\def\openone{\leavevmode\hbox{\small1\kern-3.8pt\normalsize1}}%

\usepackage{jeep}	
\mysection{section}{\large\bf}{\thesection.~}   
\mysection{subsection}{\normalsize\bf}{\thesubsection.~}
\mysection{subsubsection}{\normalsize\bf}{\thesubsubsection.~}
\mysection{paragraph}{\normalsize}{\theparagraph.}
\mysection{subparagraph}{\normalsize}{\thesubparagraph.}

\usepackage{a4wide}
\def\TUM#1{%
\dimen1=#1\dimen1=.1143\dimen1%
\dimen2=#1\dimen2=.419\dimen2%
\dimen3=#1\dimen3=.0857\dimen3%
\dimen4=\dimen1\advance\dimen4 by\dimen2%
\setbox0=\vbox{\hrule width\dimen3 height\dimen1 depth0pt\vskip\dimen2}%
\setbox1=\vbox{\hrule width\dimen1 height\dimen4 depth0pt}%
\setbox2=\vbox{\hrule width\dimen3 height\dimen1 depth0pt}%
\setbox3=\hbox{\copy0\copy1\copy0\copy1\box2\copy1\copy0\copy1\box0\box1}%
\leavevmode\vbox{\box3}}
\def\oTUM#1{%
\dimen1=#1\dimen1=.1143\dimen1%
\dimen2=#1\dimen2=.419\dimen2%
\dimen3=#1\dimen3=.0857\dimen3%
\dimen0=#1\dimen0=.018\dimen0%
\dimen4=\dimen1\advance\dimen4 by-\dimen0%
\setbox1=\vbox{\hrule width\dimen0 height\dimen4 depth0pt}%
\advance\dimen4 by\dimen2%
\setbox8=\vbox{\hrule width\dimen0 height\dimen4 depth0pt}%
\advance\dimen4 by-\dimen2\advance\dimen4 by-\dimen0%
\setbox4=\vbox{\hrule width\dimen4 height\dimen0 depth0pt}%
\advance\dimen4 by\dimen1\advance\dimen4 by\dimen3%
\setbox6=\vbox{\hrule width\dimen4 height\dimen0 depth0pt}%
\advance\dimen4 by\dimen3\advance\dimen4 by\dimen0%
\setbox9=\vbox{\hrule width\dimen4 height\dimen0 depth0pt}%
\advance\dimen4 by\dimen1%
\setbox7=\vbox{\hrule width\dimen4 height\dimen0 depth0pt}%
\dimen4=\dimen3%
\setbox5=\vbox{\hrule width\dimen4 height\dimen0 depth0pt}%
\advance\dimen4 by-\dimen0%
\setbox2=\vbox{\hrule width\dimen4 height\dimen0 depth0pt}%
\dimen4=\dimen2\advance\dimen4 by\dimen0%
\setbox3=\vbox{\hrule width\dimen0 height\dimen4 depth0pt}%
\setbox0=\vbox{\hbox{\box9\lower\dimen2\copy3\lower\dimen2\copy5%
\lower\dimen2\copy3\box7}\kern-\dimen2\nointerlineskip%
\hbox{\raise\dimen2\box1\raise\dimen2\box2\copy3\copy4\copy3%
\raise\dimen2\copy5\copy3\box6\copy3\raise\dimen2\copy5\copy3\copy4\copy3%
\raise\dimen2\box5\box3\box4\box8}}%
\leavevmode\box0}

\usepackage{fancyheadings}
\usepackage{cite}   

\advance \headheight by 3.0truept       

\newcommand{\lt}{\left}
\newcommand{\rt}{\right}
\newcommand{\ov}{\overline}
\newcommand{\nn}{\nonumber \\}
\newcommand{\no}{\nonumber }
\newcommand{\g}{\gamma }
\newcommand{\ns}{\mbox{no sum on }}
\newcommand{\lef}{\lt( 1-\g_5 \rt) }

\newcommand{\ba}{\mr{bare} }
\newcommand{\bare}{\ba}
\newcommand{\re}{\mr{ren} }
\newcommand{\ren}{\re}
\newcommand{\tr}{\mr{tr} }
\newcommand{\da}{\frac{\partial }{\partial a_{il}}}
\newcommand{\e}{\varepsilon}
\newcommand{\eps}{\varepsilon}
\newcommand{\Lagr}{{\cal L}}
\newcommand{\eq}[1]{(\ref{#1})}
\newcommand{\imag}{{\rm Im}\,}
\newcommand{\real}{{\rm Re}\,}

\newcommand{\beq}{\begin{eqnarray}}
\newcommand{\eeq}{\end{eqnarray}}
\newcommand{\mc}{m_c^2}
\newcommand{\mt}{m_t^2}
\newcommand{\ms}{m_s^2}
\newcommand{\md}{m_d^2}
\newcommand{\mw}{M_W^2}
\newcommand{\tw}{\widetilde{\openone}}
\newcommand{\1}{\openone}
\newcommand{\T}{\, \mathrm{\bf T \,}}
\newcommand{\gev}{\, \mathrm{GeV}}
\newcommand{\laMSb}{\Lambda_{\overline{\mathrm{MS}}}}
\newcommand{\laqcd}{\Lambda_{\mathrm{QCD}}}

\newcommand{\gf}{\, G_F}
\newcommand{\gft}{\, G_F^2}
\newcommand{\fig}[1]{fig.~\ref{#1}}

\newcommand{\ul}[1]{\underline{#1}}
\newcommand{\mr}[1]{\mathrm{#1}}
\newcommand{\kkm}{$\mr{K^0\!-\!\ov{K^0}}\,$--mixing\/}

\newcommand{\ddm}{$\mr{D^0\!-\!\ov{D^0}}\,$--mixing\/}
\newcommand{\kkM}{$\mr{K^0\!-\!\ov{K^0}}\,$--Mixing\/}
\newcommand{\kkmd}{$\mr{K_L\!-\!K_S}\,$--mass difference\/}
\newcommand{\rf}[1]{(\ref{#1})}
\newcommand{\dst}{$\mr{|\Delta S|\!=\! 2}$ }
\newcommand{\dso}{$\mr{|\Delta S|\!=\! 1}$ }
\newcommand{\dbt}{$\mr{|\Delta B|\!=\! 2}$ }
\newcommand{\dct}{$\mr{|\Delta C|\!=\! 2}$ }
\newcommand{\lag}{\mathcal{ L}}
\newcommand{\sla}{ \!\!\! / }
\newcommand{\txr}[1]{ \textrm{#1} }
\newcommand{\bra}[1]{\langle #1 | }
\newcommand{\ket}[1]{| #1 \rangle }
\newcommand{\dbrs}{ \delta _\mr{BRS} }
\newcommand{\msb}{$\ov{\txr{MS}}$ }
\newcommand{\mub}{\ov{\mu}}
\newcommand{\euga}{{\g_{\mr{E}}}}
\newcommand{\dmu}{\mu \frac{d}{d \mu}}
\newcommand{\pmu}{\mu \frac{\partial }{\partial \mu}}
\newcommand{\diff}[1]{ \frac{d}{d #1 }}
\newcommand{\prt}[1]{ \frac{\partial }{\partial #1 }}
\newcommand{\var}[2]{ \frac{\delta #1 }{\delta #2 }}
\newcommand{\Qh}{\widehat{Q}}
\newcommand{\Qht}{\widetilde{Q}}
\newcommand{\Tg}{\, \mathrm{\bf T_g \,}}
\newcommand{\gz}{\gamma^{(0)}}
\newcommand{\go}{\gamma^{(1)}}
\newcommand{\gzt}{\gamma^{(0)\, T}}
\newcommand{\got}{\gamma^{(1)\, T}}
\newcommand{\al}{\alpha}
\newcommand{\uz}{U^{(0)}}
\newcommand{\wt}[1]{\widetilde{#1}}
\newcommand{\wh}[1]{\widehat{#1}}
\newcommand{\uma}{\mathsf{1}}     
\newcommand{\Qheom}{\widehat{Q}_\mr{EOM}}
\newcommand{\Qhbrs}{\widehat{Q}_\mr{BRS}}
\newcommand{\Qeom}{Q_\mr{EOM}}
\newcommand{\Qbrs}{Q_\mr{BRS}}

\newcommand{\sims}{\begin{array}{c}< \\[-3mm] \sim \end{array} }

\newcommand{\oll}{\wt{Q}_{S2}}
\newcommand{\cll}[1]{\wt{C}_{S2}^{(#1)}}

\newcommand{\leso}{\lag ^\mr{|\Delta S| =  1}}
\newcommand{\leo}{\lag_\mr{eff,1}^\mr{|\Delta S| =  2}}
\newcommand{\lefft}{\lag_\mr{eff,2}^\mr{|\Delta S| = 2}}

\newcommand{\hsto}{H_1^\mr{|\Delta S| = 2}}

\newcommand{\bilo}[1]{\mathcal{O}_{#1}}
\newcommand{\bilr}[1]{\mathcal{R}_{#1}}
\newcommand{\oloc}{\wt{Q}_7}
\newcommand{\cloc}{\wt{C}_7}
\newcommand{\cvecp}{\vec{C}^{(+)}}
\newcommand{\cvecm}{\vec{C}^{(-)}}
\newcommand{\cvecpm}{\vec{C}^{(\pm)}}
\newcommand{\dvec}{\vec{D}}
\newcommand{\clocp}{\cloc^{+}}

\newcommand{\clocm}{\cloc^{-}}

\newcommand{\clocpm}{\cloc^{\pm}}
\newcommand{\gabip}{\g^{T}_{+,7}}
\newcommand{\gabim}{\g^{T}_{-,7}}
\newcommand{\gabipm}{\g^{T}_{\pm,7}}
\newcommand{\galoc}{\wt{\g}}
\newcommand{\rst}[1]{r_{#1, S2}  }

\begin{document}
\begin{titlepage}\parindent=0pt\parskip=0pt\vspace*{-11mm}
\begin{minipage}[b]{80mm} \begin{center} {\large Physik-Department}\\
                          Technische Universit\"at M\"unchen\\
                          Institut f\"ur Theoretische Physik\\
                          Lehrstuhl Univ.-Prof.~A.~J.~Buras
\end{center}
\end{minipage}
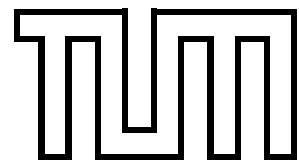
\hfill\raisebox{0mm}{\oTUM{37.5mm}}
\vspace{17mm}\par
\begin{center}\huge  Indirect CP--Violation in the \\
                     Neutral Kaon System \\
                     Beyond Leading Logarithms\\[10mm]
              \large and related topics\\[15mm]
              \Large Ulrich Nierste \end{center}
\vspace{\fill}\par
Vollst\"andiger Abdruck der von der Fakult\"at f\"ur Physik der Technischen
Universit\"at M\"unchen zur Erlangung des akademischen Grades eines\\[3mm]
\centerline{\emph{Doktors der Naturwissenschaften (Dr.~rer.~nat.)}}\\[3mm]
genehmigten Dissertation.
\vspace{5mm}\begin{center}\begin{tabular}{lll}
Vorsitzender: & & Univ.-Prof.~Dr.~F.~v.~Feilitzsch\\[1.5mm]   
Pr\"ufer der Dissertation: & 1.\ & Univ.-Prof.~A.~J.~Buras \\
                          & 2.\ & Univ.-Prof.~Dr.~W.~Weise \\
3.~Gutachter der Dissertation: & &
                         Univ.-Prof.~Dr.~W.~Buchm\"uller (Univ.~Hamburg)
\end{tabular}\end{center}\vspace{5mm}\par
Die Dissertation wurde am 25.1.1995 bei der Technischen Universit\"at
M\"unchen eingereicht und durch die Fakult\"at f\"ur Physik am 23.2.1995
angenommen.
\end{titlepage}
\thispagestyle{empty}\newpage
{}~\newpage
\pagenumbering{roman}
\setcounter{page}{1}
\chapter*{Abstract}
One of the least accurate tested sectors of the Standard Model of
elementary particles is the quark mixing mechanism. It encodes
the transitions between different quark families
in  weak charged--current decays  in terms of the
Cabibbo--Kobayashi--Maskawa--matrix $V_{CKM}$.
A complex phase $\delta$ in this matrix is the only possible source of
the violation of the CP--symmetry in the Standard Model.
This fact makes the study of $V_{CKM}$ mandatory: As soon as
the measurements of   CP--violation cannot be fitted with
the single parameter $\delta$, evidence of new laws of physics
will be found.
Today the only unambiguously measured CP--violating phenomenon
is the indirect CP--violation present in the \dst transitions
inducing the mixing of the  neutral Kaon states $K^0$ and
$\ov{K^0}$.
It is characterized by the well determined parameter $\e_K$.
Even from this single experiment on CP--violation
one can test the Standard Model, because together with
the unitarity of $V_{CKM}$ it
constrains some parameters of the Standard Model such as the mass of the
top--quark.

On the theoretical side the study of $V_{CKM}$ is challenged by
the presence of the strong interaction, which may screen or enhance
the weak transition amplitudes of quarks and which binds the quarks into
hadrons. The short distance QCD effects are comprised in the parameters
$\eta_1, \eta_2  $ and $\eta_3$ of the effective \dst--hamiltonian:
\begin{eqnarray}
\lefteqn{H^{|\Delta S|=2}(\mu)=} \nn
&& \hspace{-5ex} \frac{G_{F}^2}{16 \pi^2} \mw \!\! \lt[
                  \lambda_c^2 \eta_1
                  S(\frac{\mc }{\mw }) \! + \!
                  \lambda_t^2 \eta_2
                  S(\frac{\mt }{\mw }) \! + \!
                2 \lambda_c \lambda_t \eta_3
                  S(\frac{\mc }{\mw },
                  \frac{\mt }{\mw })
                   \rt]  \! \!
 b(\mu) \oll(\mu) + \mr{h.c.}. \; \;\no
\end{eqnarray}
For the following outline we only have to explain that
$\lambda_j = V_{jd}V_{js}^\ast$ encodes the relevant elements of $V_{CKM}$
and that $\eta_j b(\mu)  = 1$ in the absence of QCD corrections.

Now for the investigation of $\e_K$ the coefficient $\eta_2$ is most
important, but a precise determination of the subdominant term involving
$\eta_3$ is also required for a reliable analysis. The mass difference
between the two physical neutral K--mesons $\mr{K_L}$ and $\mr{K_S}$, however,
is dominated by $\eta_1$. When this work was begun, only the coefficient
$\eta_2$ was known beyond the leading log approximation \cite{bjw}.
This approximation is unsatisfactory, as it involves large errors
due to renormalization scale ambiguities and as it does not allow for
the use of the fundamental QCD scale parameter $\laMSb$.

The results of this thesis  include the following:
\begin{itemize}
\item[i)]  The calculation of $\eta_1$ and $\eta_3$ in the
          next--to--leading order (NLO) is presented in detail.
\item[ii)] An improved determination of $\imag \lambda_t$, which
          is the key quantity for CP--violation, using the
          measurement of $\e_K$ and the NLO values for
          $\eta_1$, $\eta_2$ and $\eta _3$ is  given.
          Results are tabulated as functions of the CKM--parameters
          $V_{cb}$ and $V_{ub}/V_{cb}$, the mass of the top--quark
          and of $B_K$, which parametrizes the size of the hadronic
          matrix element.
\item[iii)] The bounds on the Standard Model parameters resulting from
          the unitarity of $V_{CKM}$ are studied. We have found that
          with the new values for $\eta_1$ and $\eta_3$ a wider range
          of the parameters mentioned in ii) are allowed. In this
          sense the calculation has vindicated  the Standard
          Model.
\item[iv)] The NLO value of $\eta_1$ leads to an enhancement of the
          short distance contribution to the \kkmd. Now the short distance
          QCD contributions explain  the dominant part of the measured
          mass difference. The old leading order result has suggested
          a long--distance dominance. This has been a puzzle, because
          long--distance effects are expected to be suppressed by a factor
          of $\laqcd^2/m_c^2$ with respect to the short distance
          contributions.
\end{itemize}
A new feature of the  NLO calculation of $\eta_1$ and $\eta_3$ is the
presence of Green's functions with two  insertions of
operators (bilocal structures)
both of which have a non--zero anomalous dimension.
These Green's functions have required
a subtle analysis of the renormalization of
effective field theories.
The  discussion includes  the following points:
\begin{itemize}
\item[v)] The RG of bilocal structures is  analyzed in detail.
          Different methods for the solution are presented.
          These include the framework of an inhomogeneous
          RG equation, which is best suited for formal proofs
          like  those mentioned in vii).
          For the practical calculation a matrix formulation is
          given which in the case of $\eta_3$
          only involves one $8\times 8$--matrix
          instead of four of them as in the old LO calculation in
           \cite{gw}.
\item[vi)] The structure of the reduced effective \dso lagrangian
          is discussed. In  matrix elements with single
          operator insertions those operators which
          vanish by the field equation of motion
          or which are BRS--exact do not contribute to the matrix
          elements. But  one faces non--trivial contact terms, if the Green's
          function contains two operator insertions.
          It is well--known \cite{jl,colleom,sim} that these terms
          can be absorbed into \dst operators.  In our case
          the contact terms have been identified to correspond to
          operators describing subleading effects  in $m_s/m_c$.
\item[vii)] Likewise \dso evanescent operators  appear in the effective
          lagrangian. Here we had to develop the correct prescription
          to deal with evanescent operators in bilocal  structures.
          Just as  in the case of the operators mentioned in vi)
          non--vanishing  terms appear, which are also equivalent to
          the matrix elements of \dst operators.
           But now they can be
          absorbed into a finite renormalization of the \dst operators.
\item[viii)] The analysis of vii) has stimulated a related
          investigation: The definition of evanescent operators is
          not unique, because one can add $(D-4)$ times any physical
          operator to them. We show that any definition of them leads
          to an effective lagrangian, in which the physical sector
          is unaffected by the presence of properly renormalized
          evanescent operators.
          Hence one can drop the latter from the reduced lagrangian.
          Yet the arbitariness in the definition leads to a
          renormalization  scheme dependence in the \emph{physical}\/
          sector. This result is important for practical NLO
          calculations, because the Wilson coefficients and the
          anomalous dimension matrix   depend on the definition of the
          evanescents. Formulae to transform between the schemes are
          presented.
\item[ix)] Last but not least there are about 40 two--loop tensor
           integrals up to rank six
           involved in the calculations. The calculation
           of  $\eta_1$ has required to calculate their finite part.
           Here  we present a compact formula for the most general
           integral which can appear  in two--loop calculations of
           light hadron systems. This is the two--loop vacuum bubble
           diagram with arbitrary tensor structure,
            small infrared regulating masses on the lines flown
            through by one loop momentum and
           arbitrary heavy masses on the other lines.
           The formula holds in $D$ dimensions and is therefore
           a good investment into the future of higher order QCD
           corrections.
\end{itemize}\cleardoublepage
\tableofcontents
\cleardoublepage
\listoffigures
\cleardoublepage
\pagenumbering{arabic}\setcounter{page}{1}
\chapter{Introduction}
The aim of elementary particle physics is the discovery of new fundamental
laws of nature and their classification within a theoretical framework
which arranges the observed elementary particles into simple patterns
and describes their interactions in a unified way.

The stage for this theoretical description is the Relativistic Quantum
Field Theory incorporating both Einstein's theory of Special Relativity
and Heisenberg's, Pauli's and Schr\"o\-din\-ger's Quantum Mechanics.
Its story of success started in 1947, when the experiment of Lamb and
Re\-ther\-ford \cite{lr} gave unambigous evidence for the splitting of
the $\mathrm{2s_{1/2}}$
and the
$\mathrm{2p_{1/2}}$ energy levels of the hydrogen atom
by $\mathrm{4\, \mu eV}$.
At the same time Dirac, Feynman, Tomonaga, Schwinger and others \cite{s}
developed the theory of Quantum Electrodynamics,
which explains the Lamb shift by quantum fluctuations of the photon field.
Since then many more actors have joined the electron and the photon field
on the stage: Starting with the works of Gell-Mann and Zweig the
partonic constituents of hadrons were successfully described by
quark fields, whose strong interaction is mediated by gluons
as encoded in the Lagrangian of
Quantum Chromodynamics \cite{gz}.
And with the unification of the electromagnetic and weak interaction
by Glashow, Salam and Weinberg \cite{gsw}
the scene was entered
by three more vector bosons, which are massive owing to
their interaction with the Higgs particle \cite{hi}.
These elements form what is now called the Standard Model of elementary
particle physics. It is the most successful theory of physics:
It covers energy scales between those of the Lamb shift and the
center of mass energy of today's particle colliders such as LEP at
CERN, which are separated by 17 orders of magnitude! No experiment
has yet contradicted its quantitative predictions, although many of them
have been tested to an extraordinary precision (e.g.\ an accuracy
of $\mathrm{10^{-10}}$ for the anomalous magnetic moment of the electron
on both the experimental and the theoretical side).

On the other hand some sectors of the standard model are poorly
verified at present. One of them is the quark mixing sector, which
describes the coupling of quark fields to the charged electroweak
vector bosons
in terms of the Cabibbo--Kobayashi--Maskawa (CKM) matrix \cite{ckm}.
 Here the non--perturbative nature of the
long--distance strong interaction prevents an easy extraction of
the CKM  parameters.

Further we know that the Standard Model cannot be true for another
17 orders of magnitude in energy, as we then would reach the Planck scale,
where gravitation becomes important requiring a new theoretical concept.
Yet we know from other considerations  such as vacuum stability that
there must be new physics much below the Planck scale. How can we detect
new physics apart from building accelerators with higher and higher energy?

First one can look at rare processes which are forbidden in
the tree level approximation. Since they are suppressed in the
standard model new physics can contribute sizeably to their transition
rates. Moreover, even low--energy rare hadronic processes are sensitive
to the masses of heavy particles
appearing in intermediate states. This allows to derive bounds on the
mass of the recently discovered top--quark \cite{cdf} from these
processes \cite{blo,hn4}.
When the experimental determination of the top mass improves, these
bounds provide a consistency check of the Standard Model quark sector
over more than two orders of magnitude in energy.

Second one can look for violations of symmetries which are
respected
or weakly broken by the Standard Model. Here the CP symmetry is
of utmost importance, because
the only possible source of its violation
in the Standard Model is a
single parameter, the phase $\delta$ in the CKM
matrix.\footnote{We do not consider a possible $\theta$--term in the
QCD lagrangian here.}
The CP transformation exchanges particles and antiparticles
and reflects the spacial coordinates.
If in the future
we cannot fit all CP violating observables with this single
phase, we will have discovered new physics.
Further in the calculation of these observables the difficulties with
the long--distance strong interaction are sometimes reduced, as the strong
(as well as the electromagnetic) interaction respects CP symmetry
indicated by the smallness of the neutron electric dipole moment.

All these aspects make the study of hadronic physics mandatory.
Here a  key object is the K--meson system, to which we owe a
significant part of our present day knowledge about the Standard Model
quark sector: In 1964 Christenson, Cronin, Fitch and Turlay have
found the first evidence for CP violation by observing decays
of the long--lived neutral K--meson $\mathrm{K_L}$
into  the CP even two--pion
state \cite{ccft}. There are two potential sources for this phenomenon:
The $\mr{K_L}$ state is not a pure CP odd state but has admixtures of
the CP even neutral Kaon state or the weak interaction triggering this
decay violates the CP symmetry thereby allowing transitions from the CP
odd neutral Kaon state to CP even final states.
In the first case we speak of \emph{indirect CP violation}\/ in contrast to
\emph{direct CP violation}\/ in the second case.
The CKM mechanism of the Standard Model predicts both types, if
the phase $\delta$ is not equal to 0 or $\pi$.
While precise measurements are available for the parameter $\mr{\e_K}$
\cite{pdg} characterizing the size of the indirect CP violation in the neutral
Kaon system, there is still a controversy about the direct CP
violation \cite{pdg}.
Then in 1970 the suppression of flavour--changing neutral current
(FCNC) decays of K--mesons lead Glashow, Iliopoulos and Maiani (GIM) to
postulate the existence of the charm quark \cite{gim}.
We finally mention the prediction of the charm--quark mass from the
observed \kkmd\/ by Gaillard and Lee in 1974 \cite{gl}.
On the other hand
the theoretist's work to relate the measured quantities
to the
Standard model parameters is made difficult by the presence of the
strong interaction: Its long range interaction confines the quarks
into K--mesons and it modifies the weak CP violating amplitudes
of interest.  Since these transitions take place on short distance
scales of the order of the inverse W--boson mass
$\mr{M_W}$, the QCD corrections
can be reliably calculated perturbatively. The long--range effects,
however, must be treated by other methods such as
$\mr{1/N_c}$--expansion, sum rule techniques or lattice gauge theory.

The Technical University of Munich
has already largely
contributed to the theoretical understanding of
short distance QCD corrections to various rare hadronic processes
\cite{bw,bjw,bjlw,bjlw2,tqs,blb,bb,hn1,m,blo,stefdiss}.

In this work we will focus on short distance  QCD calculations
to the mixing of the flavour eigenstates
$\mr{K^0}$ and $\mr{\ov{K^0}}$
of the neutral
K--meson. Since the mass eigenstates
$\mr{K_L}$ and $\mr{K_S}$
are linear combinations of them, \kkm\/
encodes the information on the indirect CP violation.

The hamiltonian describing this mixing reads:
\begin{eqnarray}
\lefteqn{H^{|\Delta S|=2}(\mu)=} \nn
&& \hspace{-5ex} \frac{G_{F}^2}{16 \pi^2} \mw \!\! \lt[
                  \lambda_c^2 \eta_1
                  S(\frac{\mc }{\mw }) \! + \!
                  \lambda_t^2 \eta_2
                  S(\frac{\mt }{\mw }) \! + \!
                2 \lambda_c \lambda_t \eta_3
                  S(\frac{\mc }{\mw },
                  \frac{\mt }{\mw })
                   \rt]  \! \!
 b(\mu) \oll(\mu) + \mr{h.c.}. \; \;\label{s2}
\end{eqnarray}
Here $\mr{|\Delta S|=2}$ denotes the change in the strangeness quantum
number in the transition between $\mr{K^0}$ and $\mr{\ov{K^0}}$,
$G_{F}$
is the Fermi constant, $\lambda_j=V_{jd} V_{js}^{*}$
comprises the CKM--factors,
and
$\oll $ is  the local
four quark operator (see \fig{loc} on p.\ \pageref{loc})
\begin{eqnarray}
\oll &=&
( \ov{s}_j \gamma_\mu (1-\gamma_5) d_j)
(\ov{s}_k \gamma^\mu (1-\gamma_5) d_k) \; = \; (\ov{s} d)_{V-A}
(\ov{s} d)_{V-A}  \label{ollintro}
\end{eqnarray}
with $j$ and $k$ being colour indices.
The Inami--Lim functions $S(x)$ and $S(x,y)$ \cite{il} depend on the
masses of the charm-- and top--quark and
describe the
$\mr{|\Delta S|=2}$--transition amplitude in the absence of strong
interaction.
They are obtained by calculating the lowest order box diagrams
depicted in \fig{box} on p.\ \pageref{box}.
The  short distance part
is included in the coefficients
$\eta_1$, $\eta_2$ and  $\eta_3$ with a common factor $b(\mu)$
split off.

Here $\eta_2$ and  $\eta_3$ are most important for the size of
$\mr{\e_K}$, while
$\eta_1$ gives the dominant  contribution to the short distance part
of the \kkmd.
When  this work was begun,
a next--to--leading order (NLO) calculation for $\eta_2$ had been performed
\cite{bjw}, but
no QCD corrections beyond
leading logarithms had been known for $\eta_1$ and $\eta_3$.
The leading log (LL) approximation is very unsatisfactory, because it is
plagued with huge renormalization scale uncertainties and no use of
the QCD scale parameter $\laMSb$ can be made. In this thesis the complete
$|\Delta S|=2$--hamiltonian in the NLO will be presented \cite{hn3}.

As the importance of  $\mr{\e_K}$ has been stressed already, let us now
discuss the \kkmd : Its measured value is
$\mr{3.5\, \mu eV}$, which, by the way, is of the same order of magnitude
as the Lamb shift mentioned in the beginning of this chapter.
The prediction of the charm--quark mass, $\mr{m_c \approx 1.5 \gev }$,
from the \kkmd\/ in \cite{gl} was surprisingly good for an analysis
which neglected strong interaction effects.
The inclusion of the latter, however, made the theoretical prediction
worse, as estimates of the hadronic matrix element $\langle \oll \rangle$
by various nonperturbative methods resulted in a significant reduction
of the naive vacuum insertion value used in \cite{gl}. Further the
leading--log calculation of the short distance QCD corrections
gave an additional suppression of the predicted \kkmd\/ due to
$\eta_1 <1 $, so that the calculation reproduced less than half of the
measured \kkmd. On the other hand one also expects additive
long--distance contributions corresponding to light meson poles
which are not contained in \rf{s2}. These contributions are poorly
calculable, but should be
suppressed with respect to the short distance part
by power counting arguments. Therefore the low result of
the short distance calculation is somewhat surprising and in fact
triggered some speculation to assign the deficit to new physics.
Our NLO
calculation for $\eta_1$ \cite{hn1}, however,
together with
present day values for the input parameters such as $\laMSb$
has established $\eta_1 > 1$ implying
a sizeable increase of the short distance
contribution to the \kkmd, which is now roughly of the order
of 70\% of the observed value. This fits much better with
the expectation, especially when contrasted with an earlier incorrect
NLO calculation \cite{fky}
finding a drastical decrease of the short distance part.

In the following chapter we will summarize
some field theoretic basics needed for the subsequent chapters.
Chapter \ref{rg} will discuss the operator product expansion and
renormalization group techniques. Here we will also develop the
renormalization group formalism for Green's functions with two
operator insertions.
Chapter \ref{eva} contains new results about evanescent operators
\cite{hn2}.
They comprise the correct  treatment of bilocal structures
involving evanescent operators, which had to be developed
for the calculation of $\eta_3$.
Further a new scheme dependence associated with the definition of
the evanescents is discussed.
In chapter \ref{operator} the operator basis
involved  in  the NLO \dst lagrangian will be analyzed.
In chapter \ref{calc} the NLO calculation
of $\eta_1$ and $\eta_3$ will be described in detail, and
chapter \ref{phen} is devoted to the phenomenological analyses. Here we
will analyze the consequences of the new
NLO prediction for $\mr{\e_K}$ on the
Standard Model parameters.
We will also look at the new prediction of the short distance
contribution to the \kkmd.
\cleardoublepage
\chapter{Quantum Field Theory and the  Standard Model}\label{qft}
Our today's knowledge about the rich field of elementary particle physics
is the result of many decades of research probing the laws of nature
at small distances. The experimental effort has been paralleled
by a similarly successful  development of the theoretical framework
to understand the guiding principles of the observed phenomena.

Clearly we cannot summarize all  results of this development here,
so that we instead focus on those aspects of modern quantum field theory
which are relevant for the actual calculations concerning the \kkm\/ performed
for this thesis.
\section{Lagrangian Field Theory}
A classical field theory like electrodynamics contains a set of
functions $ \{ \! \phi_j (x)\! \}$ of the space--time variable
$x=(t,\vec{x})$\footnote{Throughout this thesis we use natural
units $\mr{c=\hbar =1}$.} as its basic building blocks.
The dynamics of these fields  $  \phi_j (x) $
is encoded in a scalar functional, the lagrangian (density)
$\lag \lt[ \phi_j (x) , \partial_\mu \phi_j (x),x \rt]$,
from which the field equations of motions can be obtained.
Here we have collectively summarized  indices related to intrinsic properties
like spin or charge quantum numbers and those distinguishing
different physical fields  into the single multi--index $j$.

When passing from classical to quantum field theory the functions
$\phi_j (x)$ are replaced by field operators $\,\widehat{\!\phi }_j (x)$.
Hence one has to deal with a quantum mechanical system with
an uncountable infinity of  degrees of freedom labeled by the
space--time coordinate $x$. Its hamiltonian density $\mathcal{H}(x)$
is a polynomial in
the elementary fields $\,\widehat{\!\phi}_j (x) $.
For non--interacting fields the field theory can be solved,
i.e.\ one can construct a Hilbert space (the Fock space)
whose elements are
eigenfunctions  of the Fourier--transformed
hamiltonian density $\widehat{\mathcal{H}} (\vec{k}) $. A short glimpse
on the simple structure of the lagrangian and on the variety
of particle phenomenology shows that it is hopeless to try
to solve the interacting quantum field theory exactly.

\section{Symmetries}\label{sym}
Before discussing the dynamics of field theory it is useful to
look for underlying patterns which may possibly restrict
the particle spectrum and the interactions.
This is indeed the case, if the total action
\begin{eqnarray}
S \lt[ \phi_j    \rt] &=& \int d^4 x
       \lag \lt[ \phi_j (x),\partial_\mu \phi_j (x),x   \rt] \label{act}
\end{eqnarray}
is invariant under some  group of  transformations of the arguments
of $\lag $.

Consider first some  group  G of linear transformations of the
space--time variable:
\begin{eqnarray}
x^\mu\rightarrow x^{\prime \mu} := \omega ^\mu_\nu x^\nu
&& \mbox{with }\; \omega \in G   \label{groupx}
\end{eqnarray}
If we can  find a (possibly multivalued)
representation $\mathcal{D}(G)$ of $G$
such that with
\begin{eqnarray}
\omega_{\nu}^{\mu} &\rightarrow &
       \lt[ d \lt( \omega \rt) \rt]_{jk} \in \mathcal{D} (G)\nn
\phi_j (x)     &\rightarrow & \phi_j^\prime (x^\prime )
 \; = \; \phi_j^\prime (\omega \cdot x)
\; := \; \lt[ d \lt( \omega \rt) \rt]_{jk} \phi_k (x)
\label{groupfield}
\end{eqnarray}
the action \rf{act} satisfies $S[ \phi_j^\prime ]=S[ \phi_j  ]$,
we speak of an
\emph{external}\/ symmetry.

A symmetry group which only transforms the fields
according to \rf{groupfield} but leaves the space--time coordinate
unchanged is likewise generating an \emph{internal}\/ symmetry.

The key importance of symmetries for a classical field theory
is due to Noether's theorem:
If a symmetry Lie group leaves $S$ invariant, one can
construct a conserved current from the fields and find a globally
conserved charge. These charges can be used to label different
solutions of the equations of motion.  A quantized theory may or may
not enjoy a certain symmetry of the corresponding classical theory.
In the second case one terms the symmetry to be broken.
Otherwise the operator corresponding to the conserved charge commutes
with the hamiltonian so that we can find a common eigenbasis of both
operators. We can use the quantum numbers of the conserved charge
to label the eigenstates of the hamiltonian  and transitions between
states with different quantum numbers are forbidden. In fact this
property also applies to discrete symmetries.
\subsection{External Symmetries}\label{extsym}
\subsubsection{Poincar\'e--Invariance}
When probing the laws of nature at distance scales
at which gravitation is unimportant
they turn out to be invariant with respect to
\begin{itemize}
\item translations in space and time,
\item rotations \\[2mm]
and
\item boosts into uniformly moving coordinate systems.
\end{itemize}
The last two types of transformations generate the proper Lorentz
group $L^\uparrow$, all of them generate the
Poincar\'e--group.
We can test the abovementioned symmetries
in a twofold way: First we can repeat the experiments
in different transformed coordinate frames\footnote{Theorists are
allowed to neglect the technical problems, if this is tried for the
boosts.}
and second we can check the associated conservation laws predicted
by Noether's theorem. They include the conservation of energy,
momentum and angular momentum.

The fact that nature is invariant under \rf{groupx} with
$G=L^\uparrow $  greatly simpifies the theorist's work to
guess the correct lagrangian describing the non--gravitational
interactions: $\lag$ must not explicitly depend on $x$ and
all fields must transform as in \rf{groupfield}
according to some (possibly multivalued) representation of the
proper Lorentz group. Since with \rf{groupx} $d^4 x= d^4 x^\prime $,
these fields must combine to $L^\uparrow\!$--singlets in $\lag$.

The proper Lorentz transformations leave the Minkowskian scalar
product $x_\mu x^\mu=x_0^2 \nolinebreak - \nolinebreak
\vec{x}^2$ invariant. Yet there are
more transformations which have this property: We may extend
$ L^ \uparrow $ by
\begin{itemize}
\item[] the \emph{parity}\/ transformation \textbf{P}:
        $\vec{x} \rightarrow  - \vec{x} $ \\[2mm] and
\item[] the \emph{time reversal}\/ transformation \textbf{T}:
        $x_0 \rightarrow - x_0 $.
\end{itemize}
$L:= \{ L^\uparrow , \textbf{P} L^\uparrow, \textbf{T} L^\uparrow,
\textbf{PT} L^\uparrow\} $
is the full Lorentz group. Unlike the transformations of $L^\uparrow $
we cannot carry out the \emph{discrete} symmetries \textbf{P} and
\textbf{T} by any active transformation of our experimental device,
but still we can test them via the conservation or nonconservation of
the associated charges. Since $\{ \textbf{1}, \textbf{P} \}$
and $\{ \textbf{1}, \textbf{T} \}$ are two--element groups, the
associated charges P and T can only assume two different values.
We may choose P and T as multiplicative quantum numbers being
$+1$ or $-1$.

There is no reason why nature should respect these  discrete
symmetries. Nevertheless it was not before 1956 that Lee and Yang
discovered the nonconservation of parity \cite{ly}. In the following
we will need the transformation property of Dirac fields
under \textbf{P} and \textbf{T}. They transform as
\begin{eqnarray}
\textbf{P}: \quad \psi(x) &\rightarrow & \psi^\prime (x^\prime ) \; = \;
   \eta_P \gamma^0 \psi(x) \label{parferm}
\end{eqnarray}
with an arbitrary  phase $\eta_P $ in any representation of the Dirac algebra.
The current
$j^\mu(x)=\ov{\psi} (x) \g^\mu \psi (x) $ is easily found to transform as a
vector field, i.e.\ like $x^\mu $, under \textbf{P}.
With $\g_5 := i \g_0 \g_1 \g_2 \g_3 $ the axial current
$j^{A\, \mu} (x) = \ov{\psi}(x) \g^\mu \g_5 \psi (x)$
likewise   transforms with an extra factor of
$(-1)$ compared to $j^\mu (x)$.
In the standard representation of the Dirac algebra one further finds:
\begin{eqnarray}
\textbf{T}: \quad \psi(x) &\rightarrow & \psi^\prime (x^\prime ) \; = \;
   \eta_T i \g ^1 \g^3  \psi^\ast (x)
\end{eqnarray}
with $\eta_T$ also being an arbitrary phase factor.

\subsubsection{Scale Transformations}\label{sctrf}
In the early 1950's St\"uckelberg and Peterman as well as Gell-Mann
and Low for the first time studied the behaviour of a quantum field
theory under dilatations of the space--time variable:
\begin{eqnarray}
x & \rightarrow & x^\prime = \lambda^{-1} x   
\end{eqnarray}
or equivalently in momentum space
\begin{eqnarray}
p & \rightarrow & p^\prime = \lambda p.
\end{eqnarray}
One might expect that for large $\lambda $ masses in $\lag$ become
negligible and any physical quantity scales according to its naive
dimension determined by power counting.
But this is not the case for a quantized theory due to the subtleties
of renormalization to be discussed in sect.\ \ref{reno}. Here
we find the powerlike scaling behaviour modified by so called
anomalous dimensions. In the context of renormalization
the symmetry under scale transformations
becomes a powerful calculational tool which will be looked at
in detail in chapter \ref{rg}.
\subsection{Internal Symmetries}
\subsubsection{Gauge Symmetries}
In view of the discussion of Quantum Chromodynamics (QCD) we first
look at an SU(N)--symmetric theory containing Dirac fermions, which
are chosen to transform according to the defining representation of
SU(N):
\begin{eqnarray}
\psi &\rightarrow &\psi _i ^\prime  = U_{ij}(\theta^a) \psi_j , \quad
                   U = e^ {-i T^a \theta^a } . \label{gaugeferm}
\end{eqnarray}
Here the $T^a$'s span the Lie algebra su(N). They satisfy
\begin{eqnarray}
\lt[ T^a , T^b \rt] = i f^{abc} T^c
\end{eqnarray}
with the structure constants $f^{abc}$.
Continous internal symmetries are called gauge symmetries.
If the gauge parameters $\theta^a$ in  \rf{gaugeferm} are independent
of $x$, one speaks of a \emph{global}\/ symmetry. \emph{Local}\/
gauge symmetries, where the $\theta^a $'s are allowed to depend on $x$,
play a key role in field theory: They incorporate a natural
description of interactions between the fermions mediated by a  vector field
$A_\mu^a (x)$. The lagrangian
\begin{eqnarray}
\lag &=& \overline{\psi}_i \lt( i D\! \sla \, _{ij} - m \delta _{ij} \rt)
\psi _j - \frac{1}{4} F_{\mu \nu}^a F^{a\, \mu \nu} \label{ginv}
\end{eqnarray}
with
\begin{eqnarray}
D\! \sla \, _{ij} &=& \partial \sla \delta_{ij}
         - i g T^a_{ij} A\sla ^a \nn
F^a_{\mu \nu} &=& \partial_\mu A_\nu ^a - \partial _\nu A_\mu ^a +
g f^{abc} A_\mu^b A_\nu ^c  \no
\end{eqnarray}
is invariant under local SU(N) transformations,
if the vector field $A_\mu^a (x)$ transforms as
\begin{eqnarray}
T^a A^{\prime a}_\mu &=& U\lt( \theta (x) \rt) \lt[
        T^a A^{a}_\mu - \frac{i}{g}
      U^{-1} \lt( \theta (x) \rt)  \partial _\mu
              U \lt( \theta (x) \rt) \rt] U^{-1} \lt( \theta (x) \rt)
. \label{gaugevec}
\end{eqnarray}
\subsubsection{Charge Conjugation}
Charge conjugation is a discrete global internal transformation.
The \textbf{C}
operation exchanges particles and antiparticles. In the standard
representation of the Dirac algebra a Dirac field transforms as
\begin{eqnarray}
\textbf{C}: \quad \psi (x)
    & \rightarrow & \psi^\prime ( x^\prime ) \; = \;
     \eta_C i \g^0 \g^2 \ov{\psi }^T (x)
\end{eqnarray}
with the usual arbitrary phase $\eta_C$.

Soon after the discovery of parity non--conservation the charged weak
lepton current was proposed to be of the type
$\ov{e} (x) \g^\mu (1-\g_5 ) \nu_e (x)$.
This
interaction also violates charge conjugation symmetry,  but it is
invariant under the combined transformation \textbf{CP}. Finally in
1964 it became clear that
the weak charged hadronic current  violates
\textbf{CP} invariance \cite{ccft}.

\subsection{CPT Theorem}\label{cpt}
After quantizing a theory one wishes to find a representation of
the symmetry transformations of interest in the space of state vectors
on which the field operators $\widehat{\phi}(x)$ act.
A theorem of Wigner states that one can always find such a
transformation, the corresponding operator is either unitary or
anti--unitary and commutes with the hamiltonian. We have already
alluded to this theorem in the discussion before
sect.\ \ref{extsym}.
In the case of \textbf{C} and \textbf{P} this operator is unitary,
while it is anti--unitary for \textbf{T}.

According to our present knowledge the electromagnetic and the strong
interaction respect \textbf{C}, \textbf{P} and \textbf{T} separately,
while the weak interaction violates all of them.
Nevertheless L\"uders and Pauli \cite{lp} have
proven that any Poincar\'e--invariant
field theory must respect the combined anti--unitary transformation
\textbf{CPT}. This has the consequences that particles and
antiparticles have the same mass and spin and that particles
with zero quantum numbers
are identical to their antiparticles.  The implications of the
\textbf{CPT} theorem are experimentally confirmed.

\section{Standard Model}
In the Standard Model the electromagnetic and the weak interaction
are unified in a $\txr{SU(2)}_\mr{L} \! \times \! \mr{U(1)}_\mr{Y}$
gauge invariant
lagrangian $\lag _{\mr{ew}}$ \cite{gsw}. The
strong interaction is comprised in the
SU(3) invariant
lagrangian of Quantum
Chromodynamics, $\lag _{\mr{QCD}}$ \cite{gz}.
Both combine to the Standard Model
lagrangian $\lag_{\mr{SM}}$.

The weak sector is the richest field of the Standard Model.
First the electroweak  gauge group
$\txr{SU(2)}_\mr{L} \times \txr{U(1)}_\mr{Y}$
is chiral, because left--handed fermion fields transform differently
from the right--handed ones implying parity violating couplings.
Second the $\txr{SU(2)}_\mr{L} \times \txr{U(1)}_\mr{Y}$ gauge symmetry
is spontaneously broken to the electromagnetic gauge group
$\txr{U(1)}_\mr{em}$.
This means that the physical vacuum state does not
have the $\txr{SU(2)}_\mr{L} \times \txr{U(1)}_\mr{Y}$--symmetry of the
lagrangian. In the Standard Model the symmetry breaking is
provided by the Higgs--Kibble mechanism \cite{hk}
introducing a scalar potential
in $\lag $ with degenerate minima which transform into each other
under $\txr{SU(2)}_\mr{L} \times \txr{U(1)}_\mr{Y}$ transformations
instead of  being individually invariant. As a consequence three gauge
bosons become massive due to their interaction with the Higgs field.
The chiral structure of the Standard Model forbids explicit mass terms
for the fermions. Fermion masses are therefore generated through
Yukawa interactions with the Higgs field.
After diagonalizing the resulting quark mass matrix the coupling
between the mass eigenstates of the quarks and the weak charged
vector bosons $W_\mu^{\pm}$ is found to be of the form:
\begin{eqnarray}
\lag _{wc} &=& - \frac{g_w}{2 \sqrt{2} } J_{cc}^{\mu}(x)
 W_\mu^{+} (x) + \mr{h.c.} \label{ccint}
\end{eqnarray}
Here $g_w$ is the weak coupling constant and $J_{cc}(x)$ is the
weak charged current:
\begin{eqnarray}
J_{cc}^{\mu}(x) &=& \overline{u}_j (x) \g^\mu (1- \g_5) V_{jk} d_k(x)\label{cc}
\end{eqnarray}
where $u_j$ denotes the up--type quarks of the three families, u,c,t,
and $d_k$ stands for the  down--type quarks d,s,b.
The Yukawa interactions in $\lag_\mr{SM}$ do not only induce fermion
masses, but also allow for family--changing couplings in \rf{cc}
described by the unitary Cabibbo--Kobayashi--Maskawa (CKM) matrix
$(V_{jk})$ \cite{ckm}. A unitary $3 \times 3$--matrix can be
parametrized by nine parameters, five of which are redundant in \rf{cc}, as
they correspond to physically irrelevant redefinition of the quark
fields by U(1) phase transformations. The other four correspond to a
complex phase $\delta$ and three rotation angles, i.e.\ $V$ is
real for $\delta =0$. If $V$ is real the charged current
interaction $\lag_{wc}$ in
\rf{ccint} is \textbf{CP}--conserving. A non--zero $\delta$
corresponds  to  \textbf{CP} violation. This can easily be seen by
recalling from sect.\ \ref{cpt} that the latter is equivalent to
\textbf{T} violation which is an antiunitary transformation involving
a complex conjugation. Hence since $\lag_{wc}$ is \textbf{T} invariant
for real $V$,
it must in general transform under \textbf{T} into an expression
with complex conjugate $V$. Since further $J_{cc}\neq J^\dagger_{cc}$
this is different from the untransformed $\lag_{wc}$.
Yet only a non--zero $\delta $ is not sufficient for \textbf{CP}
violation, we must also ensure that it cannot be absorbed into
a redefinition of the quark fields. This means that
the three up--type quarks as well as the three down--type quarks
must have
different masses.
In the lepton
sector any mixing matrix can likewise be absorbed into a redefinition
of the neutrino fields, if the neutrinos are massless.

For the weak neutral current such flavour--changing vertices  do not
occur as a consequence of the unitarity of $V$. This phenomenon is
called GIM mechanism \cite{gim} implying the suppression of
flavour changing neutral current (FCNC) processes.

The  QCD lagrangian enjoys an unbroken SU(3) gauge symmetry. It is
obtained by adding a flavour label $k$ to the fermion field in \rf{ginv}:
\begin{eqnarray}
\lag_\mr{QCD} &=&
\overline{\psi}_k \lt( i D\! \sla  - m_k  \rt)
\psi _k - \frac{1}{4} F_{\mu \nu}^a F^{a\, \mu \nu} \label{qcd}
\end{eqnarray}
with the colour indices now being suppressed.
In \rf{qcd} the fermion fields $\psi_k(x)$ correspond to quarks
with mass $m_k$ and the vector fields $A_\mu ^a (x)$ correspond to
gluons.
\section{Perturbation Theory}
When quantizing a theory we want to
preserve the symmetries of the classical theory. I.e.\ we wish to
have the vectors of the Hilbert space on which our field operators act
transform according to a representation of the symmetry group.
E.g.\ for some internal symmetry Lie group with generators
$T^a$ and structure constants $f^{abc}$ one seeks to find a set
of operators $\widehat{Q}^a$ satisfying
\begin{eqnarray}
\lt[ \widehat{Q}^a, \widehat{Q}^b \rt] &=& i f^{abc} \widehat{Q}^c
\label{lieop}
\end{eqnarray}
and
\begin{eqnarray}
e^{-i \epsilon_a \widehat{Q}^a } \widehat{\phi}_k
e^{ i \epsilon_a \widehat{Q}^a } &=&
\lt[ e^{-i \epsilon_a T^a }  \rt]_{kj}  \widehat{\phi}_j . \label{trf}
\end{eqnarray}
Here the second condition implies the consistency between the symmetry
transformation in the Hilbert space (LHS) with the corresponding
transformation of  the fields (RHS).
In \rf{trf} an additional unphysical phase factor
is allowed, too.
Further microcausality requires that
the operators corresponding to the Noether currents
commute, if the distance of their  arguments is spacelike.
All these conditions are fulfilled, if one postulates canonical
commutation relations for boson field operators and canonical
anticommutation relations for fermion field operators:
\begin{eqnarray}
\lt[ \widehat{\phi}_j (x) ,
    \frac{\delta \lag}{\delta \lt(\partial_0 \widehat{\phi }_k \rt) } (y)
  \rt]_{\pm} &=& i \delta^{(3)} (\vec{x}-\vec{y}) \quad
    \mbox{for } \; x^0=y^0 .  \label{canon}
\end{eqnarray}
The operators $\widehat{Q}^a$ are then obtained by simply replacing
the classical fields $\phi_j $ by $\widehat{\phi} _j $ in the expressions
for the classical Noether charges $Q^a$.
Finally for an unbroken symmetry the vaccuum state $\ket{0}$ is chosen
to be invariant with respect to the symmetry transformations.
\subsection{Green's Functions, S--Matrix}\label{gfsect}
In this section we consider for simplicity a  theory with
a single real scalar field $\widehat{\phi} (x) $.
The vacuum--to--vacuum amplitude of a product of $n$ field operators
is called $n$--point Green's function:
\begin{eqnarray}
G_n ( x_1, \ldots x_n ) &=& \bra{0} \T
   \widehat{\phi} (x_1) \cdot \ldots \widehat{\phi }(x_n) \ket{0}
\label{green} .
\end{eqnarray}
Here $\T$ means time ordering of the fields such that
the time coordinates increase from right to left.
For a noninteracting theory all Green's functions can be easily
expressed in terms of the simplest nonvanishing Green's function
\begin{eqnarray}
G_2 ( x_1 , x_2 ) &=&  \frac{1}{(2 \pi)^4} \int d^4 x
           e^{i k \cdot (x_1 -x_2) } \widetilde{G}_2 ( k ) \label{prop}
\end{eqnarray}
with the St\"uckelberg--Feynman propagator
\begin{eqnarray}
\widetilde{G} ( k ) &=& \frac{i}{k^2 -m^2+ i \varepsilon} . \label{freeprop}
\end{eqnarray}
\rf{canon}  implies that the Fourier components of a free field
$\widehat{\phi } (x) $ satisfy the  commutation relations of
annihilation and creation operators. Hence one can use them to create
$n$--particle states from the vacuum  state $\ket{0}$. These $n$--particle
states form the canonical basis of the Fock space $\mathcal{F}$.

For an interacting field theory we neither know the form of
$\widetilde{G}(k)$ in \rf{prop} nor can we relate the $n$--point
function in \rf{green} to the two--point function \rf{prop}.
On the other hand we  empirically know that largely separated
particles essentially behave as free particles.
Further in a scattering experiment the relevant distances
$|\vec{x}_j -\vec{x}_k|$ and $|x_j^0 - x_k^0| $ are large.
The Riemann--Lebesgue lemma states that for
$|x_j^\mu-x_k^\mu | \rightarrow \infty $
only non--integrable singularities of $\widetilde{G}(k)$
can give a contribution
to the integral in \rf{prop}. Hence one assumes
that also for an interacting theory $\widetilde{G}(k)$
develops a pole as in \rf{freeprop}:
\begin{eqnarray}
 \widetilde{G}_{2} (k) &=& \frac{R}{k^2 - m^2_{\mr{pole}}} +
          \mbox{less singular terms } . \label{intprop}
\end{eqnarray}
This makes the concept of asymptotic states and fields plausible:
One assumes that the Hil\-bert space of the interacting field theory
contains some subset of state vectors which for $t \rightarrow -\infty $
develop  into states which
correspond to $n$--particle momentum eigenstates of
the free hamiltonian.
I.e.\ for $t \rightarrow - \infty $ any of these \emph{in--states}\/
tends to an element of the canonical basis of a  Fock space
$\mathcal{F}$. These states describe  the incoming particles
in scattering experiments. The in--states are created and
annihilated by the Fourier components of
a free field operator $\widehat{\phi}_\mr{in}$.
On the subspace  spanned by the in--states
one has
\begin{eqnarray}
\widehat{\phi}(x) &
\begin{array}[t]{c}
\simeq \\[-3pt]
{\scriptstyle t \rightarrow - \infty  }
\end{array}
& R^{1/2} \widehat{\phi}_\mr{in}  (x) \label{wal}
\end{eqnarray}
in the weak sense.
Here one has to allow for a real constant factor $R$, because \rf{canon}
fixes an overall normalization of $\widehat{\phi}_{\mr{in}}$.
Next one postulates a field $\widehat{\phi}_{\mr{out}} (x) $
and a  set of
\emph{out--states}\/ which have the described property for
$t \rightarrow  \infty $ corresponding to the outgoing particles
in the particle collider. It is reasonable to assume
that every particle species which can be used to prepare
the incoming beam can possibly be observed in the final state
and vice versa, thus the out--states span the same Fock space
$\mathcal{F}$ as the in--states.
Lorentz invariance implies the invariance of the vaccuum,
$\ket{0}_\mr{interacting}=
\ket{0}_\mr{in}=  \ket{0}_\mr{out} $,
and of one--particle states,
$\ket{\vec{p}}_\mr{interacting}=
\ket{\vec{p}}_\mr{in}=  \ket{\vec{p}}_\mr{out} $.
Hence with  \rf{wal} one finally obtains
  an interacting two--point function which behaves
asymptotically for $|x_1^0-x_2^0| \rightarrow \infty $
in accordance with \rf{intprop}.

The scattering experiment
is  described by the transition amplitude
or S--matrix element
\begin{eqnarray}
S_{\alpha \beta } &=&\langle \alpha ,\, \mr{out}\,  \ket{\beta ,\, \mr{in}\,}
= \bra{\alpha, \, \mr{in}\,} S \ket{\beta ,\, \mr{in}\, }  \label{s} .
\end{eqnarray}
With the concept of asymptotic fields and states it is now possible
to relate the S--matrix elements to the Green's functions. This
relationship is given by the reduction formula of
Lehmann, Symanzik and Zimmermann (LSZ) \cite{lsz}. It is particularly simple
in terms of the Fourier transformed $(n+l)$--point Green's function
$(2 \pi) ^{4} \delta(\sum_{i=1}^{n+l} k_i ) \cdot
\widetilde{G}_{n+l} (k_1, \ldots, k_{n+l-1} )$:
\begin{eqnarray}
\lefteqn{\bra{p_1,\ldots p_n,\, \mr{out} } q_1,\ldots q_l \, \mr{in}
  \rangle \; = } \nn
&& R^{\frac{n+l}{2}} (2 \pi )^4
\delta ( \sum_{i=1}^{n} p_i - \sum_{i=1}^{l} q_i
) \cdot
\widetilde{G}_{n+l}^t \lt( -p_1, \ldots, -p_n, q_1, \ldots q_{l-1} \rt),
\label{lsz}
\end{eqnarray}
where we have for simplicity assumed that $p_i\neq q_j$ for all $(i,j)$.
In \rf{lsz} the \emph{truncated}\/ Green's function $G^t$ is defined
by
\begin{eqnarray}
\widetilde{G}_{r}^t (k_1, \ldots, k_{r-1} ) &=&
\lim_{k_1 ^2 \rightarrow m^2 } \ldots
\lim_{k_r ^2 \rightarrow m^2 } \lt.
\frac{
\widetilde{G}_{r} (k_1, \ldots, k_{r-1} ) }{\prod_{i=1}^r
      \widetilde{G}_{2} (k_1  ) }
\rt| _{k_r =-\sum_{i=1}^{r-1} k_i } \label{trunc}
\end{eqnarray}
If vector bosons or fermions are scattered the fields in \rf{lsz}
carry indices which are contracted with spinors or polarization vectors
of the external particles.

{}From \rf{lsz} and \rf{trunc} one realizes that Green's functions
carry a lot of redundant information, only their on--shell values
are related to physical observables.

Finally one is left with the evaluation of the Green's
functions $G_n(x_1,\ldots x_n) $. They can be most conveniently
expressed in terms of the path integral:
\begin{eqnarray}
G_n(x_1,\ldots x_n) &=&
\frac{\int \lt[ d \phi \rt] \phi(x_1) \cdots \phi(x_n)
      e^{i S\lt[ \phi \rt] }   }{\int \lt[ d \phi \rt] e^{i S\lt[ \phi \rt] }}
\label{path}
\end{eqnarray}
In perturbation theory
one expresses $\widetilde{G}_{n}$ by a power series in the coupling
constants of the lagrangian. In \rf{path} this is done by expanding
the kernel
\begin{eqnarray}
 e^{i S\lt[ \phi \rt] } &=& e^{i S_\mr{free}\lt[ \phi \rt] }
\sum_n \frac{i^n}{n!} \lt[ \int d^4 x
\lag_\mr{int} \lt[\phi(x) ,\partial_\mu \phi(x) \rt]
\rt]^n  \label{kernel}
\end{eqnarray}
with $\lag_\mr{int} $ being the interaction lagrangian. The individual
terms of this perturbation series are usually depicted in a collection
of \emph{Feynman diagrams}. With \rf{kernel} and \rf{path} one
realizes that one can formally write for the S--matrix in \rf{s}:
\begin{eqnarray}
S &=&
 \T e^{i \int d^4 x \lag_\mr{int} \lt[ \phi(x), \partial_\mu \phi (x) \rt] }
\label{smatrix} .
\end{eqnarray}
Within perturbation theory the assumption
\rf{intprop} is indeed verified: In any finite order of perturbation
theory the interacting two--point function is of the form
given in \rf{intprop}. $m_\mr{pole}$ and the residue $R$ of the
single particle pole are also power series in the couplings.

In gauge theories the path integral \rf{path} includes the
integration over field configurations which are related to each other
by gauge transformations. Further even the free propagator \rf{prop}
of a gauge boson when derived from \rf{ginv} does not exist.
A  remedy is the addition of a gauge fixing term
$\lag_\mr{gf}=- \lt( \partial^\mu A_\mu ^a   \rt)^2/ (2 \xi ) $
with gauge parameter $\xi$ and the introduction of
Faddeev--Popov ghost fields $\eta^a $ via
$\lag_\mr{FP} =  \lt( \partial^\mu \ov{\eta}^a \rt) D_\mu^{ab} \eta^b   $
\cite{fp}. The graphs with FP ghosts correct for the unphysical
polarizations of the gauge fields corresponding to
gauge degrees of freedom. FP ghosts are
scalar Grassmann fields meaning that they anticommute.

Hence the QCD lagrangian suitable for perturbative calculations reads:
\begin{eqnarray}
\lag_\mr{pQCD} &=&
\overline{\psi}_k \lt( i D\! \sla  - m_k  \rt)
\psi _k - \frac{1}{4} F_{\mu \nu}^a F^{a\, \mu \nu}
- \frac{1}{2 \xi} \lt( \partial^\mu A_\mu ^a   \rt)^2  +
\lt( \partial^\mu \ov{\eta}^a \rt)  D_\mu^{ab} \eta^b
\label{pqcd}
\end{eqnarray}
One terms quantities which do not depend on the arbitrary gauge
parameter $\xi$ \emph{gauge independent}. S--matrix elements and
physical observables are gauge--independent. In contrast a
field polynomial or Green's function is called
\emph{gauge invariant}, if it is invariant with respect to
the gauge transformation \rf{gaugeferm} and \rf{gaugevec}.

After all in the case of QCD the standard reduction procedure sketched
above does not seem very useful, because in the very beginning
the assumption \rf{intprop} is in contradiction with the observed
confinement property. The poles in the Green's functions are not
related to the particles described by the elementary  fields in the
lagrangian as suggested by
\rf{lsz}, but rather by bound states, hadrons, formed by them.
The asymptotic states are hadron states which we cannot simply
relate to quark and gluon fields. We may then ask  how to make use of
perturbative QCD at all.
To answer this questions  one has to mention  a further property of
perturbative quantum field theory discovered by Wilson \cite{w}:
It is possible to expand a Green's function in terms of some
ratio of mass parameters and each term of this expansion
is the product of a Green's function involving a composite operator
and a coefficient being independent of the external states.
This \emph{Wilson coefficient} comprises the short distance physics
whose scale is set by the inverse heavy  mass parameter.
Further QCD enjoys the feature of asymptotic freedom ensuring that the
QCD coupling is small when probed at small distances.
Hence for the calculation of the Wilson coefficients one can use
perturbation theory and can further take external quark states
rather than hadron states.

Sometimes people try to describe hadron properties in terms of
off--shell quark Green's functions. They, however, are unphysical,
as mentioned in the discussion after \rf{lsz}, e.g.\ they are
gauge dependent. This point will be relevant in the discussion
of the use of the field equation of motion in sect.\ \ref{eom}.
\subsection{BRS Invariance}\label{brs}
The lagrangian \rf{pqcd} is no more gauge invariant due to the
gauge--fixing term and the Faddeev--Popov term. It seems as if we
have lost all the nice features of local gauge symmetry.
This is not so, because  \rf{pqcd} is still invariant under
\emph{Becchi--Rouet--Stora} (BRS) transformations \cite{brs}.
They are constructed by factorizing the gauge parameter
in \rf{gaugeferm} and \rf{gaugevec} into the
product of the FP field  and a Grassmann parameter
$\delta \lambda $:
\begin{eqnarray}
\delta \theta ^a (x) = \eta^a (x) \delta \lambda  . \no
\end{eqnarray}
The infinitesimal transformations of the various fields read:
\begin{eqnarray}
\dbrs \psi      &=& - i g (\eta^a \delta \lambda ) T^a \psi
              \; = \; i g  T^a \eta^a \psi   \delta \lambda  , \nn
\dbrs \ov{\psi} &=& i g \ov{\psi} T^a \eta^a \delta \lambda , \nn
\dbrs A_\mu^a   &=& \lt( \partial_\mu \eta^a +g f^{abc} \eta^b A_\mu^c  \rt)
                     \delta \lambda \nn
\dbrs \eta^a &=&  - \frac{1}{2} g f^{abc} \eta ^b \eta ^c
       \delta \lambda \nn
\dbrs \ov{\eta} ^a &=& \frac{1}{\xi }
      \partial^\mu A_\mu ^a  \delta \lambda  . \label{brst}
\end{eqnarray}
The basic QCD lagrangian \rf{qcd} is clearly BRS invariant, because
for the quark and gluon
fields the BRS transformation is just a gauge transformation.
The ghost fields transform such that
$\lag_\mr{gf} + \lag_\mr{FP}$ is invariant.
The BRS symmetry is a global supersymmetry, because it involves an
$x$--independent anticommuting parameter $\delta \lambda$. Its magic
is that it encodes the original local gauge symmetry which had to
be destroyed by adding $\lag_\mr{gf}+ \lag_\mr{FP}$ to $\lag_\mr{QCD}$.

A global symmetry of $\lag $ induces relations between Green's functions
called \emph{Ward--Takahashi identities} \cite{wt}. The
Ward--\-Ta\-ka\-ha\-shi
identities implied by BRS invariance are \emph{Slavnov--Taylor}\/
identities \cite{st}. We will need them in the form
\begin{eqnarray}
\dbrs \bra{0} \T X Y \ket{0}  &=& \bra{0} \T \lt( \dbrs X \rt) Y
\ket{0} +   \bra{0} \T X \lt( \dbrs Y \rt)  \ket{0} \; =\;0 ,\label{slav}
\end{eqnarray}
where $X$ and $Y$ are product of fields whose arguments may coincide.
When making use of \rf{slav} one has to anticommute $\delta \lambda $
to the right (or to the left) thereby picking up
a relative ``--''--sign between the terms  involving
$\dbrs X$ and $\dbrs Y$, if they contain an odd number of Dirac
fields.

\subsection{Renormalization}\label{reno}
The perturbative evaluation of Green's functions involves
integrations over loop momenta. Some of these integrals
diverge when the loop  momenta get large, they are
\emph{UV divergent}.
To give a meaning to these
divergent expressions one first has to
regularize the divergence:
An additional parameter $D$ is introduced in the lagrangian
such that the latter coincides for $D=D_0$ with the original
lagrangian. The loop integrals are
finite in some range for $D$ and analytic functions in $D$.
For $D\rightarrow D_0$ one rediscovers the divergence in the
form of some singularity. Throughout this paper we use
dimensional regularization which conserves most of the interesting
symmetries. Here $D$ is the space--time dimension and $D_0=4$.
The appearance of UV divergences originates in the fact that
quantum field operators are singular distributions, see \rf{canon}.
In general  their product is ill-defined when their arguments
coincide. The renormalization process gives a meaning to
the field products in the interaction lagrangian.

The renormalization process is carried out recursively in the
power of the coupling constant $g$  or equivalently in the number
of loops. It is encoded in  \emph{Zimmermann's forest formula}
\cite{zff}. Starting from the one--loop order one finds that one can
absorb all divergences into a  multiplicative
redefinition of the parameters
of  the lagrangian such as masses, couplings and the normalization
of the fields which is connected to the quantity $R$ in \rf{wal}.
These multiplicative renormalization constants are
Laurent series in $D-4$ in dimensional regularization.
In the n--th order  all divergences of 1-- to (n--1)--loop
subdiagrams are already remedied by the earlier renormalization
steps. The remaining overall divergence can  then be absorbed into
new n--th order terms in the renormalization constants. For this
to work it is crucial that after the subtraction of subdivergences
the divergences are local, i.e.\ polynomial in the external momenta
of the Green's  function under consideration. Further the renormalization
constants must be universal meaning
they must remove the divergence of some subloop in whatever diagram
it occurs.

Yet for a predicative theory we must also ensure that the
renormalization process does not induce new couplings in any new order
of perturbation theory, because then we would end up with a theory
with infinitely many undetermined parameters. This can be ensured,
if the lagrangian contains only couplings with non--negative
dimension. In this case we term the theory to be
\emph{renormalizable by power counting}.
This criterion applies to the Standard Model.

In the following we will also be confronted with \emph{effective}
field theories which contain couplings with negative dimension.
The parameters of such effective field theories, however, will be
fixed by a \emph{matching}\/ procedure in which
the effective theory Green's functions are
related to Standard Model Green's functions, so that the theory
is predicative.

Since we will always work to lowest nonvanishing order in the weak
coupling constant, we will only have to discuss renormalization
of the QCD lagrangian. First we define
\begin{eqnarray}
\e &=& \frac{4-D}{2} . \no
\end{eqnarray}
The QCD Lagrangian now reads in terms of the unrenormalized
(\emph{bare}) parameters:
\begin{eqnarray}
\lag_\mr{pQCD} &=&
- \frac{1}{4}
 \lt( \partial_\rho A_\nu^{(0),a} - \partial_\nu A_\rho^{(0),a} \rt)
 \lt( \partial^\rho A^{(0),\nu \, a} - \partial^\nu A^{(0),\rho \,a}
 \rt) \nn
&&
- \frac{1}{ 2 \xi^{(0)} } \lt( \partial^\rho A^{(0),a}_\rho    \rt)^2
+ \overline{\psi}^{(0)}_k \lt( i \partial  \sla
+ g^{(0)} T^a A\sla ^{(0),a}  - m^{(0)}_k  \rt)\psi^{(0)} _k
\nn &&
+ \lt( \partial^\rho \ov{\eta}^{(0),a} \rt)
\lt( \partial _\rho  \delta ^{ab} -g^{(0)} f^{abc} A_\rho^c \rt)
\eta^{(0),b}
\label{bareqcd}
\end{eqnarray}
The bare fields and parameters are related to the renormalized ones
by
\begin{eqnarray}
&&A_\rho^{(0),a} = Z_A^{1/2} A_\rho^a , \quad\quad\quad
\psi^{(0)} = Z_\psi^ {1/2} \psi , \quad\quad\quad
\eta^a = Z_\eta^{1/2} \eta^a , \nn
&& g^{(0)}  = Z_g g \mu^{ \e},\quad\quad\quad
\xi^{(0)} = Z_A \xi ,\quad\quad\quad
m^{(0)} = Z_m m  .\label{renpar}
\end{eqnarray}
Here an arbitrary  dimension--1  parameter, the
\emph{renormalization scale $\mu$} has been introduced to keep  the
renormalized coupling constant $g$ dimensionless.
In dimensional regularization the
$n$--loop diagrams involve poles in $\e$ up to order $n$. Hence the
renormalization constants are of the
form:
\begin{eqnarray}
Z &=& 1+ \sum _{j} \lt( \frac{g^2 }{16 \pi ^2} \rt)^j  Z ^{(j)} ,
\quad \quad
Z^{(j)} \, = \,
        \sum _{k=0}^{j} \frac{1}{\varepsilon ^k} Z ^{(j)} _{k}
\label{exp}.
\end{eqnarray}

Now the renormalized lagrangian reads:
\begin{eqnarray}
\lag_\mr{pQCD} &=&
- \frac{1}{4} Z_A
 \lt( \partial_\rho A_\nu^{a} - \partial_\nu A_\rho^{a} \rt)
 \lt( \partial^\rho A^{\nu \,a} - \partial^\nu A^{\rho \,a} \rt)
- \frac{1}{ 2 \xi  } \lt( \partial^\rho A^{a}_\rho    \rt)^2  \nn
&&    +
Z_\psi \ov{\psi}_k i \partial \sla \psi_k
- Z_\psi Z_m m_k \ov{\psi}_k  \psi_k
+ Z_\eta
\partial^\rho \ov{\eta}^{a}
\partial _\rho  \eta^{a}  \nn
&&
- Z_g Z_A^{3/2} \frac{g}{2} \mu^{ \e} f^{abc}
  \lt( \partial_\rho A_\nu^a - \partial_\nu A_\rho^a  \rt)
  A^{b\, \rho} A^{c \, \nu}
+
Z_g Z_\psi Z_A^{1/2} g \mu^{ \e} \ov{\psi} A\sla ^a  T^a \psi \nn
&&
-
Z_g^2 Z_A^2 \frac{g^2}{4} \mu^{2 \e} f^{abe} f^{cde} A_\rho^a  A_\nu^b
A^{c \rho }   A^{d \nu }
- Z_g Z_\eta Z_A^{1/2} g \mu ^{ \e} f^{abc} \lt( \partial^\rho
\ov{\eta }^a \rt) A_\rho ^c \eta^b \label{renqcd}
\end{eqnarray}
Sometimes it is useful to decompose $\lag_\mr{pQCD}$ in \rf{renqcd} into
a sum $\lag_0 + \lag_\mr{ct}$, where $\lag_0$ has the form of
\rf{bareqcd} with    the bare quantities replaced by
renormalized ones.  $\lag_\mr{ct}$ is called
\emph{counterterm lagrangian}.

Gauge invariance has implied that all interaction terms
in $\lag_\mr{pQCD}$ involve the same coupling constant $g$.
Dimensional regularization has the advantage that it
manifestly preserves gauge invariance  in the unrenormalized
Green's functions.
As shown by 't Hooft
the subsequent renormalization  procedure does not spoil gauge
invariance,
so that $Z_g$ is the same in all these interaction terms \cite{h}.
Finally also renormalized Green's functions obey the
Slavnov--Taylor identities, but with a renormalized parameter
\begin{eqnarray}
\delta \lambda_r &=& Z_A^{-1/2} Z_\eta ^{-1/2} \delta \lambda .
\label{brstren}
\end{eqnarray}
\cleardoublepage
\chapter{Renormalization Group and Operator Product Expansion
         }\label{rg}
\section{Renormalization Schemes}\label{secrs}
When determining the renormalization constants in \rf{renqcd}
one  calculates Green's functions obtained from
$\lag_0+\lag_\mr{ct}$ and adjusts the renormalization constants
such that the result is finite. The Green's function to
some fixed order in $g$ involves
$n$--loop diagrams with interaction vertices only from $\lag_0$ and
subloop counterterm diagrams with fewer loops and
with vertices also from $\lag_\mr{ct}$.
The individual contributions contain divergent terms which
depend non--polynomial on the external momenta $p_i$, but they
cancel in the sum of every $n$--loop diagram and its sub--loop
counterterm diagrams \cite{zff}. Since these \emph{non--local poles}\/
cancel, one can absorb the remaining divergences into $n$--loop
counterterms.

Now one is free to subtract any momentum--independent
constant $Z_0^{(n)}$ together with the divergences
 order by order  in perturbation theory. Any such choice of
the $Z_0^{(n)}$'s in \rf{exp}
corresponds to a different renormalization scheme.
In the  \emph{minimal subtraction}\/ scheme (MS)  \cite{hms}
one picks $Z_0^{(n)}=0$. All schemes in which the
 $Z_0^{(n)}$'s do not depend on the masses are called
\emph{mass independent}. Due to \emph{Weinberg's theorem}\/
\cite{wei} they enjoy the property that also the divergent
parts of the $Z^{(n)}$'s are independent of the masses
(see e.g.\ \cite{coll}).
In other words in mass independent schemes the divergences
are not only polynomial in the momenta but also in the masses.
The main advantage of these schemes, however, is that they
allow for the solution of the renormalization group equations
to be discussed in sect.\ \ref{secrg}.

Since in every renormalization scheme the divergences are removed,
any two schemes can at most differ by a finite renormalization,
for example an arbitrary scheme is related to the MS scheme via
\begin{eqnarray}
Z&=& Z^\mr{MS} \lt[ 1+ Z_0^{(1)} \frac{g^2}{16 \pi^2}
      + Z_0^{(2)} \frac{g^4}{(16 \pi^2)^2} \ldots \rt] .
\label{finren}
\end{eqnarray}
{}From this one sees that the $Z_n^{(n)}$'s are scheme--independent while
the other Laurent coefficients in \rf{exp}  depend on the
lower order $Z_0^{(k)}$'s.
In fact the $Z_n^{(n)}$'s are simply related to the one--loop
$Z_1^{(1)}$'s.
The finite renormalization
$\lt[ 1+ Z_0^{(1)} g^2/(16\pi^2)
+ Z_0^{(2)} g^4/(16\pi^2)^2 \ldots \rt] $ in
\rf{finren}
may be  viewed  as a perturbative redefinition of the fields, masses
and coupling constants in $\lag$ in \rf{renqcd}.

In dimensional regularization all loop integrals involve an additive
term $\g_\mr{E} -\ln (4 \pi) $
with $\euga$ being Euler's constant.
It is reasonable to subtract this term
in every order of perturbation theory. The resulting scheme is called
\msb scheme \cite{bbdm}.
There is a simple way to relate  the \msb scheme to the
MS scheme: One first
replaces the renormalization scale $\mu$ in \rf{renpar} by
\begin{eqnarray}
\mub &=& \mu \lt( \frac{ e^\euga }{4 \pi} \rt)^{1/2} . \label{mstomsb}
\end{eqnarray}
Recalling that $\mu$ resp.\ $\mub $ has been introduced into
\rf{renqcd} to maintain integer dimensions for the renormalized
parameters one realizes that each loop integral $\int d^D q $ comes
with a factor of $\mub^{2 \e}$. Then one expands \rf{mstomsb}
to the desired order in $\e$ and multiplies this power series with
the Laurent series in $\e$ of the unrenormalized diagrams. In this way
one modifies the pole part of the unrenormalized Green's functions
such that  $\g _E$ and $\ln (4 \pi)$ vanishes. Finally one simply
applies minimal subtraction to the remaining expression.
This method, which will be used in this thesis,
has the advantage that one can avoid  Euler's constant
and  $\ln (4 \pi)$ in intermediate steps of the calculation. Further
one can easily translate relations proven in the MS scheme
 into the \msb scheme.

Yet there is another aspect of \rf{mstomsb}: When inserted into
\rf{renqcd} one may view \rf{mstomsb}  as a redefinition of
$g$   by a $g$--independent power series in $\e$,
$1+\sum_{k=1} a_k \e^k$,
or alternatively as a redefinition
of the field monomials multiplied by $g$.
Here in \rf{mstomsb} this series reads
\begin{eqnarray}
\lt( \frac{ e^\euga }{4 \pi} \rt)^{\e /2} &=&
1 + \frac{1}{2} \e \lt( \euga - \ln ( 4 \pi)   \rt) + \ldots .
\label{msbpower}
\end{eqnarray}
Clearly multiplying with such a power series does not commute
with the  renormalization process, so that it corresponds to
a change of the renormalization scheme.
Let us call this procedure  \emph{evanescent redefinition}\/ of the
coupling or of the field monomial. Such evanescent redefinitions
appear naturally in the context of effective four--fermion couplings.
We will investigate them in chapter \ref{eva} \cite{hn2}.
One task will be to translate an evanescent redefinition
into a finite renormalization \rf{finren}.
We finally remark that the change from the MS to the \msb scheme
is equivalent to a change of the scale $\mu$ as evidenced from
\rf{mstomsb}.

Any S--matrix element and any physical  observable is independent
of the chosen renormalization scheme to the calculated order.
This means, if one calculates some observable in scheme I to order
$g_\mr{I}^m$ and afterwards expresses $g_\mr{I}$ and the other parameters
such as  masses in
terms of the coupling $g_\mr{II}$ and the parameters of some
other scheme II,  the result differs at most by terms of order
$g_\mr{II}^{m+1}$ from the calculation of the observable in scheme II.

\section[Renormalization Group, $\mr{\Lambda_{QCD}}$]{Renormalization Group,
$\mr{\mathbf{\Lambda_{QCD}}}$}\label{secrg}
The invariance of the S--matrix with respect to the change of the
renormalization prescription implies a continous symmetry of the quantized
theory as noticed first by St\"uckelberg and Petermann \cite{sp}.
The group associated to this  symmetry is the \emph{renormalization
group}\/ (RG). Its power roots in the fact that it links different orders
of the perturbative series: By applying a renormalization group
transformation to some $n$--th order Green's function we can add
some logarithmic term of all uncalculated higher orders
to the result.
When the RG is applied properly,
these logarithms are the dominant contribution
of the higher orders and they are summed to all orders by the RG
transformation.

In mass independent schemes renormalized Green's functions depend
on the renormalization scale  $\mu $. Changing $\mu $ corresponds to
a very simple way of modifying the renormalization prescription
as discussed after \rf{mstomsb}.
The S--matrix is independent of $\mu $, if the bare quantities in
\rf{bareqcd} are chosen $\mu $--independent. Hence the renormalization
scale invariance is a special case of the renormalization scheme
invariance.

\subsection{Renormalization Group Functions in QCD}
In this section we will collect the important RG functions of QCD
in the \msb scheme (see e.g.\ \cite{muta}).
{}From the $\mu $--independence of the unrenormalized parameters in
\rf{renpar} one easily finds:
\begin{eqnarray}
\mu \frac{d g}{d \mu} &=& \beta (g (\mu )) ,
\quad \quad \quad \mu \frac{d m}{d \mu} \; = \; - m \g_m (g(\mu))
\label{rendef}
\end{eqnarray}
with the \emph{beta--function}
\begin{eqnarray}
\beta (g (\mu)) &=& - \e g - \frac{\mu}{Z_g}  \frac{d Z_g}{d \mu} g
\end{eqnarray}
and the \emph{anomalous dimensions} of quark mass and quark field
\begin{eqnarray}
\g_m (g(\mu))&=& \frac{\mu}{Z_m }  \frac{d Z_m }{d \mu} ,
\quad
\g_\psi (g(\mu))
\; = \; \frac{\mu}{Z_\psi ^{1/2}}  \frac{d Z_\psi ^{1/2}}{d \mu} .
\end{eqnarray}
$m$ in \rf{rendef} is the \emph{current mass}\/ appearing in the
lagrangian \rf{renqcd}. It is also called \emph{running}\/
mass, because it depends on $\mu$.
The renormalization group functions in mass independent schemes
depend on $\mu$ only through $g(\mu) $ as indicated in
\rf{rendef}. This feature allows to solve the renormalization group
equation, which reads  for a renormalized truncated
connected n--quark Green's function:
\begin{eqnarray}
\dmu \lt[ Z_\psi^{-n/2} \lt( g(\mu ) \rt)
    \widetilde{G}_n^{tc} \lt( p_i, g(\mu), m(g(\mu)),\mu  \rt)     \rt]
&=& 0 . \label{rggf}
\end{eqnarray}
Here $\dmu $ means
\begin{eqnarray}
\dmu &=& \pmu + \beta \prt{g} - \g_m m \prt{m} - n \g_\psi ,
\label{diffmu}
\end{eqnarray}
where the partial derivatives are performed with the bare quantities
kept fixed.

Let us briefly sketch the relation of \rf{rggf} to the
scale transformations alluded to in sect.\ \ref{sctrf}:
Recalling that $\mu$ has been introduced to maintain an integer
dimension $d$ for $\widetilde{G}_n^{tc}$ one easily finds
(see e.g\ \cite[p.\ 199]{muta}):
\begin{eqnarray}
\lt( \pmu +m \prt{m} + \lambda \prt{\lambda} +d  \rt)
\widetilde{G}_n^{tc}
\lt( \lambda p_i, g(\mu), m(g(\mu)),\mu  \rt) &=& 0, \no
\end{eqnarray}
which may be used to eliminate $\pmu$ in \rf{rggf} in favour of the
scaling factor $\lambda$ of the external momenta.

Now $\mu$ enters Green's functions only in powers of $\e$, thus after
expanding in $\e$ only in powers of logarithms.
Since $\beta(g) = O(g^3)$, $\g_\psi= O(g^2)$, $\g_m = O(g^2)$
and $\pmu = \prt{\ln \mu} $,
\rf{rggf} connects
the $\ln \mu $'s
of different orders in $g$ of $G_n^{tc}$ to each other.

We will need the renormalization group functions up to
the next--to leading order (NLO) \cite{muta}.
The beta--function reads:
\begin{eqnarray}
\beta (g) &=& -\beta _0^{(f)}  \frac{g^3}{16 \pi^2} -\beta _1^{(f)}
              \frac{g^5}{\lt( 16 \pi^2 \rt)^2} + O(g^7) \label{betaseries}
\end{eqnarray}
with
\begin{eqnarray}
\beta _0^{(f)} &=& \frac{11 N -2 f}{3}, \quad \quad
 \beta _1^{(f)} \; = \; \frac{34}{3} N^2 - \frac{10}{3} N f -2 C_F f
\label{beta}.
\end{eqnarray}
Here $N$ is the number of colours,
$ C_F = (N^2-1)/(2 N) $ and
$f$ is the number of quark flavours.
For the anomalous mass dimension one finds:
\begin{eqnarray}
\gamma _m (g) &=& \gamma _m^{(0)} \frac{g^2}{16 \pi^2} +
              \gamma _m^{(1)(f)} \lt( \frac{g^2}{16 \pi^2} \rt) ^2 +
              O \lt( g ^6  \rt) \no
\end{eqnarray}
with
\begin{eqnarray}
\gamma_m^{(0)}&=& 6 C_F , \quad \quad
\gamma _m^{(1)(f)} \; = \; C_F \lt( 3 C_F +
                  \frac{97}{3} N - \frac{10}{3} f \rt)
\label{gammam}.
\end{eqnarray}
Whenever we discuss leading order (LO) expressions, only the first
term is kept in \rf{betaseries} and \rf{gammam}.
The anomalous dimension of the quark field will only be needed in
the leading order:
\begin{eqnarray}
\g_\psi &=& \g_\psi^{(0)} \frac{g^2}{16 \pi^2} +
            O(g^4 )
\end{eqnarray}
with
\begin{eqnarray}
\g_\psi^{(0)} &=& C_F \xi . \label{gammapsi}
\end{eqnarray}
With these definitions it is easy to write down the solutions of the
RG equations \rf{rendef} for $\alpha(\mu)= g^2 (\mu) /(4 \pi)$:
\begin{eqnarray}
\frac{\alpha ( \mu  ) }{4 \pi } &=&
       \frac{1}{\beta_0^{(f)}\log(\mu^2/\Lambda_f^2)} -
                \frac{\beta_1^{(f)} \log \lt[ \log(\mu^2/\Lambda_f^2) \rt]}{
                 \lt( \beta_0^{(f)} \rt)^3 \log ^2 (\mu^2/\Lambda_f^2)}
      + O \lt( \frac{ \log^2[\ldots] }{
                        \log^3 (\ldots ) } \rt) . \label{runalpha}
\end{eqnarray}
The mass parameter $\Lambda_f$ has entered \rf{runalpha} as a constant
of integration. Hence the massless coupling $g$ of the classical
lagrangian  has
become a function  of the dimension--one parameter $\Lambda_f$ in the
quantized theory.
\rf{runalpha} also exhibits the property of asymptotic freedom,
$\alpha(\mu) \rightarrow 0$ for $\mu \rightarrow \infty$.
By inverting \rf{runalpha} one can define
$\Lambda_f $. For $\alpha$ defined in the \msb scheme one has
\begin{eqnarray}
\laMSb &=& \mu \lt( \beta_0^{(4)} \frac{\alpha(\mu) }{4 \pi}
               \rt)^{\frac{-\beta_1^{(4)}}{2 \lt( \beta_0^{(4)}\rt) ^2}}
       \lt[ e^{- \frac{2 \pi}{\beta_0^{(4)} \alpha( \mu) } }
        +O \lt (\alpha (\mu) \rt)     \rt] . \label{lbmsb}
\end{eqnarray}
In the limit $\mu \rightarrow \infty$   this
is a precise definition of $\laMSb$. Since the NLO coefficient
$\beta_1$ appears in an overall factor rather than in a radiative
correction, one realizes that it is mandatory to go to the NLO to
define $\laMSb$. Conversely only in NLO expressions for some
physical observable of interest one can make use of $\laMSb$.
This is an important reason to perform perturbative calculations
beyond the leading order. $\Lambda_f$ has been defined in the
MS scheme by Buras, Floratos, Ross and Sachrajda \cite{bfrs}
and the \msb scheme has been introduced by Bardeen, Buras, Duke and
Muta \cite{bbdm}.

Solving the RG equation \rf{rendef} for the running quark mass yields
for the mass at some scale $\mu$ expressed in terms of
$m ( \mu = m )$:
\begin{eqnarray}
m (\mu)&=& m ( m )
\lt( \frac{\alpha (\mu)}{\alpha (m)} \rt)^{d_m^{(f)}}
\lt( 1+ \frac{\alpha (m) - \alpha (\mu)}{4 \pi} J_m^{(f)} \rt).  \label{rum}
\end{eqnarray}
In \rf{rum} $d_m^{(f)}$ and  $J_m^{(f)}$ are defined by
\begin{eqnarray}
d_m^{(f)}  &=& \frac{\gamma_m^{(0)}}{2 \beta_0^{(f)}} \quad \mbox{and } \quad
J_m^{(f)} \; = \;
 \frac{-\gamma_m^{(1)(f)}+2 \beta_1^{(f)} d_m^{(f)}}{2 \beta_0^{(f)}} \no
\end{eqnarray}

If $\widetilde{G}_n^{tc}$
only contains a single mass parameter apart from $\mu$,
for example  some spacelike external momentum $\sqrt{-q^2}$,
$\ln \mu^2$ must necessarily appear in the form $\ln ( -q^2 /\mu^2)$
To make sense of a perturbative calculation we must choose
$\mu^2 \approx -q^2$ in order to avoid a large logarithm multiplying
the expansion parameter $\alpha (\mu)$. One may solve
\rf{rggf} to obtain $\widetilde{G}_n^{tc}$  at any other scale
$\mu$. This RG improved expression differs from a standard perturbative
result such that it includes the terms involving
$\alpha (\mu) \ln(-q^2/\mu^2 )$
to all orders in perturbation theory. The LO RG improved perturbation
theory is therefore called \emph{leading log}\/ (LL) approximation.

\section[A First Look at \kkM]{A First Look
 at $\mathbf{K^0\!-\!\ov{K^0}}\,$--Mixing}\label{firstlook}
Yet in most physical processes more than one mass scale is involved.
To make use of RG improved perturbation theory more field theoretic
tools are needed. To motivate the following sections let us have a
first look at the main subject of this thesis, the \dst  transition
inducing \kkm.

Fig.\ \ref{box} on p.\ \pageref{box}
shows the lowest order $\mr{\Delta S \!=\! -2}$
transition amplitude in the Standard Model. One--gluon radiative
corrections are depicted in fig.\ \ref{boxqcd} on p.\
\pageref{boxqcd}. Now this  amplitude is not useful to describe
physical \dst processes for the following reasons:
\begin{itemize}
\item[i)] The diagrams of fig.\ \ref{boxqcd} incorporate QCD
corrections perturbatively. While this is appropriate for short
distance effects, the long distance QCD interaction is
non--perturbative and cannot be described by the exchange of gluons.
\item[ii)] There are largely separated mass scales involved:
The radiative corrections of fig.\ \ref{boxqcd} contain
large logarithms such as $\alpha \ln (m_c/M_W)=O(1)$, so that the
radiative corrections have the same order of magnitude as the leading
term. One would like to sum this logarithm to all orders in
perturbation theory.
RG techniques applied to the Standard Model amplitude, however,
will not achieve this, because with them only logarithms of
$\mu $ can be summed to all orders.
\item[iii)] For the same reason it is not clear at which scale the
running coupling should be evaluated. Between $\mu=m_c$ and
$\mu =M_W $ the coupling $\alpha (\mu)$ changes roughly by a factor of
three.
\item[iv)] The real external states are mesons rather than on--shell
      quarks.  Off--shell quarks are an even worse description
       of mesons,
      because off--shell  amplitudes are unphysical as mentioned
      in the end of sect.\  \ref{gfsect}.
\end{itemize}
Hence to tackle the problem one wishes to separate short distance effects
from the long distance physics. This separation
is provided by Wilson's operator
product expansion \cite{w} discussed in the following section.

\section{Operator Product Expansion}
In weak processes involving hadrons one is confronted with two couplings:
While ordinary perturbation theory in the weak coupling
$g_w$ works well, we have already seen in sect.\ \ref{firstlook}
that we need a framework to include all--order effects in the
strong coupling $g$ and possibly  non--perturbative features of QCD.

For this we expand the kernel of the path integral
$\exp \lt[ i \int \lag_\mr{SM} \rt]$ in
terms of the electroweak (and Higgs)
interaction lagrangian, but keep the QCD
interaction in the exponential. I.e.\ we expand as in  \rf{kernel},
but with the QCD action added to
$S_\mr{free}$ instead of being kept in
$\lag _\mr{int}$. Hence
with \rf{ccint} a four--quark Green's function corresponding
to a weak charged current tree level process is found as
\begin{eqnarray}
G_4^{(2)\, tc} (x_1,x_2,x_3,x_4  )\!\! & = &
\frac{g_w^2}{8} \;
\bra{0} \T \ov{\psi}_1 (x_1)  \ov{\psi}_2 (x_2) \psi _3 (x_3)
\psi _4 (x_4)   \nn
&&\!\! \!\! \!\! \lt[ - \frac{1}{2!}\! \int\! d^D y \!\int\! d^D z
J_{cc}^{\dagger \, \mu}(y) W_\mu^{+} (y)
J_{cc}^{ \nu}(z) W_\nu^{-} (z) +h.c. \, \rt] \ket{0} ^{tc}.  \label{fqw}
\end{eqnarray}
Here the quark fields are free fields with respect to the electroweak
interaction, but interacting fields with respect to QCD. \rf{fqw}
represents all diagrams with one  W--boson exchange
and an arbitrary number of gluons.
The index 2 in $G_4^{(2)\, tc}$
means second order in the weak coupling constant.
The weak current $J_{cc}^{\nu}$ is the first example of a
\emph{composite operator}, as it contains interacting fields at the
same space--time point.

Now Wilson has shown that one can expand a product of field operators
in a series of composite operators with increasing dimension.
For the product of the two currents in \rf{fqw} this
reads\footnote{sum on repeated indices}:
\begin{eqnarray}
J_{cc}^{\dagger \, \mu}(y) J_{cc}^{ \nu}(z)&=&
\widehat{C}_j^{\mu \nu} (y-z) \widehat{Q}_j
\lt( \frac{y+z}{2} \rt) , \label{opej}
\end{eqnarray}
with the $\widehat{Q}_j$'s being local four--quark operators as depicted in
fig.\ \ref{ds1cclo} on p.\ \pageref{ds1cclo}. The Green's functions
on the RHS of the
\emph{operator product expansion}\/ (OPE)
\rf{opej} are the \emph{matrix elements}\/ of $\widehat{Q}_j$.
The
\emph{Wilson coefficients} $\widehat{C}_j^{\mu \nu}$ are functions
which are
singular at $y=z$. The coefficient multiplying the operator with the lowest
dimension is the most singular one by power counting. They comprise
the short distance behaviour of the current product in \rf{opej}
when $y$ approaches $z$.
After inserting \rf{opej} into \rf{fqw} one performs the path
integration over the
W--boson fields. By this the open Lorentz indices in
\rf{opej}   are contracted with a W propagator
$i \mathcal{D}^{(W)}_{\mu \nu} (y-z, M_W, \mu)  $. With
\begin{eqnarray}
C_j(M_W,\mu) &=& - \frac{g_W^2}{8}
 \int d^D \zeta \widehat{C}_j ^{\mu \nu} (\zeta)
        \mathcal{D}^{(W)}_{\mu \nu} ( \zeta , M_W, \mu) \label{wic}
\end{eqnarray}
one realizes that \rf{opej} turns into an expansion in the (inverse)
W mass $M_W$.
With \rf{wic} one finally obtains for \rf{fqw}:
\begin{eqnarray}
\lefteqn{G_4^{(2)\, tc} (x_1,x_2,x_3,x_4  ) \;=}  \nn
&&C_j(M_W, \mu ) \bra{0} \T \ov{\psi}_1 (x_1) \ov{\psi}_2 (x_2)
\psi_3 (x_3) \psi_4 (x_4)\,
i\!\! \int \!\!d^D y \widehat{Q}_j (y)  \ket{0}^{\, tc} .
\label{opeeff}
\end{eqnarray}
Here we have absorbed the weak coupling constant into $C_j$. When we
pass to the operator basis for \kkm\/ in chapter \ref{operator}
we will extract a factor of
\begin{eqnarray}
-\frac{G_F}{\sqrt{2}} &=& -\frac{g_W^2}{8 \mw}
\label{fermi}
\end{eqnarray}
out of $C_j$ to show that the operator product expansion in weak
decays replaces the Standard Model
weak interaction by a four--fermion interaction with
the Fermi constant $G_F$ as the coupling constant.
We will also sometimes leave out the integral over the space--time
coordinate $y$ of the composite operator in \rf{opeeff}. This merely
leaves out an overall momentum conserving $\delta$--function when
\rf{opeeff} is Fourier transformed into momentum space.

In momentum space \rf{opeeff}
corresponds to an expansion in $m^2_\mr{light}/\mw$, where
$m_\mr{light}$ stands collectively for the
light\footnote{lighter than $M_W$} quark masses and
the small external momenta and Mandelstam variables. The local
operators in \rf{opej}  may be obtained by expanding the momentum
space W propagator\footnote{We use the Feynman rules of \cite{muta}.}:
\begin{eqnarray}
\frac{-1}{\mw - p^2} & = & \frac{-1}{\mw} + \frac{-p^2}{M_W^4}
+ \ldots .  \label{momexp}
\end{eqnarray}
While it is trivial from \rf{momexp}
that \rf{opeeff} works for tree level diagrams with
$C_j^{(0)}= g_W^2/8$, it is not obvious that \rf{opeeff} can be
extended to diagrams with gluon loops, in which $p$ in \rf{momexp}
is a loop momentum. From Wilson's work we know that \rf{opej} holds
to any order in perturbation theory. Clearly for small loop
momenta \rf{momexp} is a good expansion, and the contributions from
small loop momenta on each side of \rf{opeeff} will match with the
tree level coefficient $C_j^{(0)}$. The contributions from large loop
momenta, however, will modify $C_j$ by adding terms
$\alpha/(4 \pi) \cdot C_j^{(1)}$\ldots  to $C_j^{(0)}$. Yet these UV momenta
are not sensitive to the small internal
masses and small external momenta.

When the heavy top quark is involved in the amplitude of interest,
it must also be integrated out like the W--boson.

In the discussion above  we have made plausible
the following properties of the operator product expansion:
\begin{itemize}
\item[i)] The $C_j$'s do not depend on the infrared structure of
         the Standard Model amplitude.
\item[ii)] The Wilson coefficients  $C_j$ depend only on the heavy
       mass scales $M_W$ and $m_\mr{top}$ and on $\mu$, while
       the matrix elements contain the light masses.
       Large logarithms such as $\ln (m_c/M_W) $ are split into
       $ \ln ( \mu/M_W ) +  \ln (m_c/\mu) $. The former logarithm
       resides
       in the Wilson coefficient and the latter is contained in the
       matrix elements.
\item[iii)] RG methods are applicable to $C_j$
         allowing for the summation of the logarithm in all orders
         of perturbation theory. With the boundary conditions of
         the RG evolution being $\mu=M_W$ and $\mu=m_c$ one
         sums the large logarithm  $\ln (m_c/M_W)$ of the Standard Model
         amplitude.
\item[iv)] The $C_j$'s are independent of the external states.
\end{itemize}
Hence we have overcome the problems i) to iv) mentioned in sect.\
\ref{firstlook}. The short distance coefficients $C_j$ are physically
sensible quantities, which can be obtained in perturbation theory.
Clearly in the end one has to cope with hadronic matrix elements,
which must be evaluated by nonperturbative methods such as lattice
gauge theory, $1/N$ expansion or sum rule techniques. Yet one
can sometimes find them from experimental data and insert the
obtained values into predictions for other observables of interest.
The dimensions of the $\widehat{Q}_j$'s in \rf{opej} now manifest themself
in powers of the hadronic scale $\laqcd$.

The price paid for the factorization of short and long distance
effects in \rf{opej} is the inclusion of only finite powers of
$m^2_\mr{light}/\mw$. These powers are modified by logarithms
which come with powers of $\alpha$ and can be  summed to all orders
in $\alpha$ by use of the RG. Since  the light quark masses are much
smaller than $M_W$ in most cases only the operator with the smallest
dimension in \rf{opej} is sufficient. When $\mu$ passes the flavour
thresholds $m_b$ or $m_c$, however, we repeat the factorization
process. Below $m_c$ the OPE is an expansion in $\laqcd/m_c$
and higher dimension operators may become important.

\subsection{Renormalization of Composite Operators}
Since composite operators involve products of interacting fields at
the same space--time point, they require a renormalization
in addition to that of the fields they are composed of. This
operator
renormalization has been worked out by Zimmermann \cite{zim},
who called renormalized operators normal products.

The unrenormalized field product composed of unrenormalized fields
will be denoted by the superscript \emph{bare}. We will frequently
be concerned with four--fermion operators:
\begin{eqnarray}
\widehat{Q}_k^\ba & = & \ov{\psi}^\ba  q_k \psi^\ba
 \cdot \ov{\psi}^\ba \tilde{q}_k \psi^\ba      ,
\quad \quad \ns k, \label{qba}
\end{eqnarray}
where $q_k$ and $\tilde{q}_k$ are strings of Dirac matrices. We will
only need four--fermion operators which are Lorentz singlets or
pseudosinglets, i.e.\ all Lorentz indices in $Q_k=q_k\otimes\tilde{q}_k$
are contracted.
Colour and flavour indices are suppressed in \rf{qba}, we will
make them explicit where necessary.
Next in the matrix elements the fields creating the external states
are conventionally not displayed. After expressing the bare fields
in \rf{qba} in terms of renormalized fields as in \rf{renpar},
the matrix elements are still divergent. The necessary operator
renormalization of $\widehat{Q}_k^\ba $
can require  counterterms proportional to other operators
$\widehat{Q}_j^\ba$. This phenomenon is called operator mixing. We say
that $\widehat{Q}_j$ \emph{mixes into}\/ $\widehat{Q}_k$, if
$\widehat{Q}_j^\ba$ requires a counterterm proportional to
$\widehat{Q}_k$. Hence the multiplicative renormalization
involves matrices renormalizing vectors of operators $\vec{\Qh}$.
The renormalized operator reads
\begin{eqnarray}
\Qh_j (x) &=& \Qh_j ^\re (x) \; = \; Z_{jk}^{-1} \Qh_k^\ba (x) .
\label{defqren}
\end{eqnarray}
As in \rf{renqcd} a quantity without superscript is understood to be
renormalized. In cases where this may be confusing (such as in
chapter \ref{eva}) we will mark renormalized quantities with
\emph{ren}. In the phenomenological sections also the caret on the
operators will be omitted.

A set of operators which renormalize each other in \rf{defqren}
is said to \emph{close under renormalization}.

In \rf{defqren} the standard definition of $\Qh_j^\re$ has been
given. It would be more useful to include the field
renormalization  into the defintion of $\Qh_j^\re$,
so that $\Qh_j^\re$ would be expressed in terms of renormalized
fields instead of bare fields. Again, in ambiguous places
completely renormalized (i.e.\ finite)
Green's functions will get a superscript
\emph{ren}:
\begin{eqnarray}
\langle \Qh_j \rangle & = & \langle \Qh_j^\re \rangle ^\re
\; = \;  Z_\psi ^2 \langle \Qh_j^\re \rangle ^\ba
\; = \;  Z_\psi ^2 Z^{-1}_{jk} \langle \Qh_k^\ba \rangle ^\ba .
\label{defrenma}
\end{eqnarray}
We will also have to discuss the mixing of four--fermion
operators with operators containing only two external
quark lines, but additional gluon or ghost legs.  They
involve the product of the corresponding   field renormalization constants
in \rf{defrenma} instead of four quark field constants
$Z_\psi^{1/2}$.

\section{Effective Field Theories}
The operator product expansion in terms of some inverse heavy mass
described in the previous section has a complementary interpretation
in terms of an effective field theory: Since the Wilson coefficients
$C_j$ on the RHS of \rf{opeeff} are independent of the external states
we may consider them as coupling constants multiplying four--fermion
couplings in some effective Lagrangian:
\begin{eqnarray}
\lag ^\mr{I}  &=& C_j (\mu )  \Qh_j^{\re} (x,\mu) \; = \;
                C_j (\mu )  Z^{-1}_{jk} (\mu ) \Qh_k^{\ba} (x) \nn
&=& C_j \, Z^{-1}_{jk} Z_\psi^2 \, \ov{\psi}_1 (x) q_{k} \psi_2 (x)
   \ov{\psi} _3 (x) \tilde{q}_{k} \psi _4 (x)  \label{eff1}
\end{eqnarray}
Clearly the S--matrix
\beq
\mr{\T \exp \lt[i \int d^4 x \lt( \lag_\mr{QCD}(x)+\lag^\mr{I} (x) \rt)\rt] }
\label{seff1}
\eeq
derived from \rf{eff1} yields to first order in the effective coupling
constants $C_j$ and to all orders in the QCD coupling constant $g$
the RHS of \rf{opeeff}. We may compare the
effective Lagrangian \rf{eff1} with the QCD Lagrangian \rf{renqcd}:
The last line in \rf{eff1} corresponds to a  renormalized
interaction vertex expressed in terms of renormalized fields.
We may view $Z^{-1}_{jk}$ as renormalizing
the effective coupling constant
$C_j$ instead of the composite operator.
Yet the concept of an effective field theory must allow for going
beyond the first order in the effective interaction in \rf{seff1}.
This is possible as we know from the work of Witten \cite{wit}.

The  Green's functions of second order in the effective interaction
involve
\beq
\langle -\frac{1}{2} \int d^D x \int d^D y \T \lag^\mr{I} (x) \lag^\mr{I} (y)
\rangle
. \no
\eeq
These \emph{bilocal structures}\/
are in general divergent even in the absence of QCD interactions
as can be seen from  \fig{bi}. Hence one must add  counterterms
proportional to  local operators to $\lag _\mr{QCD} + \lag^\mr{I}$ in order to
obtain a finite result. Although the effective theory is not renormalizable
by power counting,
this is possible in all orders of both the effective and the QCD
interaction, because after subtracting subdivergences the remaining
divergences  are polynomial in the momenta.
The non--renormalizable interaction, however, forces us to introduce
new counterterms in every order of perturbation theory. Hence we
cannot include all--order effects in the effective action as
we can do with the RG summation of the QCD interaction. Yet
ordinary perturbation theory works well for the weak interaction,
so that working to finite order in the effective coupling is
perfectly sufficient.

In this thesis we will work up to the second order in the effective
couplings. After adding the local operator counterterms
to $\lag^\mr{I}$  one ends up with:
\begin{eqnarray}
\lag^\mr{II} &=& \lag^\mr{I} + C_k (\mu)  C_{k^\prime} (\mu) \;
 Z^{-1}_{k k^\prime, l}(\mu) \; \Qht_l^\ba
+ \widetilde{C}_r(\mu) \; \widetilde{Z}^{-1}_{rl} (\mu)\; \Qht_l^\ba \nn
&=& C_j \; Z^{-1}_{jk} Z_\psi^2 \; \ov{\psi} (x) q_{k} \psi (x) \,
   \ov{\psi}  (x) \tilde{q}_{k} \psi  (x)
  \nn
&& + C_k C_{k^\prime} \;
 Z^{-1}_{k k^\prime, l} \;
 Z_\psi^2 \; \ov{\psi} (x) q^\mr{loc}_{l} \psi (x) \,
   \ov{\psi}  (x) \tilde{q}^\mr{loc}_{l} \psi  (x) \nn
&& +
\widetilde{C}_l \; \widetilde{Z}^{-1}_{lr} Z_\psi^2 \;
   \ov{\psi} (x) q^\mr{loc}_{r} \psi (x) \,
   \ov{\psi}  (x) \tilde{q}^\mr{loc}_{r} \psi  (x)
   \label{eff2}
\end{eqnarray}
with $\Qht _l =
   \ov{\psi} (x) q^\mr{loc}_{l} \psi (x)
   \ov{\psi}  (x) \tilde{q}^\mr{loc}_{l} \psi  (x)$, $\ns l$,
being the local operators needed as counterterms  to the
bilocal structures. In \kkm\/ the $\Qh_k$'s are \dso operators
and the $\Qht_l$'s are \dst operators.
In \rf{eff1} and \rf{eff2} we have for simplicity only displayed
four--fermion operator counterterms. If none of the quark fields
in $\Qh_k$, $\Qht_l$ has the same flavour quantum number as
one of the antiquark fields, only four--quark counterterms
appear in $\lag^\mr{II}$. Yet in the general case so called
\emph{penguin diagrams}\/ lead to a mixing of four--quark
operators into other operators as depicted in figs.\
\ref{glueping} (p.\ \pageref{glueping}) and
\ref{ghostping} (p.\ \pageref{ghostping}).

{}From \rf{eff2} we realize the advantage of the effective lagrangian picture:
The extra terms in \rf{eff2} exactly look like renormalization factors
for the effective couplings $\widetilde{C}_j$, but now both
in terms of the effective interaction and of QCD:
\begin{eqnarray}
\widetilde{C}_l ^\ba &=& \widetilde{Z}^{-1}_{rl} \widetilde{C}_r +
 Z^{-1}_{k k^\prime, l}  C_k C_{k^\prime} + \ldots .  \no
\end{eqnarray}
This form resembles the QCD coupling renormalization in \rf{renqcd}.
Further the effective lagrangian form allows for a simple inclusion
of the correct wave function renormalization, which is simply
taken into account by the corresponding counterterm  Feynman rules
derived from $\lag_\mr{QCD} + \lag^\mr{II}$. Finally the renormalization group
equations for the $C_k$'s and $\widetilde{C}_l$'s to be discussed
in sect.\ \ref{rgo} are very easily found from \rf{eff1} and \rf{eff2}.
\section{Appelquist--Carrazone Theorem}
The decoupling theorem  of Appelquist and Carrazone
\cite{ac} deals with the effect
of heavy virtual particles on low energy processes: If we can remove
some heavy particle field from the lagrangian without spoiling the
renormalizability\footnote{We will use the term
\emph{renormalizability} for two different properties:
It either refers to the locality of counterterms, in this sense
we also call an effective theory renormalizable, or it means
\emph{renormalizability by power counting} such that all
couplings have non--negative dimension.},
the loop effects of the heavy particle can be
absorbed in the renormalization constants of the low energy
lagrangian (without the heavy particle field) and into additive
terms which are suppressed by powers
(times logarithms) of the heavy particle mass.

The Appelquist--Carrazone theorem puts some hierarchical order into field
theory: It has allowed to discover QED without knowing about its embedding
into the Standard Model and it allows to study Standard Model predictions
at LEP energies without knowing the physics at, say, $10^{16} \gev$.

We will be concerned with the Appelquist--Carrazone theorem in a twofold
way: First we will use that quark fields and clearly also the
W field do decouple  with respect to the strong interaction:
QCD works with 5 flavours as well as with 6, hence we can remove
the top quark from the lagrangian when passing with the renormalization
scale $\mu $ below its mass.  We repeat this with the W mass and then
when crossing the b-- and c--threshold. We could even incorporate
these inverse power corrections by taking into account operators
with higher dimension in \rf{eff2} and we can use the RG to
sum the logarithms
accompanying these powers together with
powers of the strong coupling constant.

The weak interaction, however, is a prominent example for non--decoupling:
The heavy top--quark is clearly needed as an isospin partner of
the bottom quark and after removing the W--boson we are faced with
non--renormalizable effective four--fermion couplings.
The non--decoupling
allows to study
the weak interaction
in  hadronic processes, although their energy scale $\laqcd$ is
much lower than the scale $M_W$, at which the weak transition takes
place.

\section{Initial Condition for the Wilson Coefficients}\label{icwc}
RG improved perturbation theory starts with the definition of the
initial condition for the Wilson coefficients.
This is done by matching the Standard Model amplitude to the effective
theory as prescribed in \rf{opeeff}. For definiteness we will
consider some weak four--fermion amplitudes.
\subsection{Single Operator Insertion}
Let's look at some weak tree level process such as a \dso\/ transition,
which is mediated  by the exchange of one W boson. The corresponding
Standard Model amplitude factorizes
order by order in $1/\mw$
into a   Wilson coefficient vector $\vec{C}=(C_1,\ldots C_n)^T$
and matrix elements $\langle \vec{\Qh} \rangle$. For illustration we
only display the dependence on the renormalization scale
$\mu$, the W mass $M_W$ and a light quark mass $m$:
\begin{eqnarray}
-i G_4^{(2)\, tc} (M_W,\mu,m) &=&
C_j(M_W, \mu ) \langle \widehat{Q}_j \rangle (\mu,m) .\label{exma1}
\end{eqnarray}
A \dso\/ four--quark operator is depicted in fig.\ \ref{ds1cclo}
on p.\ \pageref{ds1cclo}.

To obtain \rf{exma1} one has to calculate the Standard Model
amplitude and the matrix element in perturbation theory
to the desired order in the QCD coupling $g$.
By comparing both results one finds
$C_j(M_W, \mu )$. For a meaningful
perturbative result for $C_j$
the factorization
scale  $\mu=\mu_0$ must be chosen to be of the order $M_W$ to
keep the logarithm $\al \ln ( \mu/M_W ) $ in $C_j$ small.
Thereby one obtains the initial value for  $C_j(M_W,\mu_0)$ with
respect to the RG evolution. With the RG evolution of
$C_j$ from $\mu=\mu_0$ down to
$\mu=\mu_m\approx m$
described in the
following section
we want to sum $\ln(\mu_m/\mu_0 )\approx \ln (m/M_W)$ present in the LHS of
\rf{exma1} to all orders. As indicated  one is not forced
to choose $\mu_m=m, \mu_0=M_W$ exactly. The two large logarithms
$\ln(\mu_m/\mu_0 )$ and $\ln (m/M_W)$ differ by a small logarithm
of the order of magnitude of the remaining non--summed radiative
correction. Changing $\mu_0$ and $\mu_m$ modifies the result by terms
of the order of the uncalculated higher order corrections. These
scale dependences are inherent to RG improved perturbation theory
and may be viewed as an estimate of the accuracy of the truncated
perturbation series. For the rest of this section we choose
$\mu_0=M_W$ and $\mu_m=m$ for clarity.

{}From the matching procedure one can
count the powers of the summed logarithms: In LO the matching is done
in the order $\al^0\ln^0 (m/M_W)$, and the LO RG will sum
\begin{eqnarray}
\al^n\ln^n \frac{m}{M_W} ,\quad \quad n=0,1,2,\ldots .\no
\end{eqnarray}
In NLO we will sum the log's with one additional power of $\al$.
Hence the matching is done in the order
$\al^1\ln^0 (m/M_W)$, which requires the calculation of the one--loop
diagrams of fig.\ \ref{ds1ccnlo} on p.~\pageref{ds1ccnlo}.

\subsection{Double Operator Insertion}\label{dom}
Next we will look at a process which appears in the second order
of the effective couplings. Clearly we have in mind the \dst\/
transition with light internal quarks. Its Standard Model amplitude
is described in LO by fig.\ \ref{box} (p.\
\pageref{box}) and the effective theory matrix element
is shown in fig.\ \ref{bi} (p.\ \pageref{bi}). We will often refer to
\fig{bi} as the prototype for a matrix element with two insertion
of four--fermion operators, although we will later also discuss
matrix elements containing one other operator such the one  depicted in
fig.\ \ref{glueop} (p.~\pageref{glueping})
or  \ref{ghostping} (p.~\pageref{ghostping}).

Here the matching condition reads
\begin{eqnarray}
        -i G_4^{(2)\, tc}
        &=&
        C_{k} C_{k'} \left\langle \left[\frac{i}{2} \int \T
                \widehat{Q}_{k} \widehat{Q}_{k'} \right]^{\ren}
     \right\rangle^\re
        +
        C_{l} \left\langle \widetilde{Q}_{l} \right\rangle . \label{exma2}
\end{eqnarray}
Here the square brackets around the two operators shall denote
the inclusion of the local operator counterterms in $\lag^\mr{II}$
in \rf{eff2}.

Next let us comment on  the powers of the logarithm. There are two
different cases, both of which appear in \kkm :\\[-1mm]

\ul{Case I:}

This generic case is the following situation:

The diagram in fig.\ \ref{bi} is divergent and requires local
operator counterterms depicted in fig.\ \ref{loc} (p.\ \pageref{loc}).
In the example of \kkm\/ these local operators $\wt{Q}_l$
are \dst\/ operators.
With the divergence comes  $\ln ( \mu/ m) $ in the finite part
even in the absence of QCD
corrections. Its twin in the Standard Model amplitude is
$\ln (m/M_W)$. In the LO we can simply do the matching from these
logarithms and the leading log summation comprises
\begin{eqnarray}
\al^n \ln^{n+1} \frac{m}{M_W} ,\quad \quad n=0,1,2,\ldots .\no
\end{eqnarray}
Consequently in NLO we can still  do the matching from
the one--loop graphs figs.\ \ref{box} and \ref{bi}, but now from
the $\al^0 \ln^0 (m/M_W)$ parts. Here we must also
take into account the coefficients of the $\wt{Q}_l$'s.

Case I will appear in the calculation of $\eta_3$
in \rf{s2}
in chapter \ref{calc}. Cf.\ also the NLO calculation
of rare K decays in \cite{bb}.\\[-1mm]

\ul{Case II:}

The insertion of two local operators into the
diagram of fig.\ \ref{bi} is convergent. This situation appears
when there are several contributions of the form in fig.\ \ref{bi}
combining such that the divergences of the individual contributions
cancel. With the cancellation of the divergences also the $\ln \mu$'s
will disappear, so that there is no large log in the corresponding
one--loop amplitude of fig.\ \ref{box}.
Hence the leading log summation will comprise
\begin{eqnarray}
\al^n \ln^{n} \frac{m}{M_W} ,\quad \quad n=0,1,2,\ldots \no
\end{eqnarray}
and the matching has to be done in the order
$\al^0 \ln^0 (m/M_W)$.
Consequently we are forced to perform the NLO matching in the
order $\al^1 \ln^0 (m/M_W)$. This requires the calculation of the
finite parts of the diagrams of fig.\ \ref{boxqcd} (p.\
\pageref{boxqcd}) and of those of fig.\ \ref{biqcd} (p.\
\pageref{biqcd}).

Case II appears in the calculation of $\eta_1$. The cancellation of
the large logs in the corresponding Standard Model amplitude
is caused  by the GIM mechanism: The divergence of the
diagram in fig.\ \ref{bi} is multiplied with the product of
two columns of the CKM matrix, which is zero due to the unitarity of
the latter. The finite part is of the order $m^2/\mw$ and one speaks
of a \emph{hard}\/ GIM suppression. One could even term this
suppression \emph{superhard}, because there is even no $\ln (m^2/\mw)$
multiplying $m^2/\mw$ and the  RG enhancement is thereby suppressed
by one power of $\al $ compared to case I.

\section{Renormalization Group for Operators and Coefficients}\label{rgo}
\subsection{Single Operator Insertion}\label{soi}
In this section we will set up the RG formalism for matrix elements with
single insertions of composite operators, which can be derived
from $\lag^\mr{I}$ in \rf{eff1}.
\subsubsection{Anomalous Dimension Matrix}
The anomalous dimension matrix $\gamma$ is defined by
\begin{eqnarray}
\g_{ij} (g (\mu) )&=& - \lt[ \dmu Z^{-1}_{ik} \rt] Z_{kj}
\; = \;    Z^{-1}_{ik}  \dmu Z_{kj} \nn
&=&  Z^{-1}_{ik}  \beta (g) \diff{g} Z_{kj} . \label{defandi}
\end{eqnarray}
{}From \rf{defqren} one simply obtains the RG equation for the renormalized
operator:
\begin{eqnarray}
\lt[ \dmu \delta_{jk} + \g_{jk} \rt]   \Qh_k (\mu) &=& 0. \label{rgop}
\end{eqnarray}
We expand $Z_{jk}$ as usual in $g$ and $\e$ as shown in \rf{exp}.
The anomalous dimension matrix
\begin{eqnarray}
\g_{jk} &=&  \g^{(0)}_{jk} \frac{g^2}{16 \pi^2} +
   \g^{(1)}_{jk} \frac{g^4}{\lt( 16 \pi^2 \rt)^2}  + O(g^6)
\label{andex}
\end{eqnarray}
is  related to  $Z_{jk}$ as follows:
\begin{eqnarray}
\g^{(0)} &=& -2  Z_1^{(1)} -2 \e   Z_0^{(1)} \nn
\g^{(1)} &=& -4 Z_1^{(2)} + 2 \lt\{  Z_0^{(1)}, Z_1^{(1)} \rt\}
             -2 \beta_0  Z_0^{(1)} + O(\e ) . \label{and1}
\end{eqnarray}
In \rf{andex} and in \rf{and1} we have only kept the terms relevant for
NLO calculations.
We have allowed for a finite renormalization $Z_0^{(1)}$ as well. The
structure of \rf{and1} reveals that one obtains the correct $\go$ by
inserting the finite one--loop counterterms with a factor of 1/2
instead of 1 into the one--loop counterterm diagrams.
\rf{and1} can be easily derived from the insertion of \rf{andex}
into \rf{defandi}, see e.g.\ \cite{bw}. Alternatively one can recall
that $\g_{ij}$ is related to the coefficient of $\ln \mu$ in
$\langle \Qh \rangle$. In dimensional regularization the
$\ln \mu$'s enter the finite parts of Green's functions via
\begin{eqnarray}
( \mu^{2 \e} -1) \frac{1}{\e} &=& 2  \ln \mu , \no
\end{eqnarray}
where the $-1$ stems from  counterterm diagrams.
Hence e.g.\ in lowest order the coefficient of $\ln \mu$ is
simply twice the coefficient of the divergence.

In the same way one finds for $\beta_0$ in \rf{betaseries}:
\begin{eqnarray}
\beta_0 &=& -2 Z_{g,1}^{(1)} .
\end{eqnarray}

As mentioned before, the different coefficients of $Z$ are not independent.
In NLO one has the relation
\begin{eqnarray}
4 Z_2^{(2)} + 2 \beta_0 Z_1^{(1)} - 2 Z_1^{(1)}  Z_1^{(1)}&=&0.
\label{z2nlo}
\end{eqnarray}
\rf{z2nlo} is a consequence of the renormalizability, i.e.\ of the locality
of the counterterms \cite{hoco}.
Since in mass independent schemes the divergence
of each diagram plus its subloop
counterterm diagrams is polynomial in the momenta
and masses, any  dependence
on $\ln \mu$ also cancels from the divergent parts.
The two--loop diagrams involve $\mu^{4 \e}Z_2^{(2)}/
\e^2 $, while the one--loop counterterm diagrams involve
$\mu^{2 \e} Z_1^{(1)}/\e$ times either $Z_1^{(1)}/\e$ or
$Z_{g,1}^{(1)}/\e = -\beta_0/(2 \e)$.
The vanishing of $\ln \mu$ enforces \rf{z2nlo}.
We will use this argument in a different context in a formal proof
on evanescent operators, which vanish in $D=4$ dimensions,
in chapter 4. Relation \rf{z2nlo}
is tightly related to the finiteness of anomalous dimensions, which
has been used for an elegant proof of \rf{z2nlo} in \cite{bw}.

\subsubsection{Solution of the RG Equations}\label{sre}
Let us first derive the RG equation for the Wilson coefficients:
{}From $\dmu \lag^\mr{I}=0$ in \rf{eff1} one immediately obtains:
\begin{eqnarray}
\lt[ \dmu \delta_{jk} - \g_{jk} \rt] C_j   &=& 0, \label{rgwil}
\end{eqnarray}
thus the RG equation for the
Wilson coefficient vector $\vec{C}$ involves the transpose  of $\g$.
The solution of \rf{rgwil} is given by
\begin{eqnarray}
C_j (\mu_1) &=& \lt[ U(\mu_1,\mu_0)\rt]_{jk} C_k (\mu_0 ) \label{evowil1}
\end{eqnarray}
where $U(\mu_1,\mu_0)$ is the \emph{evolution matrix}:
\begin{eqnarray}
 U(\mu_1,\mu_0) &=& \Tg \exp\lt[ \int_{g(\mu_0)}^{g(\mu_1)} d g^\prime
  \frac{\g^T (g^\prime)}{\beta(g^\prime)} \rt] . \label{evo1}
\end{eqnarray}
Here $\T_g$ means that the matrices
$\g (g^\prime ), \g (g^{\prime \prime} )\ldots $ in the expanded
exponential are ordered such that the couplings
$\lt\{\begin{array}{rl}
 \txr{increase from right to left }& \txr{for $g(\mu_0 ) < g( \mu_1) $}  \\
 \txr{decrease from right to left }& \txr{for $g(\mu_0 ) > g( \mu_1) $}
\end{array}\rt\}.$

The LO solution of  \rf{evo1} is
\begin{eqnarray}
 U^{(0)} (\mu_1,\mu_0) &=&
\exp\lt[ d \ln \frac{\al (\mu_0)}{\al (\mu_1)}  \rt]
\label{los}
\end{eqnarray}
with
\begin{eqnarray}
d &=& \frac{\gz}{2 \beta_0} . \label{defd}
\end{eqnarray}
In the diagonal basis of $d$
the LO evolution matrix $U^{(0)}$ is  simply a diagonal matrix
with diagonal elements $\lt[\alpha(\mu_0)/\alpha(\mu_1) \rt]^{d_j}$,
where the $d_j$'s are the eigenvalues of $d$.

The NLO solution of \rf{evo1} reads:
\begin{eqnarray}
 U(\mu_1,\mu_0) &=& \lt( \uma  + \frac{\alpha(\mu_1)}{4 \pi} J    \rt)
 U^{(0)} (\mu_1, \mu_0)
\lt( \uma  - \frac{\alpha(\mu_0)}{4 \pi} J    \rt) . \label{nlos}
\end{eqnarray}
Here clearly the NLO running $\alpha$ has to be used in $U^{(0)}$.
$J$ in \rf{nlos} is a solution of the matrix equation \cite{bjlw}:
\begin{eqnarray}
J+ \lt[ \frac{\gzt}{2 \beta_0} , J   \rt] &=&
- \frac{\got}{2 \beta_0} + \frac{\beta_1}{2 \beta_0^2 } \gzt . \label{defj}
\end{eqnarray}
In practical calculations it is not useful  to solve \rf{defj} by
transforming into the diagonal basis. On better solves \rf{defj}
directly: Since it is a system of linear equations for the elements
of $J$, the entries of $J$ are simply rational functions of $N$ and
$f$.

For the LO evolution matrix it is useful to transform to the diagonal
form, as this saves computation time, because $U^{(0)}$ has to be
calculated newly for every new pair $(\mu_0,\mu_1)$.
In our phenomenological analysis of
chapter \ref{phen}, however, we have simply exponentiated the matrix
in \rf{los}, because the computer algebra system \textsc{Mathematica}
provides a fairly fast numerical matrix exponentiation algorithm.

\subsection{Double Operator Insertion}\label{doi}
The local operator counterterms $ Z^{-1}_{k k^\prime, l}(\mu) \; \Qht_l $
in $\lag^\mr{II}$ do not influence the RG evolution of the coefficients
$C_k$, but they modify the running of $\wt{C}_l$. We will investigate this
in the following.

Often one does not insert the full set of operators in $\lag^\mr{I}$
into both places in fig.\ \ref{bi}, but limits oneself to a subset
which closes under renormalization. Hence $k$ and $k^\prime$
in $ Z^{-1}_{k k^\prime, l}$ may run over  different ranges. This
will be so in the case of $\eta_3$.

\subsubsection{Anomalous Dimension Tensor and RG Equations}
The evolution of the Wilson coefficients in $\lag^\mr{I}$ is
fixed with \rf{evowil1}. Now $\lag^\mr{II}$ contains new local operator
counterterms with coefficients $\wt{C}_l$. It has become standard
to define the operators $\Qht_l$
with two inverse powers of $g$, e.g.\
$\widetilde{Q}=m^2/(g^2 \ov{\mu}^{2 \e})  \cdot
        \gamma_{\mu} \left(1-\gamma_5\right) \otimes \gamma^{\mu}
        \left(1-\gamma_5\right)$,
to avoid operator mixing in the order $g^0$. Then
$Z^{-1}_{kn,l}=O(g^2)$ and the mixing matrices start in the order
$g^2$.

{}From
$\dmu \lag^\mr{I}=0$ and $\dmu \lag^\mr{II}=0$ one finds with \rf{eff2}:
\begin{eqnarray}
0 &=& \dmu \lt[ C_k (\mu)  C_{k^\prime} (\mu) \;
      Z^{-1}_{k k^\prime, l}(\mu)
      + \widetilde{C}_r (\mu)\; \widetilde{Z}^{-1}_{rl} (\mu) \rt] . \no
\end{eqnarray}
This can be compactly rewritten as
\begin{eqnarray}
\dmu \wt{C}_l (\mu) &=& \wt{C}_{l^\prime} (\mu) \wt{\g}_{l^\prime l}
               + C_k (\mu) C_{k^\prime}(\mu) \g_{kk^\prime,l}  \label{rgwil2}
\end{eqnarray}
with the \emph{anomalous dimension tensor}
\begin{eqnarray}
\g_{kn,l} &=&
        \frac{g^2}{16 \pi^2} \gamma_{kn,l}^{\left(0\right)}
        +
        \left(\frac{g^2}{16 \pi^2}\right)^{2}
                \gamma_{kn,l}^{\left(1\right)}
        +
        \ldots \nn
&=&- \lt[ \g_{kk^\prime} \delta_{n n^\prime } +
                            \delta_{kk^\prime} \g_{n n^\prime} \rt]
                     Z^{-1}_{k^\prime n^\prime ,l^\prime }
                     \wt{Z}_{l^\prime l} -
           \lt[ \dmu Z^{-1}_{k n ,l^\prime }   \rt]
                     \wt{Z}_{l^\prime l} . \no
\end{eqnarray}
The analogue of \rf{and1} is found as
\begin{eqnarray}
        \gamma^{\left(0\right)}_{kn,l}
        &=&
        2 \left[Z^{-1,\left(1\right)}_{1}\right]_{kn,l}
        + 2 \eps \left[Z^{-1,\left(1\right)}_{0}\right]_{kn,l}
\nn
        \gamma^{\left(1\right)}_{kn,l}
        &=&
        4 \left[Z^{-1,\left(2\right)}_{1}\right]_{kn,l}
        + 2 \beta_{0} \left[Z^{-1,\left(1\right)}_{0}\right]_{kn,l}
\nn
        & &
        - 2 \left[Z^{-1,\left(1\right)}_{0}\right]_{kn,l'}
                \left[\widetilde{Z}^{-1,\left(1\right)}_{1}\right]_{l' l}
        - 2 \left[Z^{-1,\left(1\right)}_{1}\right]_{kn,l'}
                \left[\widetilde{Z}^{-1,\left(1\right)}_{0}\right]_{l' l}
\nn
        & &
        - 2 \left\{
                \left[Z^{-1,\left(1\right)}_{0}\right]_{k k'}
                \delta_{n n'}
                +
                \delta_{k k'}
                \left[Z^{-1,\left(1\right)}_{0}\right]_{n n'}
        \right\}
                \left[Z^{-1,\left(1\right)}_{1}\right]_{k' n',l}
\nn
        & &
        - 2 \left\{
                \left[Z^{-1,\left(1\right)}_{1}\right]_{k k'}
                \delta_{n n'}
                +
                \delta_{k k'}
                \left[Z^{-1,\left(1\right)}_{1}\right]_{n n'}
        \right\}
                \left[Z^{-1,\left(1\right)}_{0}\right]_{k' n',l}
        +O(\e) .
\label{gamma1double}
\end{eqnarray}
\subsubsection{Solution of the RG Equations}
The solution of the inhomogeneous RG equation  \rf{rgwil} for the running
Wilson coefficient $\wt{C} (\mu)$ reads:
\begin{eqnarray}
        \widetilde{C}_{l}\!\left(g\left(\mu\right)\right)
        &=&
        \widetilde{U}^{\left(0\right)}_{l l'}
                \left(g\!\left(\mu\right),g_{0}\right)
                \widetilde{C}_{l'}\!\left(g_{0}\right)
\nn
        & &
        +
        \left[\delta_{l l'} +
                \frac{g^{2}\left(\mu\right)}{16\pi^2} \widetilde{J}_{l l'}
                \right]
        \cdot
        \int\limits_{g_{0}}^{g\left(\mu\right)} d g'\;
        \widetilde{U}^{\left(0\right)}_{l' k}
                \left(g\!\left(\mu\right),g'\right)
        \left[\delta_{k k'} -
                \frac{g'^{2}}{16\pi^2} \widetilde{J}_{k k'}
                \right]
\nn
        & &
        \hspace{1cm}
        \cdot
        \left[\delta_{n n'} +
                \frac{g'^{2}}{16\pi^2} J_{n n'}
                \right]
        U^{\left(0\right)}_{n' t}\left(g',g_{0}\right)
        \left[\delta_{t t'} -
                \frac{g'^{2}}{16\pi^2} J_{t t'}
                \right]
        C_{t'}\!\left(g_{0}\right)
\nn
        & &
        \hspace{1cm}
        \cdot
        \left[\delta_{m m'} +
                \frac{g'^{2}}{16\pi^2} J_{m m'}
                \right]
        U^{\left(0\right)}_{m' v}\left(g',g_{0}\right)
        \left[\delta_{v v'} -
                \frac{g'^{2}}{16\pi^2} J_{v v'}
                \right]
        C_{v'}\!\left(g_{0}\right)
\nn
        & &
        \hspace{1cm}
        \cdot
        \left\{
                - \frac{\gamma^{\left(0\right)}_{nm,k'}}{\beta_{0}}
                        \frac{1}{g'}
                + \left[
                        \frac{\beta_{1}}{\beta_{0}^2}
                                \gamma^{\left(0\right)}_{nm,k'}
                        -
                        \frac{\gamma^{\left(1\right)}_{nm,k'}}{\beta_{0}}
                \right]
                \frac{g'}{16\pi^2}
        \right\}
        .
\label{inhsol}
\end{eqnarray}
Here $\wt{U}^{(0)}$ and $\wt{J}$ are the RG quantities related to single
insertions of $\Qht$ as defined in \rf{los} and \rf{defj}.
Further the  Wilson coefficients and
evolution matrices have been labeled with $g(\mu)$
rather than $\mu$ to comply with the integration variable $g^\prime$.
The coupling at the scale $\mu_0$ at which the
initial condition is defined is  denoted by
$g_0=g(\mu_0)$.

The first term in \rf{inhsol}  is solely related to matrix elements
with single insertions of $\Qht$. There are no factors involving
$\wt{J}$ here, because the initial coefficients $\wt{C}_l(g_0)$
start at the order $g^2$.

\rf{inhsol} nicely reveals the structure of the RG in bilocal
matrix elements: First the two Wilson coefficients $C_{l^\prime}$
and $C_{\nu^\prime}$
run from $\mu_0$ down to the intermediate scale $\mu^\prime$ with
$g(\mu^\prime)=g^\prime$. Then they are linked by the anomalous
dimension tensor to a single coefficient $\wt{C}_{k^\prime}$,
which runs further down with the NLO evolution matrix
$\wt{U}(\mu,\mu^\prime)$. The integral sums over all intermediate
scales $\mu^\prime$.

For formal analyses as those in chapter \ref{eva} the form
\rf{inhsol} is well suited. In a practical calculation, however,
the solution of the integral in \rf{inhsol} requires to transform
the $\Qht_l$'s to the diagonal basis of $\wt{\g}$.
The same must be done for the $\Qh_k$'s of at least
one
of the two operator bases corresponding to the double
operator insertion.

Consider the case that the two operator bases feeding fig.\
\ref{bi} (p.\ \pageref{bi}) have $K$ and $K^\prime$ elements, while
$L$ linearly independent operators $\Qht_l$ are needed as counterterms
to fig.\ \ref{bi}. In the calculation of $\eta_3$ we will have
$K=2$, $K^\prime=6$ and $L=1$.
The simplest way to solve \rf{rgwil2}
is to label the product $C_k C_{k^\prime}$ with a single index
running from 1 to $K\cdot K^\prime$. We obtain the form as in
\rf{rgwil} and can proceed  as in the case of single operator
insertions. The evolution matrix has $K\cdot K^\prime +L$ rows and
columns. Yet this method hides the different sources of RG admixtures
to $\wt{C}_l$.

Next one can transform to the diagonal basis for one set of operators
$\Qh_k$: The RG evolution then decomposes into $K$ separate
sectors each involving  $(K^\prime +L) \times (K^\prime +L)$
matrices. This is similar to the method used in the LO analysis
of \kkm\/  of Gilman and Wise \cite{gw}.
If this method is used beyond the leading order, however,
\emph{the mixing matrices of different orders must commute}, if the
standard formalism of sect.\ \ref{soi} shall apply: Only if
$\lt[ \gz,\go \rt] =0 $, the NLO matrix $J$ does not induce a
mixing of the different sectors of the LO evolution.
In the calculation of $\eta_3$, where $K=2$, this is indeed the
case for the relevant LO and NLO mixing matrices given in the literature
\cite{bw}.

The discussion above stresses the advantage of commuting LO and NLO mixing
matrices, especially for the case of multiple operator insertions.
It is noticeable that one can indeed always achieve this situation
by a finite renormalization, because $\go$ is scheme dependent.
In the context of four--fermion operators scheme transformations
involving a set of continuos real parameters
naturally enter the scene due to the presence of evanescent
operators. We will look at this in chapter \ref{eva}.

Of course one can further decompose the RG evolution by transforming
the remaining operator sets to the diagonal basis, too. Yet this does
neither significantly improve the algorithm nor does it provide
more insight into physics.

The $K$ sectors each having a
$(K^\prime +L) \times (K^\prime +L)$ evolution
matrix contain some redundancy, because all $K$ of them encode
the running of the $K^\prime+L$ other operators. This redundancy can
be removed by replacing the $\wt{C}_l$'s by $\wt{C}^{(k)}_l$'s which
are obtained by
dividing the $\wt{C}_l$'s by the  $C_k$'s of the diagonalized
basis. This results in a matrix evolution for a coefficient
vector with $K^\prime +K\cdot L $ components. We have used this
method in the calculation of $\eta_3$ thereby reducing the problem
to a single $8 \times 8$ evolution matrix instead of using
four such matrices as in \cite{gw}.

We finally remark that the evolution matrix has a block--triangular
form, because the coefficients $\wt{C}_l$ of the local operators $\Qht_l$
cannot mix into  the products $C_k \cdot C_{k^\prime}$.
Since such block--triangular matrices
appear in many contexts, their discussion deserves a separate section.

\subsection{Block--triangular Mixing Matrix}\label{btmm}
We will frequently encounter block--triangular anomalous dimension
matrices:
\begin{eqnarray}
\g &=& \lt(
\begin{array}{rl}
\g_a & \g_b \\
0 & \g_c
\end{array} \rt) \label{blockg}
\end{eqnarray}
with submatrices $\g_a,\g_b$  and $\g_c$.
This form leads to a block--triangular LO evolution matrix:
\begin{eqnarray}
\lefteqn{U^{(0)} (\mu_1,\mu_0) \; = } \no \\[2mm]
&&\lt(
\begin{array}{cc}
\uz_a (\mu_1,\mu_0)
& 0 \\
U^{(0)}_\mr{inh} (\mu_1,\mu_0)  & \uz_c (\mu_1,\mu_0)
\end{array} \rt)
\; =\;
\lt(
\begin{array}{cc}
\exp\lt[ d_a \ln \frac{\al (\mu_0)}{\al (\mu_1)}  \rt]
& 0 \\
\ast & \exp\lt[ d_c \ln \frac{\al (\mu_0)}{\al (\mu_1)}  \rt]
\end{array} \rt)\label{blocku}
\end{eqnarray}
with  $d_a,d_c$ defined analogously to \rf{defd}.
Hence the diagonal blocks of the LO evolution matrix  exponentiate
separately. The RG evolution of the
Wilson coefficients of the subspace $\mr{S_a}$ corresponding
to $\g_a$ is not affected by the coefficients of the other subspace
$\mr{S_c}$. The Wilson coefficient vectors in $\mr{S_c}$, however,
satisfy a RG equation to which an inhomogeneous term is added
involving the solution of the RG equation of subspace $\mr{S_a}$.

This invariance of $\mr{S_a}$ persists in  NLO: \rf{blockg} leads to a
block--triangular $J$ matrix in \rf{defj}
protecting $\mr{S_a}$ from admixtures.

We will be confronted with the case that $\g_c$ is a $1\times 1$
matrix. Then one can also solve $U^{(0)}_\mr{inh}$ in terms of
$\g_c^{(0)}$,
the submatrix $n \times n$ matrix $\g_a^{(0)}$ and the column vector
$\g_b^{(0)}$:
\begin{eqnarray}
U^{(0)}_\mr{inh} &=& d_b  \cdot
\exp\lt[  \uz_a (\mu_1,\mu_0)  - \uma _n \uz_c (\mu_1,\mu_0)   \rt]
\cdot  \lt[ \lt( d_a - \uma _n d_c  \rt)
         \ln \frac{\al (\mu_0)}{\al (\mu_1)}  \rt]^{-1}  ,
\label{uinh}
\end{eqnarray}
which is a row vector due to $d_b = \gzt_b/(2 \beta_0)$. $\uma _n$ is the
$n \times n$ unit matrix.

A similar relation  for
\begin{eqnarray}
J&=& \lt(
    \begin{array}{cc}
       J_a
       & 0 \\
       J_b & J_c
    \end{array} \rt) \no
\end{eqnarray}
can be derived from \rf{defj}
for  the case that $J_c$ is just a number:
\begin{eqnarray}
J_b &=& \lt[- \frac{\go_b}{2 \beta_0}
            + \frac{\beta_1}{\beta_0} d_b - d_b J_a
            + J_c d_b
        \rt]  \, \lt[ \lt( 1+ d_c \rt) \uma _n - d_a   \rt]^ {-1}
\label{jinh}
\end{eqnarray}
The relations \rf{uinh} and \rf{jinh} are useful to check the
RG evolution in the calculation of $\eta_3$ in chapter \ref{calc},
because the subspace $\mr{S_a}$ is related to the Wilson coefficients
for \dso\/ transitions which is known to NLO from \cite{bjlw}.

\section{Scheme Dependence of Coefficients and Anomalous Dimension}
General aspects of renormalization scheme dependences have already
been discussed in sect.\ \ref{secrs}. In the framework of effective
field theories scheme dependences appear in a new context: Even when
keeping the scheme in the full Standard Model lagrangian fixed
there is still the freedom to invoke a finite renormalization
on the operator renormalization matrix \rf{defqren} in analogy
to \rf{finren}.
{}From now on we will only discuss such  scheme dependences related to
the effective theory.

Collecting some results of \cite{bjlw}
let us write the matrix element in the form:
\begin{eqnarray}
\langle \vec{\Qh}  \rangle &=& \langle \vec{\Qh}  \rangle^{(0)} +
    \frac{g^2}{16 \pi^2} \langle \vec{\Qh}  \rangle^{(1)} +O(g^4)
\label{matex} \\
&=& \lt[ \uma +    \frac{g^2}{16 \pi^2} r +O(g^4)     \rt]
\langle \vec{\Qh}  \rangle^{(0)} \label{matexr}
\end{eqnarray}
If the matrix $r$ obtained in two schemes $a$ and $b$ differs
by $\Delta r= r_b - r_a$, one finds the corresponding finite
renormalization as
\begin{eqnarray}
Z_a &=& Z_b \lt[ \uma + \frac{g^2}{16 \pi^2} \Delta r +O(g^4)     \rt]
+ O(\e) \label{ztrafo}
\end{eqnarray}
This leads to the scheme independence of $\gz$, while $\go$
transforms as
\begin{eqnarray}
\go_b &=& \go_a + \lt[\Delta r, \gz  \rt] +2 \beta_0 \Delta r .
\label{antrafo}
\end{eqnarray}
This yields for $J$ defined in \rf{defj}:
\begin{eqnarray}
J_b &=& J_a - \Delta r^T ,\label{jtrafo}
\end{eqnarray}
so that $J + r^T$ is scheme independent.

{}From \rf{matex} also the scheme transformation of the Wilson coefficient
is clear:
\begin{eqnarray}
\vec{C}^T_b &=& \vec{C}^T_a \lt[\uma -
         \frac{g^2}{16 \pi^2} \Delta r +O(g^4)     \rt] .
\label{witrafo}
\end{eqnarray}
With \rf{evowil1},\rf{nlos},\rf{witrafo} and \rf{jtrafo} one finds
the scheme dependence related to the upper end $\mu_0$
of the RG evolution canceled in the RG improved Wilson
coefficient \rf{evowil1}. The scheme dependence of the lower end
$\mu$ has to cancel with a corresponding dependence in the
hadronic matrix element.

Now one is often confronted with the situation that one calculates
the matrix element \rf{matex} with different
\emph{renormalization prescriptions}. This is related to the fact
that in many cases mathematical structures such as the
Dirac algebra defined in 4 dimensions
cannot unambiguosly be extended to $D$ dimensions. This phenomenon
is related to the evanescent redefinitions mentioned in sect.\ \ref{secrs}:
A priori one can modify any $D$--dimensional quantity by terms
of order $\e$ without changing its limit for $D\rightarrow 4$.
One then has to check which of the $D$--dimensional definitions leads
to a consistent renormalizable theory complying with the observed
symmetries of the described physics.
Hence one has to prove that two different renormalization
prescriptions leading to different $r$'s in \rf{matex} really
correspond to renormalization schemes. A necessary condition is that
the change in the
renormalization prescription is equivalent to a finite
renormalization as in \rf{ztrafo}.  This has been done for the
important question of how to treat $\g_5$ in $D$ dimensions
in e.g.\ \cite{bw}, \cite{bjlw} and \cite{bjlw2} and will be done
in chapter \ref{eva} for the question of how to define evanescent
operators.

\section{Unphysical Operators}\label{unphy}
Effective lagrangians also contain unphysical operators. We have
already alluded to the presence of evanescent operators, which
will be analyzed in detail in chapter \ref{eva}.

\subsection{Equation of Motion and BRS Exact Operators}\label{eom}
Another set of unphysical operators is related to the Euler--Lagrange
equation of motion (EOM) derived from the lagrangian. In classical physics
one can simply drop terms which vanish by the field equation of
motion. In a quantized field theory the issue is more difficult,
because the path integral involves the integration over all field
configurations, not just those which obey the EOM.  The second subtlety
stems from the fact that in Green's functions of operators containing
derivatives the latter must always be understood to act on the
time ordering $\T$ as well. This is necessary for the Green's
function to be a correctly defined operator--valued distribution.
(Otherwise it could not be Fourier--transformed to momentum space
giving the familiar $-i k^\mu$ for each $\partial^\mu$.)
The third difficulty concerns the  renomalization process, as two
operators whose difference vanishes by the EOM may have bare Green's
functions with a different forest structure.

The correct implementation of the EOM in quantum field theory also
depends on the regularization, in BPHZ renormalization the use of the
EOM requires oversubtractions, while this is not so for dimensional
regularization.

First analyses of the subject used the EOM derived from the lagrangian
\rf{qcd}.
Yet for perturbative calculations the lagrangian \rf{renqcd}
containing a gauge fixing term and Faddeev--Popov (FP) ghosts should
be used.
In the following we will collect the results obtained with this
modern approach \cite{jl,arz,pol,colleom}. We recommend \cite{sim} for
a nice explanation of the subject. Hence the EOM we use will always
include the terms from $\lag_\mr{gf}$ and $\lag_\mr{FP}$

The necessary theorems we will be exemplified with operators
of the forthcoming calculation of  \kkm.
We will be confronted with the following \dso operator:
\begin{eqnarray}
\Qheom &=& \ov{s} \g_\mu L T^a d \cdot
    \lt[ - \var{\lag_\mr{QCD}}{A_\mu^a} + \partial_\nu
           \var{\lag_\mr{QCD}}{\lt( \partial_\nu A_\mu^a \rt) }     \rt]
\label{defeom}
\end{eqnarray}
with $L=1-\g_5$. This operator with the
EOM of the gluon field attached to the \dso
current contains the couplings depicted in
\fig{glueop} (p.~\pageref{glueop}),
and in the right pictures of
\fig{fermping} (p.~\pageref{fermping}) and
\fig{ghostping} (p.~\pageref{ghostping}).
Four--fermion Green's functions with one insertion of $\Qheom$
vanish \cite{colleom}.
This statement holds for both  bare and renormalized Green's functions.
If external gluons are involved, the
matrix element of  $\Qheom$ equals a sum of matrix elements
of $\ov{s} \g_\mu L T^a d $ in each of which one of the external gluon
fields is missing. These \emph{contact terms}\/ vanish after
the LSZ reduction \rf{lsz}, because they miss a gluon pole.
Hence $\Qheom$ is termed \emph{on--shell equivalent}\/ to the zero
operator. The same, of course, is true for any operator vanishing
by the EOM for the fermion fields.

Another class of unphysical operators is related to $\lag_\mr{FP}$
and the BRS invariance discussed in sect.\ \ref{brs}.
We will involve the following operator:
\begin{eqnarray}
\Qhbrs &=& \frac{1}{g} \frac{1}{\xi}\,
         \ov{s} \g_\mu L T^a d \cdot
          \partial^\mu \partial_\nu A^\nu_a +
          \ov{s} \g_\mu L T^a d \cdot
     \lt( \partial^\mu \ov{\eta}_b \rt) \eta_c f_{abc}  .
\label{defbrs}
\end{eqnarray}
$\Qhbrs$ is the BRS variation of some other operator:
\begin{eqnarray}
\Qhbrs &=& \dbrs \widehat{Q}^\prime \; = \;
\dbrs
     \lt( \frac{1}{g}\, \ov{s} \g_\mu L T^a d
  \cdot \partial^\mu \ov{\eta}_a  \rt),
\end{eqnarray}
i.e.\ it is \emph{BRS--exact}.
{}From \rf{slav} one obtains:
\begin{eqnarray}
\!\!\! \bra{0} \Qhbrs^\ba     \ov{s} (x_1) d(x_2) \ov{\psi}(x_3) \psi (x_4)
\ket{0}
\!&=&\! - \bra{0} \widehat{Q}^{\prime, \ba} \dbrs   \lt(
 \ov{s} (x_1) d(x_2) \ov{\psi}(x_3) \psi (x_4) \rt) \ket{0},
\label{brszero}
\end{eqnarray}
which also does not survive LSZ reduction thereby being  zero
on--shell.

{}From the discussion it is clear that $\Qheom$ and $\Qhbrs$ do not
contribute to  the matching of Green's functions
with single operator insertions, because the matching can
(though need not) be done on--shell.

\subsection{Mixing}\label{eommix}
In order for the coefficients of  $\Qheom$ and $\Qhbrs$ to be irrelevant
one has to verify the block--triangular form of the mixing
matrix to ensure that $\Qheom$ and $\Qhbrs$ do not mix into physical
operators.

To discuss the mixing it is useful to introduce three classes of operators:

\ul{${\cal P}$}: The set of gauge invariant operators which do not
vanish by the EOM. Hence {\cal P} contains
physical and evanescent operators.

\ul{${\cal  E}$}: The set of operators vanishing by
the EOM such as $\Qheom$.

\ul{${\cal  B}$}: The set of BRS--exact operators. The
operator $\Qhbrs$ belongs to this class.

Now the renormalization matrix has the following form
\cite{colleom,jl}:
\begin{eqnarray}
Z^{-1} &=& \lt(
\begin{array}{ccc}
\;\; \ast \;\; & \;\; \ast \;\; & \;\; \ast \;\; \\[2mm]
0              & \;\; \ast \;\; & \;\; \ast \;\; \\[2mm]
0              & 0              & \;\; \ast \;\;
\end{array} \rt)
\quad \quad \mbox{corresponding to} \quad \quad
\lt(
\begin{array}{ccc}
{\cal P} \\[2mm]
{\cal B} \\[2mm]
{\cal E}
\end{array} \rt) ,
\label{nomixeom}
\end{eqnarray}
i.e.\ the renormalized version of \rf{brszero} involves operators from
${\cal B} \oplus {\cal E}$ as counterterms.

The block--triangular form of \rf{nomixeom} allows to ignore the
operators from ${\cal E}$ and ${\cal B}$, because they neither
contribute to the matching nor do their coefficients mix
into the coefficients of the operators from ${\cal P}$. The
lagrangian with the renormalized operators of
${\cal B} \oplus {\cal E}$
removed is sometimes called \emph{reduced}\/ lagrangian. In a
calculation beyond one--loop, however, the operators of
${\cal B} \oplus {\cal E}$
appear in subloop counterterm
diagrams. On can even avoid operators from ${\cal B}$ altogether
by using a \emph{background field gauge}. In our case of
\kkm\/ the use of a background field gauge does not simplify
the calculation.

Of course there are other non--gauge invariant operators which   belong
to none of the three classes. But since the initial values of their
coefficients are zero and none of the operators from
${\cal P} \oplus {\cal B} \oplus {\cal E}$
mixes into them, they clearly do not affect any physical coefficient.

\subsection{Double Insertions}\label{seceomdouble}
For double insertion the issue is more complex, because now the operators
in ${\cal B} \oplus {\cal E}$ give a non--zero contribution.
For a four--fermion matrix element containing one insertion of
$\Qheom$ one finds \cite{colleom}:
\begin{eqnarray}
\lefteqn{
\langle \int\!\!\!\! \int d^D x d^D y \T \Qheom (x) \wh{Q}(y)  \rangle
\; = } \nn
&& i  \sum_{\Lambda = 1 , \partial_\nu, \partial_\nu \partial_\rho
  }
(-1)^{P_\Lambda}
\langle \int\!\!\!\! \int d^D x d^D y \T
\Lambda \lt[ \ov{s} \g_\mu L T^a d (x) \rt]
\var{}{\lt( \Lambda \lt[ A_\mu^a (x)  \rt] \rt)} \wh{Q}(y)  \rangle
. \label{eomdouble}
\end{eqnarray}
Here the sum extends over all gluon field derivatives present
in $\Qheom$. $P_\Lambda$ means the number of derivatives in
$\Lambda $.
The main point in \rf{eomdouble} is that the variation
of $\wh{Q} (y)$
with respect to $A_\mu^a (x)$
is proportional to $\delta^{(D)} (x-y)$. Hence the matrix element
in \rf{eomdouble}
is identical to the matrix element of a single local operator
$\Qht_k(y)$. Since this holds for both  bare and renormalized
operators, we can
reduce the effective lagrangian  $\lag^\mr{II}$ by
substituting $\Qht_k$ for $\Qheom$.
 If we kept both of them,
the anomalous dimension
tensor defined in sect.\ \ref{doi} would have a degenerate eigenvalue.
The product of the two Wilson coefficients corresponding
to the operators of the RHS of \rf{eomdouble} would evolve parallely
to the coefficient $\wt{C}_l$ of $\Qht_l$
under the RG flow, and we could do the reduction of the effective
lagrangian at any scale. The method to perform the reduction,
i.e.\ to construct the local operator $\Qht_k$ is decribed e.g.\
in \cite{sim}. In a practical calculation, however, it is sufficient
to know that we can drop $\Qheom$ from the operator basis, because
we have to do the matching and mixing calculation
anyway with all operators of the desired dimension
in $\lag^\mr{II}$,
including the $\Qht_l$'s.

Next for the operators from ${\cal B}$ one finds with \rf{slav}:
\begin{eqnarray}
\langle   \Qbrs^\ba  \Qh^\ba    \rangle &=&
\langle \lt( \dbrs \Qh^{\prime \, \ba}  \rt) \Qh^\ba  \rangle \; = \;
- \langle  \Qh^{\prime\, \ba}  \dbrs \Qh^\ba    \rangle \nn
&&+ \mbox{on--shell vanishing terms} .  \label{brsdouble}
\end{eqnarray}
Here the RHS immediately vanishes, if
$\Qh \in {\cal P} \oplus  {\cal E}$,
because gauge--invariant operators are BRS--invariant.
For $\Qh \in {\cal B}$ we involve
$\dbrs^2$ on the RHS, which is either zero or yields an operator
vanishing by the EOM of the ghost field.
Recalling further that the renormalized version of \rf{brsdouble}
involves only operators from ${\cal B}\oplus {\cal E}$ one
finally realizes that one can also trade the renormalized
operators from ${\cal B}$ for suitable $\Qht_l$'s.
\cleardoublepage
\chapter{Evanescent Operators}\label{eva}
\section{Motivation}\label{Sect:Intro}
When  effective field theories containing four--fermion interactions
are regularized dimensionally,
evanescent operators, which vanish for $D=4$, enter the scene.
A prominent example is connected to the Fierz transformation:
Since  it is  allowed only for $D=4$,
the difference of some $D$--dimensional
operator and  its Fierz transformed one is  an evanescent
operator.
Since the lagrangian is a D-dimensional object, it
also contains these evanescent operators. The whole issue would be
trivial, if the evanescent sector was decoupled from the physical
sector in the matching and mixing process.

Yet
when calculating radiative corrections in a SU(N) gauge theory
to some matrix element of
a physical operator one has to face evanescent operators both in the
matching procedure \rf{exma1} and among the operator counterterms in
\rf{defqren}.   Let us exemplify this with the current--current
operator $\Qh$ corresponding to the Dirac structure
\begin{eqnarray}
Q&=&\g_{\mu } \lef \otimes \g^{\mu } \lef.
\label{q}
\end{eqnarray}
We will use this example frequently in this section together with
an anticommuting $\g_5$ (NDR scheme). The consistency of this scheme
in NLO QCD short distance calculations  has been demonstrated  in
\cite{bw}, \cite{bjlw} and \cite{bjlw2} for cases without closed
fermion loops. In the diagrams of the effective field theory the
latter can be avoided by passing to a Fierz transformed operator basis.
The use of an anticommuting
$\g_5$ in a renomalizable field theory  has been justified in \cite{krei}.
The general arguments in this chapter, however, are not restricted
to the NDR scheme.

\begin{figure}[htb]
\caption[One--loop current--current type radiative
corrections to a four--quark operator]{One--loop current--current type
           radiative corrections to
$\protect\langle \protect\widehat{Q}_k  \protect\rangle $.}
\label{ds1ccnlo}
\end{figure}
Now consider the one--loop matrix elements of $\Qh$ depicted in
\fig{ds1ccnlo}. The Dirac structure of the result is easily
expressed in terms of
a linear combination of $Q$ and an evanescent
Dirac structure:
\begin{eqnarray}
E_1\lt[ Q \rt]&=&\g_{\mu} \g_{\nu} \g_{\vartheta} (1-\g_5) \otimes
\g^{\vartheta} \g^{\nu} \g^{\mu } (1-\g_5) -
(4+a \varepsilon)  \g_{\mu} (1-\g_5) \otimes \g^{\mu} (1-\g_5) .
\label{exeva}
\end{eqnarray}
{}From the basis decomposition of the four--dimensional Dirac algebra
we know that \rf{exeva} vanishes in four dimensions. Nevertheless
\rf{exeva} appears with a factor of $1/\e$ in bare matrix elements
and therefore in counterterms to physical operators.
Depending on the flavour structure of $\Qh$ one must
take into account also the penguin--type diagrams of \fig{fermping}
(p.\ \pageref{fermping}). They do not contribute to the coefficients
of  evanescent operators in the one--loop order.
Hence $\Qh$ mixes into some evanescent operator $\wh{E}_1[Q]$. After
inserting $\wh{E}_1 [Q]$ into the diagrams of \fig{ds1ccnlo} another
evanescent operator  $\wh{E}_2 [Q]$ with five Dirac matrices
in each string comes into play. Hence one
realizes that one must deal with an infinite set of
evanescent operators, because the Dirac algebra is
infinite dimensional in dimensional regularization.
With the real parameter $a$ in \rf{exeva}
we  have displayed the arbitrariness in the definition of $E_1 [Q]$.
One can a priori add $\e$ times any physical operator (not only
proportional to $\Qh$) to any evanescent operator. Indeed, in the
literature one finds different definitions of the evanescents.

Let us display the evanescent
operators in the effective lagrangian \rf{eff1} in the following way:
\begin{eqnarray}
	\Lagr^{\rm I}
	&=&
	C_{k} Z^{-1}_{kl} \widehat{Q}^{\bare}_{l}
	+
	C_{k} Z^{-1}_{k E_{rl}} \widehat{E}_{r}\left[Q_{l}\right]^{\bare}
\nn
	& &
	+
	C_{E_{jk}} Z^{-1}_{E_{jk} l} \widehat{Q}^{\bare}_{l}
	+
	C_{E_{jk}} Z^{-1}_{E_{jk} E_{rl}} \widehat{E}_{r}\left[Q_{l}\right]^{\bare}
	.
\label{Lagr1}
\end{eqnarray}
Here and in the following we will distinguish the renormalization
constants related to some evanescent operator $E_j[Q_m]$ by denoting
the corresponding index with $E_{jm}$.

Buras and Weisz were the first to recognize
that evanescents cannot simply be neglected but
influence physical quantities \cite{bw}. They focused on the
evanescent operators' contribution to the matching \rf{exma1}
and found that the one--loop matrix elements of $E_1[Q]$ involve
finite components proportional to physical operators. In \cite{bw} it has
been proposed to cancel these components, which are local, by a finite
renormalization of $E_1[Q]$. By this the evanescent operators do not
contribute to the matching in \rf{exma1} anymore and their
undetermined coefficients $C_{E_{jk}}$ are irrelevant at the matching
scale $\mu_0$.
{}From \rf{and1} the authors of \cite{bw} found the influence
of this finite renormalization on the physical part of the
NLO anomalous dimension matrix $\go$.

To insure that evanescent operators remain irrelevant at any scale
one must insure
that the evanescents do not mix into physical operators.
As discussed in sect.\ \ref{btmm} such a
block--triangular
anomalous dimension matrix
protects the physical coefficients $C_k$ in \rf{Lagr1}
from admixtures of the  $C_{E_{jk}}$.
This has been noticed first by Dugan and Grinstein in \cite{dg}.
They, however, have used a definition of the evanescents different
from that of \cite{bw}, i.e.\ a different $a$ in \rf{exeva}.
We do  not need to repeat the definitions used in \cite{bw}
and \cite{dg} in this introductory section to understand the issue.
The authors of \cite{dg}
proved that the
anomalous dimension
matrix indeed has the desired block--triangular form
using their special definition of the evanescents, if
the finite renormalization of \cite{bw}   is performed.
Yet it is not clear at all that these features holds for any definition
of the evanescents: An evanescent redefinition of e.g.\  $E_1[Q]$ in
\rf{exeva} by adding $\e \cdot \delta a \cdot Q$ to it
modifies the components
of the
divergent parts of the bare two--loop matrix elements
proportional to physical operators.
Hence we are immediately confronted with the question:

\emph{Does the definition of the evanescents employed in \cite{bw}
also yield a block--triangular anomalous dimension matrix?}

We will answer this question affirmatively in sect.\ \ref{Sect:Triang}.
There we will also generalize the method of \cite{bw}.
Especially one is not forced to
use the definition of the evanescent operators proposed in \cite{dg},
whose implementation is quite cumbersome.

Then the next point to discuss is:

\emph{Is the dependence on $a$ in \rf{exeva} spurious or does it
      influence the physical Wilson coefficients $C_k$ and the
      physical portion of the anomalous dimension matrix $\g$?}

We will find in sect.\ \ref{Sect:Scheme} that the latter is the case.
Here we will also find that evanescent redefinitions corresponding
to a change of $a$  in
\rf{exeva} imply a renormalization scheme dependence  in the
physical sector, i.e.\ we will be able to find a quantitity
$\Delta r$ such that the Wilson coefficients and the mixing matrix
transforms as in \rf{witrafo} and \rf{antrafo}. This touches a
non--trivial aspect: It is not clear from the beginning
that the change of the
definition of the evanescents does not spoil renormalizability
of the theory, i.e.\ the locality of the divergences. Yet in
NLO it is trivial that renormalizability holds for any definition
of the  $E_{jk}$'s, because
a redefiniton as in \rf{exeva} only reshuffles local terms in
the divergences of the bare matrix elements. When going beyond
NLO terms of order $\e^2$, etc., in \rf{exeva} become important,
and these terms might be constrained by the condition of
maintaining renormalizability. We will not discuss this point.
In fact we will extend the discussion only beyond NLO when addressing
the all--order argument of \cite{dg} for the block--triangular
mixing matrix.

A  main subject of this thesis is the correct renormalization of
bilocal structures with two coloured operators. Clearly for this
we have to answer the question:

\emph{What is the correct method to handle evanescent operators in
      matrix elements with two operator insertions?}

This point will be the subject of sections \ref{Sect:Double}
and \ref{Sect:Inclusive}, where four--fermion processes and inclusive decays
will be discussed.

\section{Preliminaries and Notation}
\label{Sect:Prelim}
Let us consider a set of physical four--quark operators
$\{ \widehat{Q}_k=\ov{\psi} q_k \psi \cdot \ov{\psi} \widetilde{q}_k \psi,
\; k=1,2,3,\ldots \}$ as shown in \rf{qba}. For simplicity
we restrict the basis to dimension-six operators.
As usual we are interested in the  Green's functions
of a SU(N) gauge theory with insertions of $\widehat{Q}_k$
renormalized in a mass-independent scheme. For simplicity of the
notation we pick
minimal subtraction (${\rm MS}$).
The arguments are easily generalized to other schemes
like $\overline{\rm MS}$.
The dirac structures $Q_k = q_k \otimes \widetilde{q}_k$ corresponding to
$\widehat{Q}_{k}$ are considered to form a basis of the space of Lorentz
singlets and pseudosinglets for $D=4$.
Neither the Lorentz indices of $q_k$ and $\widetilde{q}_k$ are
displayed nor any flavour or colour indices, which are irrelevant for the
discussion of the subject.
$\lt[ \Gamma \otimes 1 \rt] Q_k \lt[ 1 \otimes \Gamma ^\prime \rt]$
means
$\Gamma q_k \otimes \widetilde{q}_k \Gamma ^\prime$.

For the purpose of this chapter we define the perturbative expansion
of the  matrix elements of  $\widehat{Q}_k$ differently from \rf{matex}:
\begin{eqnarray}
Z_\psi ^2
\langle \widehat{Q}_k^{\ba}  \rangle
  &=&
 \sum _{j \geq 0} \lt( \frac{g^2}{16 \pi ^2}  \rt)^j
\langle \widehat{Q}_k ^{\ba} \rangle ^{(j) }. \label{me}
\end{eqnarray}
We have extracted the wave function renormalization here, because
it is not related to the issue of the mixing of evanescent operators.

Now the insertion of $\widehat{Q}_k$ into the one-loop diagrams of
fig.~\ref{ds1ccnlo} yields a linear combination of the $\widehat{Q}_l$'s and
a new operator with the Dirac structure
$Q_k^\prime= \lt[ \g_\rho \g_\sigma  \otimes 1 \rt]
  Q_k \lt[ 1 \otimes \g^\sigma \g^\rho  \rt]$:
\begin{eqnarray}
\langle \widehat{Q}_k ^{\ba} \rangle ^{(1) } &=&
d^{(1)}_{kl} \langle \widehat{Q}_l \rangle ^{(0)} +
d^{(1)}_{k , Q^\prime _k}
   \langle \widehat{Q}^\prime _k \rangle ^{(0)} \quad \ns k \label{d1},
\end{eqnarray}
where $\langle \ldots \rangle^{(0)}$ denote tree level matrix elements.
Both coefficients have a term proportional to $1/\varepsilon $ and a finite
part. $\widehat{Q}_k^\prime$ is now decomposed into a linear combination of
the $\widehat{Q}_l$'s and an evanescent operator:
\begin{eqnarray}
\widehat{Q}_k^{\prime \ba} &=& \lt( f_{kl} + a_{kl} \varepsilon \rt)
   \widehat{Q}_l ^{\ba }
      + \widehat{E}_1[ Q_k]^{\ba } \label{e1}
+O(\varepsilon ^2).
\label{DefEvan1}
\end{eqnarray}
Here the $f_{kl}$'s  are uniquely determined by the
Dirac basis decomposition
in $D=4$ dimensions. The
$a_{kl}$'s, however, are arbitrary, and a different choice for the
$a_{kl}$'s  corresponds to a different definition of
$\widehat{E}_1 [ Q_k ]=\widehat{E}_1 [ Q_k, \{ a_{rs} \} ]$.
When going beyond the one-loop order new evanescent operators
$\widehat{E}_2 [ Q_k ], \widehat{E}_3 [ Q_k ], \ldots $
will appear. Their precise definition is irrelevant for the moment and
will be given after (\ref{defe2}).

Now in the framework of dimensional regularization the renormalization
of some physical operator $\widehat{Q}_k$ requires counterterms
proportional to physical and evanescent operators:
{}From \rf{Lagr1} we find
\begin{eqnarray}
Z_\psi ^2 \langle \widehat{Q}_k ^{\ba} \rangle   &=& Z_{kl} \langle
      \widehat{Q}_l ^{\re} \rangle  +
  Z_{k,E_{jm}} \langle \widehat{E}_j [Q_m] ^{\re} \rangle  \quad.
\label{z}
\end{eqnarray}
(\ref{d1}) and (\ref{e1}) imply that $Z_{kl}^{(1)}$ depends on the
$a_{rs}$'s,
while $Z_{k,E_{jm}}^{(1)}$ is independent of
them. The definition of   the coefficients in the expansion of
$Z$ in terms of the gauge coupling constant $g$ has been given in
\rf{exp}.

Next we describe the method of
Buras and Weisz \cite{bw} for the treatment of evanescent operators:
They have determined the  $a_{kl}$'s by choosing some
set of Dirac structures $M= \{ \g^{(1)} \otimes \widetilde{\g} ^{(1)}, \ldots
 \g^{(10)} \otimes \widetilde{\g} ^{(10)}  \} $,
which forms a basis for $D=4$, and contracting all elements in $M$ with
$Q_k^\prime$ and $Q_l$ in (\ref{e1}):
\begin{eqnarray}
\tr \left( \g^{(m)} q_k ^\prime \widetilde{\g} ^{(m)} \widetilde{q}_k ^\prime
\right)  &=&
(f_{kl} + a_{kl} \varepsilon ) \,
\tr \left( \g^{(m)} q_l \widetilde{\g} ^{(m)} \widetilde{q}_l  \right)
+O(\varepsilon ^2) , \no \\
&& \quad \quad \quad
 \ns k \mbox{ and on $m$\/=1,\ldots , 10}. \label{gp}
\end{eqnarray}
The solution of the  equations (\ref{gp}) uniquely defines the
$f_{kl} + a_{kl} \varepsilon $.
In other words, $E_1[Q_k ]$ obeys the equations:
\begin{eqnarray}
E_1[Q_k]_{ijrs} \g^{(m)}_{si} \widetilde{\g} ^{(m)}_{jr} &=&
 O ( \varepsilon^2 ) \quad \quad \quad
\mbox{ for $m$\/=1, \ldots ,10},
\end{eqnarray}
where $i,j,r,s$ are Dirac indices.

Our arguments will not depend on the scheme used for the treatment of
$\g_5$.
The use of a totally anticommuting $\g_5$
 does not cause any ambiguity in the trace operation in
(\ref{gp}), because all Lorentz indices are contracted, so that the
traced Dirac string is a linear combination of $\g_5$
and the unit matrix.

E.g.\ the choice of
\begin{eqnarray}
M&=& \{ 1\otimes1, 1\otimes \g_5, \g_5 \otimes 1, \g_5 \otimes \g _5 ,
   \g_{\mu} \otimes \g ^{\mu},
   \g_{\mu} \otimes \g ^{\mu} \g_5, \nn
&& \quad
   \g_5 \g_{\mu} \otimes \g^{\mu},
   \g_5 \g_{\mu} \otimes \g^{\mu} \g_5,
   \sigma_{\mu \nu } \otimes \sigma ^{\mu \nu },
   \g_5 \sigma_{\mu \nu } \otimes \sigma ^{\mu \nu }
         \}  \label{m}
\end{eqnarray}
gives  for $Q$ in (\ref{q})
\begin{eqnarray}
Q^\prime &\! = \!  &
 \g_\rho \g_\sigma
\g_{\mu } \lef \otimes \g^{\mu } \g^\sigma   \g^\rho \lef  =
( 4- 8 \varepsilon ) Q + E_1[Q]
 +O(\varepsilon^2) \label{ex}
\end{eqnarray}
as in \cite{bw}. We remark that this choice $a=-8$ respects the Fierz
symmetry, which relates the first to the second diagram in
fig.~\ref{ds1ccnlo}.

A basis different from $M$ in (\ref{m}) yields the same
$f_{k,l}$'s, but different $a_{kl}$'s. For example
by replacing the sixth and eighth element of $M$ in (\ref{m}) by
$\g_\alpha \g_\beta \g_\delta \otimes \g ^\alpha \g^\beta \g^\delta $
and
$\g_5 \g_\alpha \g_\beta \g_\delta \otimes \g ^\alpha \g^\beta \g^\delta $
one finds
\begin{eqnarray}
Q^\prime &=&  4  Q + 16 \varepsilon
   (1+\g_5) \otimes (1-\g_5) + E_1^\prime [Q_k] +O(\varepsilon^2), \no
\end{eqnarray}
instead of (\ref{ex}),
i.e.\   a different evanescent operator.
The Dirac algebra is infinite dimensional for non-integer $D$ and is
spanned by $M$ and an infinite set of evanescent Dirac structures.
Hence one can reverse the above procedure  and first arbitrarily choose the
$a_{kl}$'s and then add properly  adjusted
 linear combinations of the evanescent
structures  to the elements of $M$ such as to obtain the chosen $a_{kl}$'s.

Yet the so defined evanescent operators do not decouple from the
physics in four dimensions:
In \cite{bw} it  has    been observed that their one-loop
 matrix elements generally have nonvanishing components proportional
to the physical operators $Q_k$:
\begin{eqnarray}
\langle \widehat{E}_1 [Q_k ] ^{\ba} \rangle ^{(1)}  &=&
     \lt[ Z^{(1)}_0 \rt]_{E_{1k},l}  \langle \widehat{Q}_l \rangle ^{(0)} +
       \frac{1}{\varepsilon }
     \lt[ Z^{(1)}_1 \rt]_{E_{1k},E_{1k} }
   \langle \widehat{E}_1 [Q_k ] \rangle ^{(0)} \nn
&&   +  \frac{1}{\varepsilon }
     \lt[ Z^{(1)}_1 \rt]_{E_{1k},E_{2k} }
   \langle \widehat{E}_2 [Q_k ] \rangle ^{(0)}
  +O(\varepsilon) \label{e2}.
\end{eqnarray}
Here  a second evanescent operator $\widehat{E}_2$, which will be discussed in
a
moment,  has appeared.
Clearly no sum on $k$ is
understood in (\ref{e2}) and in following analogous places.
In (\ref{e2}) $[Z^{(1)}_0]_{E_{1k},l}$ is
local, because it originates from the local  $1/\varepsilon$--pole
of the tensor integrals and a term proportional to $\varepsilon$
stemming from the evanescent Dirac algebra.
For the same reason there is no divergence in the term proportional to
$\langle \widehat{Q}_l \rangle ^{(0)} $.
Now in \cite{bw} it has been proposed to renormalize $\widehat{E}_{1}$ by a
finite amount to cancel this component:
\begin{eqnarray}
  \widehat{E}_{1} [Q_k ]  ^{ \re }    &=&
  \widehat{E}_{1} [Q_k ]  ^{ \ba }  + \frac{g^2}{16 \pi ^2 } \lt\{
     - \lt[ Z^{(1)}_0 \rt]_{E_{1k},l}   \widehat{Q}_l  \rt.
\nn
&& \quad \quad
-     \frac{1}{\varepsilon }
     \lt[ Z^{(1)}_1 \rt]_{E_{1k},E_{1k} }
     \widehat{E}_1 [Q_k]      \nn
&& \quad \quad \lt.  -    \frac{1}{\varepsilon }
     \lt[ Z^{(1)}_1 \rt]_{E_{1k},E_{2k} }
    \widehat{E}_2 [Q_k ]  \rt\} +O \lt( g^4 \rt).
\label{fin}
\end{eqnarray}
With (\ref{fin}) the renormalized matrix elements
of the evanescent operators are $O(\varepsilon )$, so that they
do not contribute to the one-loop matching of some  Green's function
$G^{\re}$ in the full renormalizable theory
with matrix elements in the effective theory:
\begin{eqnarray}
 i G^{\re } &=& C_l \langle \widehat{Q}_{l}  \rangle ^{\re}
     + C_{ E_{1k} } \langle \widehat{E}_{1} [Q_k]  \rangle ^{\re}
     + O \lt( g^4  \rt), \label{match}
\end{eqnarray}
i.e.\ the coefficients $C_{ E_{1k} }$ are irrelevant, because they
multiply  matrix elements which vanish for $D=4$.
In \cite{bw} it has been further noticed
that $ Z ^ {(1)}_0 $ in
(\ref{fin}) influences the two-loop anomalous dimension matrix of the
{\em physical} operators, so that the presence of evanescent operators
indeed has an  impact on physical observables.

Next we discuss $\widehat{E}_2[Q_k]$, which has entered the scene in
(\ref{e2}): When inserting
$\widehat{E}_1[Q_k]$ defined in (\ref{e1})
into the one-loop diagrams of fig.~\ref{ds1ccnlo}, one involves
\begin{eqnarray}
Q_k^{\prime \prime} &=&
\lt[ \g_\rho \g_\sigma  \otimes 1 \rt]
  Q_k^\prime \lt[ 1 \otimes \g^\sigma \g^\rho  \rt]
\nn
&=&  \lt[ f +a \varepsilon  \rt]^2_{\, kl}
                       Q_l
      + \lt( f_{kl} + a_{kl} \varepsilon \rt) E_1[Q_l] \nn
&&  +  \lt[ \g_\rho \g_\sigma  \otimes 1 \rt]
  E_1 [ Q_k ] \lt[ 1 \otimes \g^\sigma \g^\rho  \rt]
 \label{ins} \\
&=&      \lt\{ \lt[ f + a \varepsilon \rt]^2 _{kl} + b_{kl}
              \varepsilon \rt\} Q_l
  + E_2 [Q_k]
  + O\lt( \varepsilon ^2   \rt)
    , \label{defe2}
\end{eqnarray}
which defines $E_2[Q_k]= E_2[Q_k,\{ a_{rs} \} , \{ b_{rs} \}   ]$.
Only the last term in
(\ref{ins}) can contribute to the
new coefficients $b_{kl}$. If the projection is performed
with e.g.\ $M$ defined in  (\ref{m}), one finds
$b_{kl}=0$\footnote{This is the case for any basis $M$ in which
for each
$\g^{(m)} \otimes \g^{(m)} \in M$ the quantity
$\g_\rho \g_\sigma \g^{(m)} \g^\sigma \g^\rho    \otimes \g^{(m)} $
is a linear combination of the elements in $M$.}.
In our discussion we will keep $b_{kl}$ arbitrary.
Clearly, one has a priori to deal with the mixing of an infinite
set of evanescent operators $\lt\{ \widehat{E}_j[Q_k] \rt\}$ for each physical
operator $\widehat{Q}_k$,
where $\widehat{E}_{j+1}[Q_k]$ denotes the new evanescent operator
appearing first in the one-loop matrix elements of $\widehat{E}_j[Q_k]$.

With the finite renormalization of $\widehat{E}_1[Q_k]$ in
(\ref{fin}) the evanescent operators  do not affect the
physics at the matching scale,
at which   (\ref{match}) holds.
In order that this will be true at any scale $\mu $, however,
one must
also ensure that the evanescent operators do not mix into the physical ones,
as noticed first by Dugan and Grinstein in \cite{dg}.
In their analysis they have introduced another way to define
the evanescent operators,
which is also frequently used:
It is easy to see that one can restrict the operator basis
$ \{  Q_k \} $
to the set of operators whose Dirac structures
$q_k, \widetilde{q}_k$
are completely antisymmetric in their
Lorentz indices.
Dirac strings being antisymmetric in more than four indices vanish in
four dimensions and are therefore evanescent.
Operators with five
antisymmetrized indices correspond to  $\widehat{E}_1$
in our notation, and $\widehat{E}_2$ would be expressed
in terms of a linear combination of
Dirac structures with seven and with five antisymmetrized indices.
Clearly this method also
corresponds to some special choice for the $a_{kl}$'s and $b_{kl}$'s
in (\ref{e1}) and (\ref{defe2}).
Now in \cite{dg}
the authors have proven that with the use of those definitions and a finite
renormalization analogous to (\ref{fin}) the anomalous dimension matrix
indeed has the desired block-triangular form, so that the
evanescent operators do not mix into the physical ones.
While the  anomalous dimension matrix is trivially block-triangular
at one-loop level, the proof for the  two-loop level
was given in \cite{dg} by the use of the abovementioned
special definition of the evanescent operators. The latter, however,
has some very special features, which are absent for the general case
with arbitrary  $a_{kl}$'s and $b_{kl}$'s, e.g.\ the definition
used in  \cite{dg} automatically yields an anomalous dimension matrix
which is tridiagonal in the evanescent sector.

This immediately raises the question whether the more general method
of \cite{bw}
also yields a block-triangular anomalous dimension matrix
for any chosen projection basis, i.e.\ any chosen set
$\{ a_{ kl} \}$. In the following section we will prove that this is
indeed the case and, more generally,  that one may also choose
the $b_{kl}$'s in (\ref{defe2}) completely arbitrary.

\section{Block Triangular Anomalous Dimension Matrix}
\label{Sect:Triang}
Consider some set of physical operators $\{ \widehat{Q}_k \}$ which closes
under renormalization together with the corresponding evanescent
operators $\{ \widehat{E}_j[Q_k] : j \geq 1 \}$. Their $O(\varepsilon)$--parts
$a_{rs},b_{rs}, \ldots $ are chosen arbitrarily.  We want to show that
the block of the anomalous dimension matrix describing the mixing of
$\widehat{E}_j[Q_k]$ into $\widehat{Q}_l$ equals zero,
\begin{eqnarray}
\lt[ \g \rt]_{E_{jk},l} &=& 0,   \label{nomix}
\end{eqnarray}
provided one uses the finite renormalization described in (\ref{fin}).

Our sketch will follow the outline of \cite{dg}, where
(\ref{nomix}) has been proven by complete induction.
At the one-loop level (\ref{nomix}) is trivial, and the induction starts
in two-loop order:
The next-to-leading order contribution to the anomalous dimension
matrix has been stated in \rf{and1}.
The nonzero  contributions to (\ref{nomix}) in two-loop order are
\begin{eqnarray}
	\lt[ \g^{(1)} \rt]_{ E_{jk},l }
	&=&
	-4  \lt[ Z_1^{(2)} \rt]_{ E_{jk},l }
	-2 \beta_0 \lt[ Z_0^{(1)} \rt] _{ E_{jk},l }
\nn
	& &
	+ 2 \lt\{
		\lt[ Z_1^{(1)} \rt]_{ E_{jk},E_{rs} }
		\lt[ Z_0^{(1)}\rt] _{ E_{rs},l }
		+
		\lt[ Z_0^{(1)}  \rt] _{ E_{jk},m }
		\lt[ Z_1^{(1)}  \rt] _{ m l  }
	    \rt\}.
\label{twoloop}
\end{eqnarray}
Here (\ref{twoloop}) contains terms which are absent when the  special
definition of the evanescent operators in \cite{dg} is used:
In \cite{dg} one has $\lt[ Z^{(1)} \rt]_{E_{jk},l}=0 $
for $j \geq 2$ contrary to  the general case, where any evanescent
operator can have counterterms proportional to physical operators.

Next we look at $\lt[ Z_1^{(2)} \rt]_{E_{jk},l}$, which stems from the
$1/\varepsilon$--term of the $O(g^4)$--matrix elements
of $\widehat{E}_j[Q_k]$. As discussed in \cite{dg}, these
$1/\varepsilon$--terms
originate from
$1/\varepsilon^2$--poles in the tensor integrals
multiplying a factor proportional to $\varepsilon$ stemming from the
evanescent Dirac algebra.
Now in each two-loop diagram the former are related to the
corresponding one-loop counterterm diagrams by a factor of 1/2,
because the non-local $1/\varepsilon$--poles cancel in their sum
\cite{hoco}.
For this  it is crucial that the one-loop counterterms are
properly adjusted, i.e.\ that they cancel the
$1/\varepsilon$--poles in the one-loop tensor integrals.
In the one-loop matrix elements of evanescent operators the latter
are multiplied with $\varepsilon$ originating from the Dirac algebra.
Hence the proper one-loop renormalization of the evanescent operators
must be such as to give matrix elements of order $\e$, as shown
for $E_1[Q_k]$ in (\ref{fin}).

{}From the  one-loop counterterm graphs one simply reads off:
\begin{eqnarray}
\lt[ Z_1^{(2)}  \rt]_{E_{jk},l}
\!\!\!  &=& \!\!
   \frac{1}{2}  \lt\{
      \lt[ Z_0^{(1)}  \rt]_{E_{jk},m}  \!
      \lt[ Z_1^{(1)}  \rt]_{ ml} \! \!
+  \lt[ Z_1^{(1)}  \rt]_{E_{jk},E_{rs}} \!
\lt[ Z_0^{(1)}  \rt]_{E_{rs},l} \! \!
- \beta_0
\lt[ Z_0^{(1)}  \rt]_{E_{jk},l}  \rt\},    \no
\end{eqnarray}
which yields the desired result when inserted into (\ref{twoloop}).
Here the first two terms stem from insertions of physical and
evanescent counterterms to $E_j[Q_k]$, while the term involving
the coefficient of the one-loop $\beta$--function
$\beta (g) = - g^3/(16 \pi^2) \beta_0 $
originates from the diagrams with coupling constant counterterms.
The terms involving the  wave function renormalization constants
cancel with those stemming from the factor  $Z_\psi^2$ in
(\ref{z}).

The inductive step in \cite{dg} proving
(\ref{nomix}) to any loop order does not use any special definition
of the evanescent and therefore applies unchanged here.

\section{Evanescent Scheme Dependences}
\label{Sect:Scheme}
In this section we will analyze  the dependence of the
physical part of $\g^{(1)}$ given in (\ref{and1})
and of the one-loop Wilson coefficients on
$a_{il}$ and $b_{il}$.
In practical next-to-leading order calculations one often has to
combine Wilson coefficients and anomalous dimension matrices
obtained with different definitions of the evanescent operators
and it is therefore important to have formulae allowing to
switch between them (see e.g.\ appendix B of \cite{m}).

We start with the investigation of the dependence of
$\g ^{(1)}$ on $a_{il}$.
The bare one-loop matrix element
\begin{eqnarray}
\langle \widehat{Q}_k ^{\ba} \rangle^{(1) }  &=&
\lt\{
  \frac{1}{\varepsilon}  \lt[ Z_1^{(1)}  \rt]_{kj}
  +  \lt[ d_0^{(1)} \rt]_{kj} \rt\}  \langle  \widehat{Q}_j \rangle^{(0)}
+ \frac{1}{\varepsilon} \lt[ Z_1^{(1)}  \rt]_{k,E_{1k}}
  \langle  \widehat{E}_1 [Q_k]   \rangle ^{(0)} \nn
&&+ O( \varepsilon)
\label{q1}
\end{eqnarray}
is independent of $a_{il}$, which is evident from
(\ref{d1}).
$E_1[Q_k]$ depends linearly on $a_{il}$ through its definition
(\ref{e1}) with the coefficient
\begin{eqnarray}
\da \widehat{E}_1[Q_k] &=& - \varepsilon \, \delta _{ki} \,  \widehat{Q}_l,
\label{DiffE1}
\end{eqnarray}
so that  (\ref{q1}) gives:
\begin{eqnarray}
\da \lt[ d_0^{(1)} \rt]_{kj}  &=& \lt[ Z_1^{(1)} \rt]_{k,E_{1k}}
\delta _{ki} \delta_{lj}, \label{d0}
\end{eqnarray}
while $Z_1^{(1)}$ is independent of $a_{il}$.

In the same way on can obtain the $a_{kl}$--dependence
of $Z_1^{(2)}$. (\ref{z}) reads to two-loop order (cf.~(\ref{me})):
\begin{eqnarray}
\langle \widehat{Q}_k ^{\ba} \rangle^{(2) }
&=& Z_{kj}^{(2)}
    \langle \widehat{Q}_j  \rangle^{(0)}
+   Z_{k,E_{1m}}^{(2)}  \langle \widehat{E}_1 [Q_m]   \rangle^{(0) }
+   Z_{k,E_{2m}}^{(2)}  \langle \widehat{E}_2 [Q_m]   \rangle^{(0) }
\nn
&&   + Z_{kr}^{(1)}  \langle \widehat{Q}_r ^{\re} \rangle^{(1) }
+  Z_{k,E_{1k}}^{(1)} \langle \widehat{E}_1 [Q_k] ^{\re}  \rangle^{(1)}
+   \langle \widehat{Q}_k ^{\re} \rangle^{(2) }.
\label{q2}
\end{eqnarray}
{}From (\ref{defe2}) we know
\begin{eqnarray}
\da \widehat{E}_2 [Q_m] &=& - \varepsilon \lt[ f_{mi} \delta _{lj} +
\delta_{mi} f_{lj}    \rt] \widehat{Q}_j ,
\end{eqnarray}
and from (\ref{q1}) one reads off:
\begin{eqnarray}
    \langle \widehat{Q}_r ^{\re} \rangle^{(1) }  &=&
 \lt[ d_0^{(1)} \rt]_{rj}     \langle \widehat{Q}_j  \rangle^{(0)}
\label{qren}.
\end{eqnarray}
These relations and (\ref{d0}) allow to calculate the derivative
of  (\ref{q2}) with respect to $a_{il}$.
Keeping in mind that the evanescent matrix elements are $O(\e )$
the $O(1/ \e)$--part of the derivative yields:
\begin{eqnarray}
\da \lt[ Z_1^{(2)}  \rt] _{kj}  &=&
-  \lt[ Z_1^{(1)}  \rt]_{ki}
  \lt[ Z_1^{(1)}  \rt]_{i,E_{1i}} \delta_{lj} +
  \lt[ Z_2^{(2)}  \rt]_{k,E_{1i}}  \delta_{lj} \nn
&& + \lt[ Z_2 ^ {(2)}  \rt]_{k,E_{2m}}
 \lt( \delta_{mi} f_{lj} + f_{mi} \delta_{lj}    \rt)
,\quad \ns i. \label{z2}
\end{eqnarray}
We pause here to stress that \rf{z2} is local showing that an
evanescent redefinition does not spoil the renormalizability of the
physical sector in NLO.

Next we extract  $\lt[ Z_2^{(2)} \rt]$
from the one-loop counterterm diagrams as described
in the preceding section:
\begin{eqnarray}
\lt[ Z_2^{(2)}  \rt]_{k,E_{1i}}
&=&  \frac{1}{2}
\lt[ Z_1^{(1)}  \rt]_{ki} \lt[ Z_1^{(1)}  \rt]_{i,E_{1i} } +
 \frac{1}{2}
\lt[ Z_1^{(1)}  \rt]_{i,E_{1i}} \lt[ Z_1^{(1)}  \rt]_{E_{1i},E_{1i} }
\delta _{ki} \nn
&& - \, \frac{1}{2} \beta _0
\lt[ Z_1^{(1)}  \rt]_{i,E_{1i} } \delta_{ki} , \quad \quad \ns i  \nn
\lt[ Z_2^{(2)}  \rt]_{k,E_{2m}}
&=&
 \frac{1}{2}
\lt[ Z_1^{(1)}  \rt]_{k,E_{1k}} \lt[ Z_1^{(1)}  \rt]_{ E_{1k},E_{2k} }
\delta_{km}
 , \quad \quad \ns k.
\label{coun}
\end{eqnarray}
After inserting (\ref{coun}) into (\ref{z2}) we want to
substitute the last term in (\ref{z2}).  For this we
derive both sides of (\ref{e2})
with respect to $a_{il}$
giving:
\begin{eqnarray}
\lefteqn{ \lt[ Z_1^{(1)}   \rt]_{E_{1k} ,E_{2k} }
\lt( \delta_{ki} f_{lj} + f_{ki} \delta_{lj}    \rt)
\; =} \nn
&& \da \lt[ Z_0^{(1)}   \rt]_{E_{1k},j}
 +  \lt[ Z_1^{(1)}   \rt]_{lj} \delta_{ki}  -
\lt[ Z_1^{(1)}   \rt]_{E_{1k} ,E_{1k} } \delta_{ki} \delta_{lj}, \label{z0}
\quad  \ns k.
\end{eqnarray}
Finally one has to insert the expression for  (\ref{z2})
obtained by the described substitutions
into
\begin{eqnarray}
\da \lt[ \g^ {(1)} \rt]_{kj} &=&
-4 \da \lt[ Z_1^{(2)}   \rt]_{kj}
+ 2 \lt[ Z_ 1 ^ {(1)}  \rt]_{k,E_{1k}}
\da \lt[ Z_0^{(1)} \rt] _{E_{1k},j}, \quad \ns k , \no
\end{eqnarray}
which follows from (\ref{and1}).
The result reads:
\begin{eqnarray}
\da \lt[ \g ^{(1)} \rt]_{kj} &=&
- 2 \lt[ Z_1^{(1)} \rt]_{lj} \lt[ Z_1^{(1)} \rt]_{i,E_{1i}} \delta _{ki}
 + 2 \lt[ Z_1^{(1)} \rt]_{ki} \lt[ Z_1^{(1)} \rt]_{i,E_{1i}}
  \delta_{lj} \nn
&& +
 2 \beta_0 \lt[ Z_1^{(1)} \rt]_{i,E_{1i}} \delta_{ki} \delta_{lj},
      \quad \quad \ns i.  \label{dag}
\end{eqnarray}
Since the quantities on the right hand side of (\ref{dag}) do not
depend on $a_{il}$, one can easily integrate (\ref{dag}) to find
the desired relation between two $\g$'s corresponding to  different
choices for $a_{kl}$
in (\ref{e1}).
To write the result in matrix form we recall the expression for the
physical one-loop anomalous dimension matrix
\begin{eqnarray}
\lt[ \g ^{(0)} \rt]_{lj} &=& -2 \lt[ Z_1^{(1)} \rt] _{lj} \no
\end{eqnarray}
and introduce the diagonal matrix $D$ with
\begin{eqnarray}
D_{im} &=&  \lt[ Z_1^{(1)} \rt] _{i,E_{1i}} \delta_{im}
   , \quad \quad \quad \ns i.
\label{DiagMatrix1}
\end{eqnarray}
Hence
\begin{eqnarray}
\g ^{(1)} ( a^\prime)&=&
\g ^{(1)} (a ) + \lt[ D \cdot ( a^\prime -a ), \g^{(0)} \rt]
       + 2 \, \beta_0 \, D \cdot (a^\prime -a)  ,   \label{result}
\end{eqnarray}
where the summation in the row and column indices only runs over
the physical submatrices.

(\ref{result}) exhibits  the familiar structure of the scheme
dependence of $\g ^{(1)}$ as given in \rf{antrafo}. The matrix $J$
in \rf{jtrafo} consequently reads:
\begin{eqnarray}
	J(a^\prime)
	&=&
	J(a)
	- \left[ D \cdot (a^\prime -a) \right]^{T}.
\label{SchemeJ1}
\end{eqnarray}
Hence we have found a scheme dependence which is caused by
an evanescent redefinition of a
{\em bare} operator basis
(i.e.\ of the bare evanescent operators).

The dependence of the one-loop matrix elements on $a$ can be found
easily from (\ref{qren}) and (\ref{d0}):
\begin{eqnarray}
\langle \vec{Q} \rangle ^{\re}  (a^\prime) &=&
 \lt[ 1+ \frac{g^2}{16 \pi^2} D \cdot (a^\prime-a)    \rt]
  \langle \vec{Q}  \rangle ^{\re} (a) + O( g^4).
\label{depma}
\end{eqnarray}
Since in (\ref{match}) $G$ does not depend on $a$ and the evanescent
matrix element is $O(\e )$, the corresponding relation for the
Wilson coefficients at the matching scale reads:
\begin{eqnarray}
 \vec{C}^T  (a^\prime) &=&
   \vec{C}^T  (a)
 \lt[ 1 -  \frac{g^2}{16 \pi^2} D \cdot (a^\prime-a)    \rt]
+ O( g^4).
\label{SchemeWC1}
\end{eqnarray}
Thus also the Wilson coefficient transforms correctly as in
\rf{witrafo} and the scheme dependence  cancels between
coefficients and anomalous dimension matrix.

In the same way one can investigate the dependence of
$\g^{(1)}$ on $b_{il}$ given in (\ref{defe2}): While $Z_1^{(2)}$
and $Z_0^{(1)}$ depend on $b_{il}$, this dependence cancels in
(\ref{and1}). Hence  neither $\g^{(1)}$ nor the one-loop
Wilson coefficient are affected by the choice of $b_{il}$.
Thus in NLO the $b_{il}$'s do not induce a scheme dependence.
By keeping $b_{il}$ arbitrary in a practical calculation one
has a non--trivial  check of the correct handling of the evanescents:
The $b_{il}$'s disappear from the result
only  after inserting the evanescent counterterms
with a factor of 1/2.

The nice feature of \rf{depma},
\rf{SchemeWC1} and \rf{result} is that the summed
matrix indices only run over the physical subset. From
\rf{antrafo} and \rf{witrafo} we could have expected
a formula where the summation runs over the whole operator basis
including the evanescents. This is why we could not simply deduce
\rf{result} from \rf{depma} using \rf{ztrafo} and \rf{antrafo}.
Possible contributions from summations over
evanescent operator indices  cannot be inferred from
\rf{depma}, because there they would multiply vanishing matrix
elements.

We have already stressed in the end of
sect.\ \ref{sre} that it is advantageous
to have a $\g^{(1)}$ which commutes with $\gz$.
Now one can use (\ref{result}) to  simplify $\g^{(1)}$: By going to
the diagonal basis for
\mbox{$\g^{(0)}=\mbox{diag}\left(\g^{(0)}_i\right)$}
one can easily find solutions for $a^\prime - a$ in (\ref{result})
which even give \mbox{$\g^{(1)}(a^\prime)=0$} provided that all
$Z_{k,E_{1k}}$'s are nonzero and all eigenvalues of $\g^{(0)}$
satisfy \mbox{$\left| \g^{(0)}_i -\g^{(0)}_j \right| \neq 2 \beta_0$}.
We will exemplify this in a moment.

A choice for $a_{il}$ which leads to a $\g^{(1)}$ commuting with
$\g^{(0)}$ has been done implicitly  in \cite{bw}:
There the mixing of the two operators
$Q_+ = Q \lt( {\rm 1} + \widetilde{\rm 1} \rt) /2 $ and
$Q_- = Q \lt( {\rm 1} - \widetilde{\rm 1} \rt) /2 $ has been considered,
where ${\rm 1}$ and $\widetilde{\rm 1}$ denote colour singlet
and antisinglet and $Q$ was introduced in (\ref{q}).
Now $Q_+$ is self-conjugate under the Fierz transformation,
while $Q_-$ is anti-self-conjugate, so that
$\g^{(0)}$ is diagonal to maintain the
Fierz symmetry in the leading order renormalization group
evolution.
As remarked after (\ref{ex}), the definition of $E_1[Q]$
in (\ref{ex}) is necessary to ensure the  Fierz symmetry
in the  one-loop matrix elements.  Consequently
with (\ref{ex}) also $\g^{(1)}$ has to obey the Fierz symmetry
preventing the mixing of $Q_+$ and $Q_-$, i.e.\ yielding
a diagonal $\g^{(1)}$. A different definition of $E_1[Q]$
would  result in non-Fierz-symmetric matrix elements,
but in renormalization scheme independent expressions
they would combine with  a non-diagonal $\g^{(1)}$
such as to restore Fierz symmetry.

Let us consider the example above to demonstrate that one can
pick $a^\prime$ such that $\gamma ^{(1)} (a ^\prime ) =0$:
In  \cite{bw} the definitions
\begin{eqnarray}
 E_1[ Q_\pm] & =&
 \lt(   \pm
1- \frac{1}{N} \rt)
   \lt[  \g_\rho \g_\sigma \g_\nu (1-\g_5) \otimes
     \g^\nu \g^\sigma \g^\rho (1-\g_5) \rt.  \nn
&& \quad \quad \lt.   -  (4 - 8 \varepsilon)\;\;
  \g_\nu (1-\g_5) \otimes \g ^\nu (1-\g_5)
    \rt], \no
\end{eqnarray}
i.e.\ $a_{++}=8 (1/N-1), a_{--}=8 (1/N+1), a_{+-}=a_{-+}=0 $,  were
adopted yielding a diagonal
$\g^{(1)}(a)=$diag$\lt(\g^{(1)}_{+}(a),\g^{(1)}_{-}(a)\rt)$.
{}From the insertion of $Q_+$ and $Q_-$ into the diagrams of
fig.~\ref{ds1ccnlo} one finds $Z_{+,E_{1+}}=Z_{-,E_{1-}}=1/4$.
Hence if we pick
$a^\prime_{\pm \pm}=a_{\pm \pm} - 2/\g^{(1)}_{\pm} (a)/\beta_0$ and
$a^\prime_{\pm \mp}=0$, (\ref{result}) will imply
$\g^{(1)}(a^\prime)=0$.

\section{Double Insertions}
\label{Sect:Double}
\subsection{Motivation}
\label{Sect:DoubleMotiv}
{}From the
preceding sections we know that the
coefficients $C_{E_{jk}}$
of properly renormalized
evanescents in \rf{Lagr1}
are  irrelevant for physical quantities and
remain undetermined.
In the following we will
extend these results to 4-fermion
Green's functions with two insertions of local operators.

For the discussion of Green's functions with insertion of two local
operators as depicted in
fig.~\ref{bi}
we first display the evanescents in
$\lag^\mr{II}$  in \rf{eff2}:
\begin{eqnarray}
	\Lagr
	&=&
	\Lagr^{\rm I}
	+
	\Lagr^{\rm II}
\label{Lagr}
\\
	\Lagr^{\rm II}
	&=&
	C_{k} C_{k'}
	\left\{
		Z^{-1}_{kk',l} \widetilde{Q}^{\bare}_{l}
		+
		Z^{-1}_{kk',E_{rl}}
			\widetilde{E}_{r}\left[\widetilde{Q}_{l}\right]^{\bare}
	\right\}
\no \\
	& &
	+
	C_{k} C_{E_{j' k'}}
	\left\{
		Z^{-1}_{k E_{j' k'},l} \widetilde{Q}^{\bare}_{l}
		+
		Z^{-1}_{k E_{j' k'},E_{rl}}
			\widetilde{E}_{r}\left[\widetilde{Q}_{l}\right]^{\bare}
	\right\}
\no \\
	& &
	+
	C_{E_{j k}} C_{E_{j' k'}}
	\left\{
		Z^{-1}_{E_{jk} E_{j' k'},l} \widetilde{Q}^{\bare}_{l}
		+
		Z^{-1}_{E_{jk} E_{j' k'},E_{rl}}
			\widetilde{E}_{r}\left[\widetilde{Q}_{l}\right]^{\bare}
	\right\}
\no \\
	& &
	+
	\widetilde{C}_{k} \widetilde{Z}^{-1}_{kl} \widetilde{Q}^{\bare}_{l}
	+
	\widetilde{C}_{k} \widetilde{Z}^{-1}_{k E_{rl}}
		\widetilde{E}_{r}\left[\widetilde{Q}_{l}\right]^{\bare}
\no \\
	& &
	+
	\widetilde{C}_{E_{jk}} \widetilde{Z}^{-1}_{E_{jk} l} \widetilde{Q}^{\bare}_{l}
	+
	\widetilde{C}_{E_{jk}} \widetilde{Z}^{-1}_{E_{jk} E_{rl}}
		\widetilde{E}_{r}\left[\widetilde{Q}_{l}\right]^{\bare}
\label{Lagr2}
\end{eqnarray}
Since the $\Qh_k$'s have dimension six,
the $\widetilde{Q}_{l}$'s are dimension--eight operators.
For simplicity, we assume the $\widetilde{Q}_{l}$'s to be linearly
independent from the $\widehat{Q}_{k}$'s,
e.g.\ the $\widehat{Q}_{k}$'s represent $\Delta S = 1$
	operators and  the $\widetilde{Q}_{l}$'s denote $\Delta S = 2$
	operators.
The $E_{r}\left[\widetilde{Q}_{l}\right]$ in \eq{Lagr2} are defined
analogously to \eq{DefEvan1} with new coefficients $\widetilde{f}_{kl}$,
$\widetilde{a}_{kl}$, $\widetilde{b}_{kl}$, etc.
Hence new arbitrary constants $\widetilde{a}_{kl}$, $\widetilde{b}_{kl}$
potentially causing scheme dependences enter the scene.

Clearly the following questions arise here:
\begin{enumerate}
\item
\label{Q1}
Are the coefficient functions $C_{E_{jk}}$ irrelevant also for the
double insertions;
i.e.\ do
\begin{eqnarray}
\left\langle \int \T \widehat{E} \widehat{Q} \right\rangle
\hspace{1cm} \mbox{and} \hspace{1cm}
\left\langle \int \T \widehat{E} \widehat{E} \right\rangle
\label{DoubleWithEva}
\end{eqnarray}
contribute to the matching procedure and the operator mixing?
\item
\label{Q2}
Does one need a {\em finite} renormalization in the evanescent sector
of double insertions;
if yes, how does this affect the anomalous dimension tensor?
\item
\label{Q3}
How do the $\widetilde{C}_{l}$ and anomalous dimension matrices depend on
the $a_{kl}$, $b_{kl}$, $\widetilde{a}_{kl}$, $\widetilde{b}_{kl}$ ?
\item
\label{Q4}
Are the RG improved observables scheme independent?
\end{enumerate}

\subsection{Scheme Consistency}
\label{Sect:DoubleConsist}
In this section we will carry out the program of
section~\ref{Sect:Triang} for the case of double insertions to answer
questions~\ref{Q1} and \ref{Q2} of page~\pageref{Q1}.

We have mentioned already in sect.\ \ref{dom} that two
cases have to be distinguished:
The matrix element of the double insertion of the two local
renormalized operators can be divergent or finite:
\begin{eqnarray}
	\left\langle \frac{i}{2} \int \T \widehat{Q}_{k} \widehat{Q}_{k'}
		\right\rangle
	&=&
	\left\{
	\begin{array}{lcl}
	\mbox{divergent} & , & \mbox{case 1} \\
	\mbox{finite} & , & \mbox{case 2}
	\end{array}
	\right.
	.
\label{cases}
\end{eqnarray}
Therefore we need or do not need a separate renormalization for the
double insertion
\begin{eqnarray}
	Z^{-1}_{k k',l}
	\left\{
	\begin{array}{clcl}
	\neq & 0 & , & \mbox{case 1} \\
	=    & 0 & , & \mbox{case 2}
	\end{array}
	\right.
	.
\end{eqnarray}

Let us start the discussion with the matching procedure \ref{exma2}:
\begin{eqnarray}
	-i G^{\ren}
	&=&
	C_{k} C_{k'} \left\langle \left[\frac{i}{2} \int \T
		\widehat{Q}_{k} \widehat{Q}_{k'} \right]^{\ren} \right\rangle
	+
	C_{k} C_{E_{i' k'}} \left\langle \left[i \int \T
		\widehat{Q}_{k} \widehat{E}_{i'}\left[Q_{k'}\right]
		\right]^{\ren} \right\rangle
\no \\
	& &
	+
	C_{E_{ik}} C_{E_{i' k'}} \left\langle \left[\frac{i}{2} \int
		\T \widehat{E}_{i}\left[Q_{k}\right]
		\widehat{E}_{i'}\left[Q_{k'}\right]
		\right]^{\ren} \right\rangle
	+
	C_{l} \left\langle \widetilde{Q}_{l} \right\rangle
\no \\
	& &
	+
	\widetilde{C}_{E_{jl}} \left\langle
		\widetilde{E}_{j}\left[\widetilde{Q}_{l}\right]
		\right\rangle
	,
\end{eqnarray}
where $G^{\ren}$ corresponds e.g.\ to the  box--function
shown in \fig{box} and the radiative corrections
depicted in \fig{boxqcd}.

Since the coefficients $C_{E_{jk}}$ must be irrelevant for this
matching procedure, one must have
\begin{eqnarray}
	Z_{\psi}^{2}
	\left\langle \left[ \frac{i}{2} \int \T
	\widehat{E}_{j}\left[Q_{k}\right] \widehat{Q}_{k'} \right]^{\ren}
	\right\rangle
	&\stackrel{!}{=}&
	\left\{
		\begin{array}{l@{\hspace{1cm}}l}
		O\left(\eps^{0}\right) & \mbox{in case 1 (LO)} \\
		O\left(\eps^{1}\right) & \mbox{in case 1 (NLO and higher)} \\
		O\left(\eps^{1}\right) & \mbox{in case 2}
		\end{array}
	\right.
\no \\
	& &
\label{CondMatchDouble}
\end{eqnarray}
and the analogous relation for two insertions of evanescent operators.
Recall from sect.\ \ref{dom} that
in case~1 the LO matching is performed by
the comparison of the coefficients of logarithms of the full theory
amplitude and the effective theory matrix element \eq{CondMatchDouble}
(the latter being trivially related to the coefficient of the
divergence),
while the NLO matching is obtained from the finite part and also
involves the matrix elements of the local operators.
In case~2 the matching is performed with the finite parts in all
orders.
In both cases the condition \eq{CondMatchDouble} is trivially
fulfilled in LO, because the evanescent Dirac algebra gives an
additional $\eps$ compared to the case of the insertion of two
physical operators.
Therefore a finite renormalization for the double insertion turns out
to be unnecessary at the LO level.
This statement remains valid at the NLO level only in case~2, in
case~1 condition \eq{CondMatchDouble} no longer holds if one only
subtracts the divergent terms in the matrix elements containing a
double insertion.
With the argumentation preceding \eq{fin} one finds that in this case
the finite term needed to satisfy the condition \eq{CondMatchDouble}
is local and therefore can be provided by a finite counterterm.

The operator mixing
can be analyzed analogously to section~\ref{Sect:Triang}.
Using the finite renormalization enforcing
\eq{CondMatchDouble} and the locality of counterterms, one
shows for the anomalous dimension tensor \rf{gamma1double}:
\begin{eqnarray}
	\gamma^{\left(0\right)}_{E_{rk}l,n}
	&=&
	\gamma^{\left(1\right)}_{E_{rk}l,n}
	=
	0
	\hspace{1cm}
	\mbox{and}
	\hspace{1cm}
	\gamma^{\left(0\right)}_{E_{rk}E_{sl},n}
	=
	\gamma^{\left(1\right)}_{E_{rk}E_{sl},n}
	=
	0
	,
\end{eqnarray}
i.e.\ a double insertion containing at least one evanescent operator
does not mix into physical operators.
Together with the statement that evanescent operators do not
contribute to the matching this proves our method to be consistent at
the NLO level.
As in the case of single insertions one can pick the $\widetilde{a}_{kl}$,
$\widetilde{b}_{kl}$,\ldots
completely arbitrary and then has to perform a finite
renormalization for the double insertions containing an evanescent
operator in \eq{DoubleWithEva}.
This statement remains valid also in higher orders of the SU(N)
interaction, which can be proven analogously to the proof given
by Dugan and Grinstein \cite{dg} for the case of single insertions.

Now we use the findings above to show the nonvanishing terms
in \eq{gamma1double} explicitly for the physical submatrix:
\begin{eqnarray}
	\gamma^{\left(1\right)}_{kn,l}
	&=&
	4 \left[Z^{-1,\left(2\right)}_{1}\right]_{kn,l}
	- 2 \left[Z^{-1,\left(1\right)}_{1}\right]_{kn,E_{1 l'}}
		\left[\widetilde{Z}^{-1,\left(1\right)}_{0}\right]_{E_{1 l'}l}
\no \\
	& &
	- 2 \left[Z^{-1,\left(1\right)}_{1}\right]_{k E_{1 k'}}
		\left[Z^{-1,\left(1\right)}_{0}\right]_{E_{1 k'} n,l}
	- 2 \left[Z^{-1,\left(1\right)}_{1}\right]_{n E_{1 n'}}
		\left[Z^{-1,\left(1\right)}_{0}\right]_{k E_{1 n'},l}
\no \\
	& &
\label{physdoub}
\end{eqnarray}
The last equation encodes the following rule for the correct
treatment of evanescent operators in NLO calculations:
{\em The correct contribution of evanescent operators to the NLO
	physical anomalous dimension tensor is obtained by inserting
	the  evanescent one-loop counterterms with a factor of $\frac{1}{2}$
	instead of\/ $1$ into the counterterm graphs.}
Hence the finding of \cite{bw} for a single operator insertion
generalizes to Green's functions with double insertions.
Here the second term in \eq{physdoub} corresponds to the graphs with
the insertion of a local evanescent  counterterm into the graphs
depicted in fig.~\ref{ds1ccnlo}, while the last to terms correspond to
the diagrams of fig.~\ref{bi} with one physical and one evanescent
operator.

\subsection{Double Insertions: Evanescent Scheme Dependences}
\label{Sect:DoubleScheme}

In this section we will answer questions~\ref{Q3} and \ref{Q4} on
page~\pageref{Q3}.
Let us first look at the dependence of the anomalous dimension tensor
$\gamma_{kk',l}$ on the coefficients $a_{rs}$.
First one notices, that the LO $\gamma^{\left(0\right)}_{kk',l}$ is
independent of the choice of the $a_{rs}$.
Similarly to the
procedure of section~\ref{Sect:Scheme} one derives
the following NLO relation:
\begin{eqnarray}
	\gamma^{\left(1\right)}_{kk',l}\left(a'\right)
	&=&
	\gamma^{\left(1\right)}_{kk',l}\left(a\right)
	+
	\left[D \cdot \left(a'-a\right)\right]_{ks}
		\gamma^{\left(0\right)}_{sk',l}
	+
	\left[D \cdot \left(a'-a\right)\right]_{k' s}
		\gamma^{\left(0\right)}_{sk,l}
\label{SchemeAnomBi1}
\end{eqnarray}
with the diagonal matrix $D$ from \eq{DiagMatrix1}.
Note again that the indices only run over the physical subspace.

The variation of the anomalous dimension tensor $\gamma_{k k',l}$ with
the coefficients $\widetilde{a}_{rs}$ again vanishes in LO, in NLO we find
the transformation
\begin{eqnarray}
	\gamma^{\left(1\right)}_{kk',l}\left(\widetilde{a}'\right)
	&=&
	\gamma^{\left(1\right)}_{kk',l}\left(\widetilde{a}\right)
	+
	\gamma^{\left(0\right)}_{kk',i}
		\left[\widetilde{Z}^{-1,\left(1\right)}_{1}\right]_{i E_{1i}}
		\left[\widetilde{a}'-\widetilde{a}\right]_{il}
	- 2 \beta_{0}
		\left[Z^{-1,\left(1\right)}_{1}\right]_{kk',\widetilde{E}_{1i}}
		\left[\widetilde{a}'-\widetilde{a}\right]_{il}
\no \\
	& &
	+ \left[
		\gamma^{\left(0\right)}_{kj}
		\delta_{k' j'}
		+
		\delta_{kj}
		\gamma^{\left(0\right)}_{k' j'}
		\right]
		\left[Z^{-1,\left(1\right)}_{1}\right]_{jj',\widetilde{E}_{1i}}
		\left[\widetilde{a}'-\widetilde{a}\right]_{il}
\no \\
	& &
	- \left[Z^{-1,\left(1\right)}_{1}\right]_{kk',\widetilde{E}_{1i}}
		\left[\widetilde{a}'-\widetilde{a}\right]_{is}
		\widetilde{\gamma}^{\left(0\right)}_{sl}
\label{SchemeAnomBi2}
\end{eqnarray}

As in the case of single insertions, up to the NLO level there exists
no dependence of $\gamma$ on the coefficients $b_{rs}$ and also no one
on the $\widetilde{b}_{rs}$.
This provides a nontrivial check of the treatment of evanescent
operators in a practical calculation, when the $b_{rs}$,
$\widetilde{b}_{rs}$ are kept arbitrary:
the individual renormalization factors $Z$ each exhibit a dependence
on the coefficients $b_{rs}$, $\widetilde{b}_{rs}$ but all this dependence
cancels, when the $Z$'s get combined to $\gamma$.
This has been done in the calculation  of $\eta_3$ in chapter
\ref{calc}.

Next we verify the scheme independence of RG improved
physical observables.
{}From the solution of the inhomogeneous RG equation \eq{inhsol}
we find the local operators' Wilson coefficients $\wt{C}_l$
independent under the transformations \rf{SchemeWC1},
\rf{SchemeJ1} and  \rf{SchemeAnomBi1}.

In a similar way one treats
the scheme dependence stemming from the coefficients $\widetilde{a}_{kl}$.
Here some work
has been necessary to prove the cancellation of the scheme dependence
connected to $g^{\prime 2} \wt{J}_{kk^\prime}$ and $\go_{nm,k^\prime}$
in \rf{inhsol}:
Although it is not possible to perform the integration in \rf{inhsol}
without transforming some of the operators to
the diagonal basis, one can do the integral for the scheme dependent
part of \rf{inhsol}, because the part of the integrand
depending on $\wt{a}_{kl}$'s
is a total  derivative with respect to $g$.
There is one important difference compared to the dependence
on the $a_{kl}$'s: A  scheme
dependence of the Wilson coefficient  related  to the lower end of
the RG evolution $\mu$ in \rf{inhsol} still remains.
This residual $\widetilde{a}_{kl}$ dependence must be canceled by a
corresponding one in the hadronic matrix element.
If the matrix elements are obtained in the parton model,
the dependence of the $\widetilde{C}_{l}$'s on $\widetilde{a}_{kl}$
cancels due to \rf{depma}.
Finally, as in the case of single insertions \cite{bjlw}, one can
define a scheme--independent Wilson coefficient for the local operator
\begin{eqnarray}
	\overline{\widetilde{C}_{l}\!\left(\mu\right)}
	&=&
	\left[
		\delta_{l l'}
		+
		\frac{g^2\left(\mu\right)}{16\pi^2}
		\cdot
		\widetilde{r}_{l' l}
	\right]
	\widetilde{C}_{l'}
	+
	\frac{g^2\left(\mu\right)}{16\pi^2}
	\cdot
	\widetilde{r}_{nm,l}
	\cdot
	C^{\left(0\right)}_{n}\!\left(\mu\right)
	C^{\left(0\right)}_{m}\!\left(\mu\right)
	+
	O\left(g^4\right)
	,
\no \\
	& &
\end{eqnarray}
which multiplies a scheme independent matrix element defined
accordingly.
It contains the analogue of $\widehat{r}$ in \rf{matexr} for the double
insertion
\begin{eqnarray}
	\left\langle \frac{i}{2} \int \T Q_{n} Q_{m}
		\right\rangle^{\left(0\right)}
	&=&
	\frac{g^2}{16\pi^2}
	\cdot
	\widetilde{r}_{nm,l}
	\cdot
	\left\langle \widetilde{Q}_{l} \right\rangle^{\left(0\right)}
	.
\label{defrtw}
\end{eqnarray}

\section{Inclusive Decays}
\label{Sect:Inclusive}

Inclusive decays are calculated either by calculating the exclusive
process and performing a subsequent phase space integration and a
summation over final polarizations etc.\ (referenced as method 1)
or by use of the optical theorem, which corresponds to taking the
imaginary part of the self-energy diagram depicted in fig.~\ref{Fig:3}
(method 2).
This figure shows that inclusive decays are in fact related to double
insertions, but in contrast to the case of section~\ref{Sect:Double}
they do not involve local four-quark operators as counterterms for
double insertions.
In fact, even local two-quark operator counterterms would only be
needed to renormalize the real part, but the imaginary part of their
matrix elements clearly vanishes.
The only scheme dependence to be discussed is therefore the one
associated with the $a_{kl}$'s, $b_{kl}$'s, etc., as there are no
$\widetilde{a}_{kl}$'s $\widetilde{b}_{kl}$'s, etc.\ involved.

To discuss the dependence on the $a_{kl}$'s it is nevertheless
advantageous to consider method~1, i.e.\ the exclusive process plus
the subsequent phase space integration.
{}From section~\ref{Sect:Scheme} we already know most of the properties
of the exclusive process:
At the upper renormalization scale the properly renormalized
evanescent operators do not contribute and the scheme dependence
cancels.
Further we know from \rf{SchemeJ1}
the scheme dependence of the (RG improved) Wilson
coefficients at the lower renormalization scale.
What we are left with is the calculation of the properly renormalized
operators in perturbation theory, i.e.\ with on-shell external
momenta.
Clearly the form of the external states does not affect the scheme
dependent terms of the matrix elements, they are again given by
\eq{depma} and therefore cancel trivially between the Wilson
coefficients and the matrix elements, because the scheme dependent
terms are independent of the external momenta.
Since we now have a finite amplitude which is scheme independent, we
may continue the calculation in four dimensions and may therefore forget
about the evanescent operators.
The remaining phase space integration and summation over final
polarizations does not introduce any new scheme dependence, therefore
we end up with a rate independent of the $a_{kl}$'s, $b_{kl}$'s, etc.
\footnote{We discard problems due to infrared singularities and the
	Bloch-Nordsiek theorem.
	At least in NLO one can use a gluon mass, because no
	three-gluon vertex contributes to the relevant diagrams}

Alternatively one may use the approach via the optical theorem
(method~2).
Then one has to calculate the imaginary parts of the diagram in
fig.~\ref{Fig:3} plus gluonic corrections.
\begin{figure}[htb]
	\caption[Inclusive decays and the optical theorem]{The
        lowest order self-energy diagram needed for the
	calculation of inclusive decays via the optical theorem
	(method~2).}
\label{Fig:3}
\end{figure}
Of course the properly renormalized operators have to be plugged in:
\begin{eqnarray}
	\imag \left\langle
	\widehat{O}^{\ren}_{k} \widehat{O}^{\ren}_{l}
	\right\rangle
\end{eqnarray}
One immediately ends up with a finite rate.
What is left to show is the consistency of the optical theorem
with the presence of evanescent operators and with their arbitrary
definition proposed in \eq{DefEvan1}, \eq{defe2}:
The result must not depend on whether the evanescent operators are
kept in the basis inserted into the diagram of \fig{Fig:3} or whether they
have been removed knowing the above result of method 1.
This means that evanescent operators must not contribute to the rate,
i.e.\ diagrams containing an insertion of one or two evanescent
operators must be of order $\eps$
\begin{eqnarray}
	\imag \left\langle
	\widehat{E}_{i}\left[O_{k}\right]^{\ren} \widehat{O}^{\ren}_{l}
	\right\rangle
	=
	O\left(\eps\right)
	\hspace{0.2cm}
	&\mbox{and}&
	\hspace{0.2cm}
	\imag \left\langle
	\widehat{E}_{i}\left[O_{k}\right]^{\ren}
	\widehat{E}_{j}\left[O_{l}\right]^{\ren}
	\right\rangle
	=
	O\left(\eps\right)
	.
\label{CondInclNoEva}
\end{eqnarray}
As in the previous sections one can discuss tensor integrals and Dirac
algebra separately leading to \eq{CondInclNoEva}.

\section{Summary and Outlook}
\label{Sect:Concl}
Let us list the results of this chapter:
\begin{itemize}
\item[i)]
It is allowed to redefine any  \emph{bare}\/
evanescent operator by $(D-4)$
times any physical
operator without affecting the block-triangular form of the
anomalous dimension matrix, which ensures that
properly renormalized
evanescent operators (as described in \cite{bw})
do not mix into physical ones.
\item[ii)]
We have analyzed the
renormalization scheme dependence associated with the
redefinition transformation in the next-to-leading order in
renormalization group improved perturbation theory. The
formulas transforming between different schemes have been derived.
It is meaningless to give some anomalous dimension
matrix or some Wilson coefficients beyond leading logarithms without
specifying the definition of the evanescent operators used during the
calculation.
In physical observables, however, this renormalization scheme dependence
cancels between Wilson coefficients and the anomalous dimension matrix.
One may take advantage of this feature by defining the
evanescent operators such as to achieve a simple form for the
anomalous dimension matrix.
\item[iii)]
We have extended the work of \cite{bw} and \cite{dg} and the
results of i) and ii)
to the case of Green's functions with two operator insertions.
This analysis is necessary for the correct treatment of evanescents
in particle--antiparticle mixing and rare decays.
Scheme-independent Wilson coefficients have been defined.
\item[iv)]
Finally we have analyzed the role of evanescents in
inclusive decay rates.
\end{itemize}

The issue of the renormalization of composite operators in dimensional
regularization is not restricted to four--quark operators.
We suggest that one may investigate to what extent evanescent
redefinitions are useful to gain insight into the
renormalization scheme
dependences of other operators. If, for example, some current $j$
is handled with two renormalization prescriptions denoted by $j_1$ and
$j_2$, one could try to reabsorb the scheme dependence  by
an evanescent redefinition $j_1 \rightarrow j_1^\prime =
(1 + a  \e + \ldots ) j_1 + \e \cdot (\mr{other\; currents})$. I.e.\
$j_1^\prime$ yields with prescription 1
the same finite parts of physical amplitudes  as $j_2$ renormalized
with prescription 2.
\cleardoublepage
\chapter{Effective Lagrangian for Flavour Changing Processes}\label{operator}
In this chapter we  will set up the operator basis needed for the
calculation of the \kkm\/  described in chapter \ref{calc}.
We will also explain the general structure of the effective lagrangian
and of the calculations which are necessary to obtain its RG improved
coefficients.
The basic
framework is very similar for the general case of
particle--antiparticle mixing. We will sometimes comment on
differences between the considered \dst transitions and
\dbt  or \dct amplitudes.

First the Standard Model diagrams contributing to the \kkm\/  will be
analyzed. Since \dst  transitions involving light quarks contain
\dso processes as substructures, sect.\ \ref{ds1} will discuss
them in detail. Here we will notice that the definition of the correct
\dso operator basis  involved in \dst amplitudes is much more
complicated than the one needed to describe the four quark amplitudes
appearing in \dso weak decays. The analysis of the operator basis
requires the careful application of the theorems of sect.\
\ref{unphy} on the use of the field equations of motion
and on BRS--exact operators.

\section{\textbf{\dst} transition in the Standard Model}
The \dst transition is a flavour--changing neutral current
(FCNC) process and therefore forbidden at tree level.
The lowest order contribution to it 
is depicted in \fig{box}.
\begin{figure}[ht]
\caption[$\Delta S \!\!=\!\!2$ box diagram]{
        The lowest order box diagrams mediating a \dst transition.
        The zigzag lines stand for W--bosons or
        ficticious Higgs particles. The diagrams rotated by
        $90^\circ$ must also be considered.}
\label{box}
\end{figure}
\subsection{Notations and Conventions}
Before writing down the result for the diagram of \fig{box},
we set up the notations used in the following.

The different contributions from the internal quarks involve different
CKM factors  $\lambda_j=V_{jd} V_{js}^{*}$. The GIM mechanism
$\lambda _t+\lambda _c +\lambda _u=0$ allows for the elimination
of  $\lambda _u $. Hence we can split up the Standard Model
Green's function as
\begin{eqnarray}
\wt{G} &=& \lambda_c^2 \wt{G}^c + \lambda_t^2 \wt{G}^t
 + 2   \lambda _c \lambda _t  \wt{G}^{ct} ,
\label{smbox}
\end{eqnarray}
which is understood to be truncated, connected and Fourier--transformed
into momentum space.
We will sometimes use the abbrevation $x_i=m_i^2/\mw $ for the
squared ratio of some quark mass and the W mass.
For the  W propagator the 't Hooft--Feynman gauge will be used,
while we keep the QCD gauge parameter $\xi$ arbitrary.
As only open fermion lines  appear in the calculation,  we
can safely use a naive anticommuting $\g_5$ (NDR scheme)
with $L=1-\gamma _5$ and $R=1+\gamma _5$.
$N$ is the number of colours,
and $\1$ and $\tw$ denote colour
singlet and antisinglet\footnote{This is clearly a
misnomer from a group theoretical point of view.},
i.e.\ $(L \otimes R) \cdot \tw$ stands for
$\ov{s}_i (1-\gamma _5)  d_j \cdot
\ov{s}_j (1+\gamma _5)  d_i$
with $j$ and $k$ being colour indices.
The $SU(N)$ Casimir factor involved will be $C_F=(N^2-1)/(2 N)$.

The result of \fig{box} is proportional to the tree--level
matrix element of
\begin{eqnarray}
\oll \; = \; (\ov{s} d)_{V-A}
(\ov{s} d)_{V-A} \1 &=&
( \ov{s} \gamma_\mu (1-\gamma_5) d)
(\ov{s} \gamma^\mu (1-\gamma_5) d) \cdot \1
 \label{oll}
\end{eqnarray}
shown in \fig{loc}.
In four dimensions the Fierz transformation maps $\oll$
onto an operator with the same Dirac structure, but $\1$ replaced
by $\tw$.
Hence if we substituted $\1$ by $\tw$ or by $(\1 +\tw)/2$
in the definition \rf{oll}, we would redefine
$\oll$ by an evanescent operator proportional to
$(\ov{s} d)_{V-A} (\ov{s} d)_{V-A} (\tw - \1) $.
If the evanescent operator $E_1[ \oll ]$ is defined as
in \rf{ex}, the Fierz symmetry is maintained on the loop level and
the replacement $\1 \rightarrow \tw $ does not change the two--loop
anomalous dimension.

The expansion
of matrix elements in terms of $\alpha$ has already been defined in
\rf{matex}. Analogously we will expand the $\wt{G}$'s in \rf{smbox},
\begin{eqnarray}
\wt{G}^{j} &=& \wt{G}^{j,\, (0)}
      + \frac{\alpha}{4 \pi} \wt{G}^{j,\, (1)} + O( \alpha^2 ),
\label{greenex}
\end{eqnarray}
and the Wilson coefficients:
\begin{eqnarray}
C_k &=& C_k^{(0)} + \frac{\alpha}{4 \pi} C_k^{(1)}
+ O( \alpha^2 ) . \label{wilsex}
\end{eqnarray}

The  $\wt{G}^{j,\, (1)}$'s  involve infrared
(mass) singularities, which will be regularized by small quark masses.

Finally the weak coupling constant $g_w$
will be traded for the Fermi constant $\gf$
according to \rf{fermi}.

\subsection{Zeroth Order Amplitude}
In the leading order of $m_\mr{light}/m_\mr{heavy}$, where
$m_ \mr{heavy}$ stands for $m_t$ or $M_W$ and $m_\mr{light}$
denotes any other massive parameter, one can neglect the
external momenta in \rf{smbox}.

One finds for the mixed top--charm contribution
in \rf{smbox}:
\begin{eqnarray}
i \wt{G}^{ct,(0)} &=& \frac{\gft}{16 \pi^2} \mw S(x_c,x_t)
\langle \oll \rangle^{(0)} .
\label{gct}
\end{eqnarray}
Here the \emph{Inami--Lim function} \cite{il} $S(x_j,x_k)$
is defined as
\begin{eqnarray}
S(x_j,x_k) &=& \wt{S}(x_j,x_k) -   \wt{S}(x_j,0) -   \wt{S}(x_k,0) +
\wt{S}(0,0)  ,
\end{eqnarray}
where the result of the box diagram with internal quarks $j$ and $k$
is denoted by $\wt{S}(x_j,x_k)$ and the u--quark mass is set to zero.
For $j=c,t$ one finds:
\begin{eqnarray}
i \wt{G}^j &=& \frac{\gft}{16 \pi^2} \mw S(x_j)
\langle \oll \rangle^{(0)} ,
\label{gj}
\end{eqnarray}
with $S(x_j)=S(x_j, x_j)$. Here one realizes that the effect of the GIM
mechanism is not only to forbid FCNC's at tree level, but also
to cancel the constant terms in the $\wt{S}$'s and to
nullify  \kkm\/ in the case of degenerate quark masses.

The reminder of the thesis will deal with
RG improved short distance QCD corrections in
\rf{smbox}, which are parametrized by
\begin{displaymath}
\begin{array}{rccl}
\eta_1 &
\mbox{on the RHS of \rf{gj}} & \mbox{for } &  \wt{G}^{c},  \\
\eta_2 &  \mbox{on the RHS of \rf{gj}}& \mbox{for } &  \wt{G}^{t}
\quad \mbox{and }\\
\eta_3 & \mbox{on the RHS of \rf{gct}} & \mbox{for } &  \wt{G}^{ct}.
\end{array}
\end{displaymath}
I.e.\ we want to derive the low energy hamiltonian given in \rf{s2}.

Let us first  look  at the three contributions
\rf{gct} and \rf{gj} to \rf{smbox}
with emphasis on large logarithms:
\begin{eqnarray}
S(x_t) &=& x_t \lt[ \frac{1}{4} + \frac{9}{4}
\frac{1}{1-x_t} - \frac{3}{2} \frac{1}{(1-x_t)^2}      \rt]
- \frac{3}{2} \lt[ \frac{x_t}{1-x_t} \rt]^3 \ln x_t
\label{sxt}
\end{eqnarray}
clearly involves no large logarithm because of
$\ln x_t \approx 1.5$. After multiplying with
$\alpha(M_W) /(4 \pi)$ this is of the order $10^{-2}$.
\begin{eqnarray}
S(x_c) &=& x_c + O(x_c^2)
\label{ilgim}
\end{eqnarray}
Here we have only kept  terms which are larger than those
of order $(m_s m_c)/\mw$ neglected  by setting the external
momenta to zero.
As we will see in the following section
one would naturally expect  a large logarithm $\ln x_c$
multiplying $x_c$
here. Its absence is due to the GIM mechanism.
\begin{eqnarray}
S(x_c,x_t) &=& - x_c \ln x_c
+x_c \lt[ \frac{x_t^2-8 x_t+4 }{4 (1-x_t)^2} \ln x_t
          + \frac{3}{4} \frac{x_t}{x_t-1}   \rt] + O(x_c^2 \ln x_c)
\label{sxcxt}
\end{eqnarray}
Here we encounter a large logarithm $|\ln x_c| \approx 8$.
According to our discussion in sect.\ \ref{dom} we can
sum $\ln x_c (\alpha \ln x_c)^n, n=0,1,2,\ldots$ with
the help of factorization and RG techniques.

$S(x_t)$ is much larger than $S(x_c)$ and $S(x_c,x_t)$, but it
is CKM suppressed in \rf{smbox}.
Yet this is not so in
\dbt\/ transitions, where $S(x_t)$
and its radiative corrections comprised in $\eta_2$
are clearly dominating.
\begin{figure}[tbp]
\caption[QCD corrections to the box diagram]{\protect
        The classes of diagrams constituing  the
        $O(\alpha)$--contribution
        to $\wt{G}$ in \rf{smbox};
        the remaining diagrams are obtained by left--right and
        up--down reflections.
        The curly lines denote gluons.
        Also QCD counterterm diagrams have to be included.
        Diagram ${\rm F}_8$ equals 0 for zero external momenta.
}
\label{boxqcd}
\end{figure}
\begin{figure}[btp]
\caption[The \dst operator]{
        The diagram for the matrix element of $\oll$
        in the effective three--quark theory to order
        $\alpha^0$.
        The cross denotes the insertion of the effective
        $\Delta S = 2$ operator.
        }
\label{loc}
\end{figure}\clearpage

\subsection[$O(\alpha)$ Radiative Corrections]{$\mathbf{O(\alpha)}$ Radiative
Corrections}
Since we are worried about the absence of the logarithm in
$S(x_c)$, let us look at the one--loop radiative corrections
to $\wt{G}^c$ depicted in \fig{boxqcd}.

These diagrams
have been calculated for arbitrary internal quark masses in
\cite{bjw}:
\begin{eqnarray}
i \wt{G}^{c \,(1)} (\mu ) \! \! &=& \! \! \frac{G_F^2}{16 \pi^2}
                \lambda_c^2 \mc (\mu ) \lt\{ \!
          \langle \oll \rangle ^{(0)} \wt{h} \lt( \mu \rt)+
      \langle T \rangle ^{(0)} h_T  +
       \langle U \rangle ^{(0)} h_U  \! \rt\}.  \label{h1}
\end{eqnarray}

In (\ref{h1}) new operators have emerged:
\begin{eqnarray}
\hat{T}&=& ( L \otimes L + R \otimes R  - \sigma _{\mu \nu}
         \otimes \sigma ^{\mu \nu} )\cdot \frac{N-1}{2 N} \1 , \nn
\hat{U} &=& \frac{1}{2} (  \gamma _\mu L \otimes \gamma ^\mu R +
               \gamma _\mu R \otimes \gamma ^\mu L ) \cdot
            \lt( \frac{N^2+N-1}{2 N} \1 - \frac{1}{2 N} \tw \rt) \nn
&& - (  L \otimes R + R \otimes L ) \cdot
       \lt( \frac{N^2+N-1}{2 N} \tw - \frac{1}{2 N} \1 \rt) , \label{u}
\end{eqnarray}
We have written $\hat{U}$ and $\hat{T}$
in a manifestly Fierz self--conjugate way.
The
functions in (\ref{h1}) are:
\begin{eqnarray}
\wt{h} (\mu )\! &=& \! C_F \lt[- 1- 6 \log\frac{\mc}{\mu^2 }
  +  \xi \lt(2-2 \frac{\ms \log (\ms/\mu^2) -\md
                             \log (\md/\mu^2)}{\ms-\md}
               \rt)        \rt]     \nn
&& + \frac{N-1}{2 N} \lt[- 11+\frac{4}{3} \pi^2
                       + 12 \log\frac{\mc}{\mw}
                       +3  \log\frac{\md \ms}{\mu^4}
                       - 6 \log\frac{\mc }{\mu ^2} \rt. \nn
 && \quad \hspace{1ex} \lt.   + \xi \lt( 2 + \log \frac{\md \ms}{\mu^4} -
                                   2 \frac{\ms \log(\ms/\mu^2) -
                                \md \log(\md/\mu^2) }{\ms-\md} \rt)
                                  \rt]  \nn
h_T  \! &=& \!  (-3 - \xi ) \nn
h_U  \! &=& \!
 \frac{3+\xi }{2} \frac{m_d m_s}{\ms-\md} \log \frac{\ms}{\md} ,\label{ltu}
\end{eqnarray}
Let us discuss \rf{h1} in more detail: First \rf{h1} is obviously unphysical,
because it is gauge dependent. This is an artefact of the use of small
quark masses to regularize the infrared singularities while
at the same time using zero on--shell quarks for the external states.
For the same reason we encounter the new operators $T$ and $U$.
When factorizing \rf{h1} to extract the Wilson coefficients $C_k$,
these unphysical parts will match their counterparts in the
effective theory with the zeroth order coefficient $C_k^{(0)}$.
Hence they do not contribute to the $C_k$'s. Yet we can anticipate this
by noticing that the terms involving $T,U$ and $\xi$ are independent
of $M_W$, which sets the scale for the short distance physics.
An alternative method to treat the infrared singularities is
to keep the quark masses exactly equal to zero and to rely on
dimensional regularization to cope with both IR and UV singularities.
This gives a different result, as the limit $m \rightarrow 0$ is
non--uniform owing to the fact that $m^\e/\e$ has a  zero
mass limit, while the expanded version $1/\e +\ln m$ diverges
in this limit. If  we had used such an on--shell IR regularization,
the additional operators $T$ and $U$ would be absent. Since  further
the wave function renormalization constant $Z_\psi$ equaled zero then,
\rf{h1} would turn out to be gauge independent, too.
The use of small quark masses for regularization has the advantage
that the vanishing of the otherwise invisible  IR terms from the
$C_k$'s provides a check of the calculation. Yet one may argue that this
method requires a more complicated calculation. In fact this is
not so as we will see in chapter \ref{calc}.

Having in mind the RG evolution down to scales $\mu_c$ of order $m_c$
we have arranged the logarithms in
$\wt{h}$ in (\ref{ltu})
such that one can easily distinguish
those which are small for $\mu \simeq m_c$
from the large logarithm
$\log( \mc /\mw )$. For this we have abstinated from using $x_c$,
but have turned to $m_c$. We will do this again without
warning.
Due to the magic of factorization the coefficient of $\ln (\mc/\mw)$
is a linear combination of the anomalous dimensions of
composite operators of some effective field theory.

Let us finally remark on the $\mu$--dependent terms in $\wt{h}$:
The colour octet term proportional to $(N-1)/(2N)$ is $\mu$-independent.
In the colour singlet part proportional to $C_F$ one identifies
$\gz_m=6 C_F$ as the coefficient of $\ln \mu^2$, which
stems from the running charm quark mass in \rf{gj}.
Likewise $2 C_F \xi \ln \mu^2 $ originates from the wave function
renormalization.

\section{General Structure of the RG Improvement}\label{general}
After integrating out the heavy top quark and the W--boson
one is left with an effective field theory containing
a lagrangian of the form \rf{eff2}.
Now we extract the Fermi constant out of the Wilson coefficients
to render them dimensionless:
\begin{eqnarray}
\leo &=& - \frac{\gf}{\sqrt{2}} V_{CKM,F} C_k Q_k^{(F)} -
     \frac{\gft}{2} V_{CKM,F}^\prime  \wt{C}_l^{(F)}
       \wt{Q}_l  \label{leff1} .
\end{eqnarray}
Here $V_{CKM,F}$ and $V_{CKM,F}^\prime$ stand for products of CKM
elements and $F$ is  a flavour index.
The light internal
quarks in \fig{box} and \fig{boxqcd} require to match the
corresponding parts of $G$ to matrix elements with two insertion
of \dso operators $Q_k$. Fig.\ \ref{bi} and \fig{biqcd}
show such diagrams for the case of two insertions of
current--current operators.
\begin{figure}[htb]
\caption[Matrix element of two \dso operators]{Diagram
      ${\rm D}_0$ in the effective five--quark theory.
        The cross denotes the insertion of a
        $\Delta S = 1$ current--current operator.
        }
\label{bi}
\end{figure}
\begin{figure}[htb]
\caption[Radiative Corrections to \fig{bi}]{
        The classes of diagrams
        giving the $O(\alpha)$ contribution to
        $\langle \bilo{il} \rangle$ and $\langle \bilr{ij} \rangle$
        for $j=1,2$.
        The other members of a given class are obtained by
        left--right and up--down reflections.
        Also QCD counterterms have to be included.
        Diagram ${\rm D}_8 = 0$ for zero external momenta.
}
\label{biqcd}
\end{figure}
The $\Qht_l$'s in \rf{leff1} correspond to \dst operators. In general
their matrix elements also participate in the matching at the
initial scale $\mu_W$.
We will be interested  in the contribution of the lowest dimension
operators: The $Q_k$'s in \rf{leff1} have dimension six
and the $\wt{Q}_l$'s are dimension--eight operators.
The remaining sections of this chapter will concern the correct reduced
operator basis needed in \rf{leff1}.

Next one has to carry out the program of sect.\
\ref{unphy}, \ref{icwc} and \ref{rgo}
consisting of the following steps:
\begin{itemize}
\item[i)] Find a reduced operator basis for \rf{leff1}, from
  which renormalized operators of the types described in
  sect.\ \ref{unphy} and properly renormalized evanescent operators
  have been removed.
\item[ii)] Match the Standard Model Green's function $\wt{G}$ in \rf{smbox}
  to the   matrix elements derived from \rf{leff1}
  as shown in sect.\ \ref{icwc}. This is to be done
  at a scale $\mu_W$ satisfying $M_W \sims \mu_W \sims m_t$, but being
  arbitrary within this range.
\item[iii)] Follow sect.\ \ref{rgo} to perform  the RG evolution
  down to $\mu_c \approx m_c$ thereby
  summing $\ln (\mu_c/\mu_W)$ to all orders.
  When passing the b--quark threshold remove this degree
  of freedom from the calculation. This simply modifies the running
  coupling  and trivially changes the penguin operators to be
  introduced in sect.\ \ref{peng}. Strictly speaking one matches
  \rf{leff1} to another lagrangian with the b--quark integrated out,
  but the structure of both lagrangians is essentially that of
  \rf{leff1}.
\end{itemize}
Hence $\leo$ in \rf{leff1}  describes the physics between the scales $\mu_c$
and $\mu_W$.
At the scale $\mu_c$ one has to match the matrix elements derived from
\rf{leff1} to those of a new effective lagrangian
$\lefft$ with the c--quark
integrated out and to repeat steps i) to iii).
This amounts to the matching of diagrams of the
type of \fig{bi}, \fig{biqcd} to those with a single
local four--quark   operator as depicted in \fig{loc}  and
\fig{locqcd}.
\begin{figure}[tb]            
\caption[Radiative corrections to the \dst operator]{
        The classes of diagrams in the effective three--quark theory
        contributing to $\langle \oll \rangle$ to order $\alpha$.
        The other members of a given class are obtained by
        left--right and up--down reflections.
        QCD counterterms have to be included.
        }
\label{locqcd}
\end{figure}
In fact the reduced $\lefft$ contains only
a single physical dimension--six operator, which
we  have already met in \rf{oll}:
\begin{eqnarray}
\lefft &=& - \frac{\gft}{16 \pi^2}
 \lt( \cll{c} \lambda_c^2  +\cll{t} \lambda_t^2 +
   \cll{ct} \lambda_c \lambda_t
      \rt) \oll
     \label{leff2}
\end{eqnarray}
In the matching we equate dimension--eight Green's functions
derived from $\leo$ with di\-men\-sion--six Green's functions
from $\lefft$. Hence by power counting the $\cll{j} $'s in \rf{leff2}
pick up two powers of $m_c(\mu_c)$. The subsequent RG improvement
of the $\cll{j} $'s from $\mu_c$ down to some scale $\mu$
will sum $\ln \mu_c/ \mu$. The final scale $\mu $ must be
large enough to trust into perturbation theory, this is
roughly of the order of $1 \gev$. The matrix element of
$\oll (\mu)$ must be evaluated between neutral K--meson states.
The calculation of such QCD binding effects can be
done in the framework of lattice gauge theory, $1/N$--expansion
or with sum--rule techniques.

The observation that we need two powers of $m_c$ in the matching
greatly simplifies the finding of the correct set of dimension--eight
operators $\Qht_l$ needed in $\leo$. The only candidate
is
\begin{eqnarray}
\oloc &=& \frac{\mc}{g^2 \ov{\mu} ^{2 \e}}
     \ov{s} \g_\mu L d \cdot \ov{s} \g^\mu L d ,
\label{defqseven}
\end{eqnarray}
which we already know from an example in sect.\ \ref{doi}.
To understand this recall that any other
dimension--eight operator lacks the factor of $\mc$ and
involves  derivatives and/or
gluon fields instead\footnote{Operators containing only one power of
$m_c$ can be removed by the use of the EOM for the fermion field
according to the procedure described in sect.\ \ref{unphy}
giving $m_s m_c / g^2 \cdot \g_\mu L \otimes \g^\mu L$. }.
The matrix elements between s--quarks and d--quarks evaluated
for the matching at $\mu_c$ cannot produce a factor of $m_c^2$:
The c--quark can only enter  matrix elements with single
insertions of such operators through
loops in gluon propagators, but the result is always
proportional to $k^2 g^{\mu \nu} - k^\mu k^\nu$
rather than $m_c^2$, where
$k^\mu$ is the momentum flowing through the propagator.
Hence the QCD gauge structure keeping the gluon massless
provides us with unexpected help here.
For the same reason other dimension--eight operators cannot mix into
$\oloc$ during the RG evolution from $\mu_W$ to $\mu_c$.
In principle operators with $\ov{\mu}^{-2 \e} m_c^2/g^2$
multiplying other Dirac
structures can also appear, but we will see that
this is not the case due to the
flavour structure of the problem.
Let us drop some words on the factor $\ov{\mu}^{-2 \e}$ in
\rf{defqseven}: Bare quantities must
not depend on $\mu $. Therefore the bare operator reads
\begin{eqnarray}
\oloc^\ba &=& \frac{ m_{c,\,\ba }^{2} }{g_{\ba }^{ 2}} \lt[
     \ov{s} \g_\mu L d \cdot \ov{s} \g^\mu L d \rt]^{\ba} .
\no
\end{eqnarray}
With \rf{renpar} and \rf{mstomsb} one is therefore lead to
\rf{defqseven}. It is also clear from \fig{bi} that the
dimension of $\oloc$ is larger than the dimension
of the $Q_k$'s by $2- 2 \e$, because the loop integration gives a power
of $D-2$. We have refrained from defining powers of
$\mu ^{2 \e}$ into the $C_k$'s, because this  would lead to
an unnecessary complication in the RG equations \rf{rgwil2}.

The origin of these simplifications of $\leo $
in \rf{leff1} is the
fact that the internal c--quark is heavier than the external
s--quarks and d--quarks. The situation is therefore
\emph{completely different}\/ for the case of the
c--quark contributions to \dbt transitions or the effect
of internal s--quarks in \dct amplitudes.
Here the external quarks are heavier and the dimension--eight
operators containing derivatives  dominate
over the effect of
$\ov{\mu}^{-2 \e}m_c^2/ g^2 \cdot \g_\mu L \otimes \g^\mu L $
in \dbt amplitudes
resp.\
$\ov{\mu}^{-2 \e}m_s^2/ g^2 \cdot \g_\mu L \otimes \g^\mu L $
in \dct amplitudes.
Fortunately
the former are phenomenologically uninteresting,
as they are suppressed with  $m_b^2/m_t^2$, and in the
case of \dct transitions there are other badly
controllable  long--distance effects in the game which
most likely make short distance calculations of
\ddm\/ useless.

\section{\textbf{\dso}--Operator Basis}\label{ds1}
Let us now find a basis for the \dso operators $Q_k$ in
\rf{leff1}, which closes under renormalization.
\subsection{Current--Current Operators}
Consider first the matching procedure for the LO diagram
\fig{box} with only light quarks in the intermediate state.
Contracting the W boson lines yields  the diagram of \fig{bi}
with two insertions of the current--current operator of \fig{ds1cclo}:
\begin{eqnarray}
Q_2^{kl} &=& \ov{s} \g_\mu L  k  \cdot \ov{l} \g_\mu L d \cdot \1 \label{ccq2},
\end{eqnarray}
where $k$ and $l$ are $u$ or $c$.
\begin{figure}[ht]
\caption{\dso current--current operator}
\label{ds1cclo}
\end{figure}

For $k \neq l$ $Q_2^{kl}$ only mixes with
\begin{eqnarray}
Q_1^{kl} &=& \ov{s} \g_\mu L  k  \cdot \ov{l} \g_\mu L d \cdot \tw .
\label{ccq1}
\end{eqnarray}
The corresponding diagrams are shown in \fig{ds1ccnlo} on p.\
\pageref{ds1ccnlo}.
The $2 \times 2$ mixing matrix in LO and NLO has been derived in
\cite{bjw}. If the corresponding
evanescents are defined as in \rf{ex},
the NLO anomalous dimension matrix commutes with the LO one.
These matrices are  diagonal in the basis containing the
operators $Q_+^{kl}$ and $Q_-^{kl}$ defined by
\begin{eqnarray}
Q_\pm^{kl} &=& \frac{1}{2} \lt( Q_2^{kl} \pm Q_1^{kl} \rt) . \no
\end{eqnarray}
Their anomalous dimensions in the NDR scheme with \rf{ex} are
\begin{eqnarray}
\gamma _\pm^{(0)}&=& \pm 6 \frac{N \mp 1}{N}, \quad
\gamma _\pm^{(1)(f)} \; = \;   \frac{N \mp 1}{2 N}
    \lt(-21 \pm \frac{57}{N} \mp \frac{19}{3} N \pm \frac{4}{3} f  \rt)
\label{gammapm} .
\end{eqnarray}

\subsection{Penguin Operators}\label{peng}
We will focus first on the LO operator mixing.
For $k=l$ the current--current operators mix with penguin operators.
The simplest divergent one--particle irreducible penguin  diagrams,
into which $Q_{1/2}^{kk}$ can be inserted, are depicted in \fig{glueping}.
\begin{figure}[ht]
\caption[Mixing into gluon--foot operators]{The mixing
of a \dso current--current operator into
gluon--foot penguin operators.
Diagrams with crossed gluons compared to G$_2$ and G$_3$ must be included,
too.
}
\label{glueping}
\end{figure}
These diagrams require the gluon--foot penguin
operators $Q_{g1},Q_{g2}$ and $Q_{g3}$  of \fig{glueop} as counterterms.
\begin{figure}[ht]
\caption[Gluon--foot penguin operators]{The
gluon--foot penguin operators}
\label{glueop}
\end{figure}
We will
encounter the basic penguin diagram $G_1$ later as a subdiagram, so that
we comment on it here:  Although it is a dimension--two diagram, its
result contains no term proportional to the square of the
internal mass. One has instead
\begin{eqnarray}
G_1  &\propto & \ov{s} \g_\mu L T^a d \cdot
  \lt( p^\mu p^\nu -p^2 g^{\mu \nu}  \rt) , \label{g1prop}
\end{eqnarray}
where $\nu$ and $a$ are the open Lorentz and colour
indices of the gluon line and $p$ is the
momentum flowing into the gluon vertex. \rf{g1prop} holds to
all orders in $D-4$.
Especially the inserted operators $Q^{kk}_{1/2}$ do not mix
into an operator of the form $m^2 \ov{s} A\sla^a L T^a d $.
Likewise the dimension--one diagram $G_2$ does not give a positive power
of the internal mass.

In order to find a closed operator basis one now has to investigate
into which operators $G_1$, $G_2$ and $G_3$ mix. One readily
finds that e.g.\ $G_1$ mixes via  box diagrams into new  four--fermion
operators. A box diagram with two fermion lines containing one
$Q_{g1}$ vertex and three QCD gluon--fermion vertices is divergent,
because $Q_{g1}$ involves two powers of the loop momentum due to
\rf{g1prop}. These new operators are fermion--foot penguin operators:
\begin{eqnarray}
Q_3 &=& \ov{s} \g_\mu L d \cdot \sum_{q=d,u,s,c,b} \ov{q}\g^\mu L q
\cdot \1 \nn
Q_4 &=& \ov{s} \g_\mu L d \cdot \sum_{q=d,u,s,c,b} \ov{q}\g^\mu L q
\cdot \tw \nn
Q_5 &=& \ov{s} \g_\mu L d \cdot \sum_{q=d,u,s,c,b} \ov{q}\g^\mu R q
\cdot \1 \nn
Q_6 &=& \ov{s} \g_\mu L d \cdot \sum_{q=d,u,s,c,b} \ov{q}\g^\mu R q
\cdot \tw \label{pengop}
\end{eqnarray}
We will be more interested in the mixing of $Q_{1/2}$ into them, which
is shown in \fig{fermping}.
\begin{figure}[ht]
\caption[Fermion foot penguin operator]{The mixing of a
 \dso current--current operator into
fermion--foot penguin operators depicted on the right.}
\label{fermping}
\end{figure}
At this point we realize that our simple minded approach to the
operator mixing leads to problems, as we could renormalize the
divergence in \fig{fermping} either by a counterterm proportional
to $Q_{g1}$ or by one involving fermion foot operators.
Further we obtain mixing into a third type of operator as depicted
in \fig{ghostping}:
\begin{eqnarray}
Q_\mr{FP} &=& \ov{s} \g_\mu L T^a d \cdot
        \lt( \partial^\mu \ov{\eta}_b \rt) \eta_c f_{abc}. \label{qfp}
\end{eqnarray}
\begin{figure}[ht]
\caption[Ghost--foot penguin operator]{The mixing of a \dso
 current--current operator into
ghost--foot penguin operators depicted on the right.}
\label{ghostping}
\end{figure}

Fortunately we can put order into the penguin zoo with the theorems of
sect.\ \ref{unphy}. For this we first construct the
dimension--six operators (with total ghost number zero)
from class ${\cal B}$: The only  possibility is given by $Q_\mr{BRS}$
from \rf{defbrs}. It is related to $Q_\mr{FP}$ via
\begin{eqnarray}
Q_\mr{BRS} &=& Q_\mr{FP} + Q_\mr{gf}, \no
\end{eqnarray}
where $Q_\mr{gf}$ is the term on the LHS of \rf{defeom} stemming
from the gauge fixing part of the QCD lagrangian:
\begin{eqnarray}
Q_\mr{gf}&=& \frac{1}{g} \frac{1}{\xi}\,
         \ov{s} \g_\mu L T^a d \cdot
          \partial^\mu \partial_\nu A^\nu_a .
\label{qgf}
\end{eqnarray}
Next we arrange the operators found so far into combinations belonging
to class ${\cal E}$. Here only the EOM for the gluon field can give
dimension--six operators
related to the penguin zoo. The relevant operator has been
introduced in \rf{defeom}. In terms of the operators defined
in \rf{pengop}, \rf{qfp} and \rf{qgf} it reads:
\begin{eqnarray}
\Qeom \! &=& \!
 \frac{1}{g} \cdot \ov{s}\g_\mu L T^a d \cdot D_{\nu\, a} F^{\mu \nu
   \, a} - \frac{1}{4} \lt( Q_4 + Q_6 \rt) + \frac{1}{4 N}
    \lt( Q_3 +Q_5   \rt) - Q_\mr{gf} + Q_\mr{FP}   \label{qeom}
\end{eqnarray}

Now all other operators involved in the mixing
must combine to gauge invariant operators, because they belong
to class ${\cal P} $. We have already listed $Q_1^{kl},\ldots Q_6$,
the only other gauge invariant dimension--six operator
is
\begin{eqnarray}
 \frac{1}{g} \cdot \ov{s}\g_\mu L T^a d \cdot D_{\nu\, a} F^{\mu \nu
   \, a} &=& Q_{g1}+Q_{g3}+Q_{g3}, \label{combine}
\end{eqnarray}
which we can drop from the basis, because it is a linear combination
of $\Qeom$, $\Qbrs$ and the other physical operators.

Hence the unreduced \dso  operator basis consists of $Q_1^{kl}, \ldots
Q_6$, $\Qeom$, $\Qbrs$ and of course of evanescent operators.
This structure implies that the mixing of any operator into
$Q_{g1},Q_{g2}$ and $Q_{g3}$ must be the same.
Indeed, the divergent parts
of $G_1,G_2$ and $G_3$ all involve the same coefficient,
as required by \rf{combine}. The theorems of sect.\ \ref{unphy} also
explain the observation of \rf{g1prop}, because there is no  dimension--six
operator  in ${\cal P} \oplus {\cal E} \oplus {\cal B}$
involving $m^2 \ov{\psi} A\sla \psi$.

Due to sect.\ \ref{eommix} and chapter \ref{eva}
the reduced lagrangian for \dso transitions is simply obtained by
dropping the renormalized $\Qeom$, $\Qbrs$ and the renormalized
evanescents from the effective lagrangian.

The on--shell vanishing
of $\langle \Qeom \rangle$ can be easily understood in terms of
diagrams \cite{sim}. The Feynman rule for the
one--gluon piece of $\Qeom$ depicted in $Q_{g1}$ of \fig{glueop}
involves an inverse gluon propagator:
\begin{eqnarray}
\frac{1}{g} T^a \g_\mu L \lt[ p^2 g^{\mu \nu}
                     - \frac{\xi - 1}{\xi} p^\mu p^\nu \rt] . \label{invg}
\end{eqnarray}
Hence in any diagram in which the gluon line of $G_1$ ends up in an internal
vertex the gluon line is effectively shrunk to a point, because
\rf{invg} is contracted with a gluon propagator.
The resulting diagram is identical with
a diagram containing one of the other  operators in \rf{qeom} and
these contributions cancel. If the gluon of $G_1$ is an external line,
the diagram vanishes after LSZ reduction. Care is necessary, however,
if renormalization constants are calculated off--shell, because then
$\Qeom$ and $\Qbrs$ give non--zero contributions, which must be
properly taken into account. For example in \fig{fermping} one finds
off--shell terms involving $\ov{s} \g_\mu L d \cdot
\lt( p^\mu p^\nu/p^2  \rt)/\e $ which require a counterterm proportional
to $\Qbrs$.

Now for the case of the \dso substructure of
\dst diagrams the situation is more complicated,
because here the gluon of $G_1$ can end up in another $Q_{g1}$--vertex.
Then we have one propagator hitting two inverse propagators and the
result is non--vanishing. But from sect.\ \ref{seceomdouble} we know that
this contact term is  identical to the matrix element of a local
dimension--eight \dst operator $\wt{Q}_l$. Hence we can still restrict
the renormalized $Q_k$'s in \rf{leff1} to the set
$\{Q_1^{kl},\ldots Q_6\}$.

\section{Operator Basis for \textbf{\dst} transitions}
Let us now assign the correct CKM factors to the effective lagrangians
$\leo$ in \rf{leff1} and $\lefft $ in  \rf{leff2}.
\subsection{Above $\mathbf{\mu_c}$}\label{above}
The reduced \dso lagrangian can be taken from \cite{bjlw}:
\begin{eqnarray}
\leso &=& - \frac{\gf}{\sqrt{2}}
\lt[ \sum_{i=1}^2 \sum_{k,l =u,c} V_{ks}^\ast  V_{ld} C_i
Q_i^{kl} - \lambda_t \sum_{j=3}^6 C_j Q_j \rt].
\end{eqnarray}
Here the coefficient $\lambda_t$ originates from the fact that
the matching of the penguin operators and the
mixing of any four--quark operator into penguin operators
involves the diagram of \fig{fermping} and
radiative corrections to it all being proportional to
$\lambda_u + \lambda_c = - \lambda_t $.

With the findings of sect.\ \ref{general} the effective lagrangian
$\leo$ of \rf{leff1} is found as
\begin{eqnarray}
\leo &=& \leso - \frac{\gft}{2} \frac{\mc}{g^2} \cloc \oloc  \nn
&=& - \frac{\gf}{\sqrt{2}}
\lt[ \sum_{i=1}^2 \sum_{k,l =u,c} V_{ks}^\ast  V_{ld} C_i
Q_i^{kl} - \lambda_t \sum_{j=3}^6 C_j Q_j \rt]
- \frac{\gft}{2} \frac{\mc}{g^2} \cloc \oloc   . \no
\end{eqnarray}
We may ask where the contact terms from the reduction of the
\dst lagrangian have gone. Since diagrams with two insertions
in which the two \dso vertices are connected by a gluon line or by a
ghost--loop cannot produce a factor of $\mc$, these terms are
absorbed into other dimension--eight operators not containing
$\mc$, which have been found to be irrelevant in sect.\ \ref{general}.

Since there are no more derivative couplings involved, we can pass to
the more common hamiltonian formalism by simply flipping the sign:
$H = -\leo$. We will do further and define the \dst hamiltonian
such that it only contains terms of order $\gft$. For this we equate
the first order term of the
Gell-Mann--Low series $\,\exp[- i \int d^D x \hsto (x)]\, $ with the
corresponding series involving $\leo $:
\begin{eqnarray}
\lefteqn{\langle -i\int d^D x \hsto (x) \rangle |_{|\Delta S| =2 } \; = \;
 \langle e^{ i \int d^D x \leo (x) } \rangle |_{|\Delta S| =2 }
  +  O(\gf ^3 ) }\nn
 & =& \langle \frac{-1}{2} \int \!\!\int d^D x d^D y  \T \leso (x) \leso (y)
   - i \frac{\mc}{g^2} \int d^D x \cloc \oloc(x) \rangle |_{|\Delta S| =2 }
    + O(\gf ^3 ) . \no
\end{eqnarray}
This notation is a bit sloppy, but useful.
Let us write
\begin{eqnarray}
\hsto &=& H^{c} + H^{t} + H^{ct} . \no
\end{eqnarray}
One then obtains
\begin{eqnarray}
H^c (x) &=& \lambda_c^2 \frac{\gft}{2}
       \sum_{i,j=+,-} C_i C_j \bilo{ij} (x) \nn
H^t (x) &=& \lambda_t^2 \frac{\gft}{16 \pi^2} \cll{t} \oll (x) \nn
H^{ct}(x)  &=& \lambda_c \lambda_t  \frac{\gft}{2}
           \sum_{i=+,-} \sum_{j=1}^6 C_i C_j \bilr{ij}(x) +
           \lambda_c \lambda_t
           \frac{\gft}{2}  \cloc \oloc(x)
\label{thehs}
\end{eqnarray}
with
\begin{eqnarray}
\bilo{ij} (x)\!\!&=&\!\!   \frac{-i}{2} \int d^4 y \T
      \lt[  Q_i^{cc} (x) Q_j^{cc} (y) +Q_i^{uu} (x) Q_j^{uu} (y)
          -   Q_i^{cc} (x) Q_j^{uu} (y)           \rt. \nn
&& \lt. \quad \quad \quad \quad \quad
   - Q_i^{uu} (x) Q_j^{cc} (y)
   - Q_i^{uc} (x) Q_j^{cu} (y) -Q_i^{cu} (x) Q_j^{uc} (y) \rt]
      \; , \no \\[3mm]
\bilr{ij} (x) \!\!\!\!  &=& \!\! \!\!
\lt\{ \!\!\! \!\!\!
\begin{array}{l}
\lt.
\begin{array}{l}
 \frac{-i}{2}\! \int \! d^4 y \T
      \lt[-  Q_i^{uc} (x) Q_j^{cu} (y) - Q_i^{cu} (x) Q_j^{uc} (y)
          -  Q_i^{uu} (x) Q_j^{cc} (y)           \rt. \\[1mm]
 \lt. \quad  \quad \quad \quad \quad
   - Q_i^{cc} (x) Q_j^{uu} (y)
   + 2 Q_i^{uu} (x) Q_j^{uu} (y)  \rt]
\end{array}\!\! \rt\} \mbox{for j=1,2}, \\[8truemm]
\lt.
\begin{array}{l}
 \frac{i}{2} \! \int \! d^4 y \T
      \lt[ \lt(  Q_i^{cc} (x) -Q_i^{uu} (x) \rt)
         Q_j  (y)
             \rt. \\[1mm]
 \lt. \quad  \quad \quad \quad \quad
  + Q_j  (x) \lt(  Q_i^{cc} (y) -Q_i^{uu} (y) \rt)  \rt]
\end{array} \!\! \rt\} \mbox{for j=3,\ldots 6}.
\end{array}
\rt. \label{bilocs}
\end{eqnarray}
The structure of \rf{thehs} requires some explanation:
\subsubsection{The top--top contribution $\mathbf{H^t}$}
First there is
only a \dst operator involved in $H^t$. We have only
listed the terms relevant for the leading term in the expansion in
$1/m_\mr{heavy}$.
In the case of  $H^t$ this contribution comes from the internal
top--quarks in \fig{box}, which are matched at $\mu_W$ to the
local operator $\oll$. From \rf{gj} and \rf{sxt} we know that
$\cll{t}$ is proportional to $x_t$, while the double
penguin contributions can at most give a factor of $x_b$.
Also  the dimension--eight operators stemming from the contact terms
in \rf{eomdouble} and \rf{brsdouble} can only give contributions
proportional to $m_\mr{light}^2/\mw$, because there is no more
top--quark in the effective theory. The calculation of $\eta_2$
therefore only requires the consideration of matrix elements with a
single \dst operator. Yet the NLO matching has to be done from the
finite parts of the two--loop diagrams in \fig{boxqcd}. This
calculation has been carried out by Buras, Jamin and Weisz \cite{bjw}.

\subsubsection{The charm--charm contribution $\mathbf{H^{c}}$}
Second there is no local \dst operator involved in $H^c$. This feature
is related to the absence of  a large logarithm in \rf{ilgim}.
The GIM combination in $\bilo{ij}$ in \rf{bilocs} leads to
the cancellation of the divergences and thereby of the $\ln (\mu/M_W)
$'s in the diagrams \rf{bi}. The same holds true for the diagrams of
\rf{biqcd} after the inclusion of counterterms for the \dso operators
and in fact to all orders in perturbation theory.
This has been first observed by Witten \cite{wit}.
Hence $H^c$ belongs to case II of sect.\ \ref{dom}. Yet we may
conclude from sect.\ \ref{dom} only that there is no mixing into a local
operator, but there could be a local operator involved in the matching
at $\mu_W$, which could evolve down to $\mu_c$ unaffected by
the presence of the bilocal structures $\bilo{ij}$. But in the case at
hand the matching at
$\mu_W$ can be made by the replacement $1/(\mw-k^2)\rightarrow 1/\mw$
in the diagrams of \fig{box} and \fig{boxqcd}.
This is allowed, if the renormalized Standard Model
diagram does not become UV--divergent by shrinking the W line to
obtain the effective diagrams of \fig{bi} and \fig{biqcd}, because
then the contribution of the UV momenta is suppressed.
In $H^c$ the GIM mechanism cancels the UV divergences in the effective
theory, so that there is no room for a non--zero coefficient of
$\oloc$. This fact becomes very clear if one splits the loop integral
connecting the two \dso vertices into a part with momenta smaller
than $m_c$ and an UV part comprising loop momenta above $m_c$.
For the first part we can use the factorization of the \dso
substructure yielding \rf{bilocs} up to terms of order $\mc/\mw$,
and the UV momenta cannot give non--negative powers of $\mw$ to
spoil this factorization, because the
integral from $m_c$ to $\infty$ vanishes in the limit $M_W \rightarrow
\infty$ after forming the GIM combination $(c,c)-(c,u)-(u,c)+(u,u)$.
Hence we only need to know the \dso coefficients $C_+$ and $C_-$,
which have been determined in \cite{bw}. No extra \dst matching in
necessary.
{}From sect.\ \ref{dom}, however,  we know that we need the finite parts of
the diagrams of \fig{biqcd} for the NLO matching at the scale $\mu_c$,
so that we can check the absence of a local operator in
$H^c$ in \rf{bilocs}
by comparing the matrix element of $H^{c}(\mu_W)$ in \rf{thehs} with
\rf{gj} and \rf{h1}:
\begin{eqnarray}
i \wt{G}^c (\mu_W) + O(x_c^2) &=& \langle H^c (\mu_W) \rangle \; = \;
      \lambda_c^2 \frac{\gft}{2} C_i (\mu_W)C_j(\mu_W)
         \langle \bilo{ij}(\mu_W) \rangle .
\label{matchcup}
\end{eqnarray}

On the other hand the operator mixing is simple and does not require
new calculations. We have put the current--current operators in
$\bilo{ij}$ into the diagonal basis to simplify the RG evolution.
The \dso Wilson coefficients $C_+$ and $C_-$ in \rf{matchcup}
simply evolve down to $\mu =\mu_c$ according to \cite{bjlw}.

\subsubsection{The charm--top contribution $\mathbf{H^{ct}}$}
$H^{ct}$ involves the familiar large logarithm in \rf{sxcxt}, so that
it belongs to case I of sect.\ \ref{dom}. Hence both the LO and the NLO
matching can be done from the one--loop graphs \fig{box} and \fig{bi}.
Again we have chosen the diagonal
basis in $\bilr{ij}$ for the current--current vertex. For the other
vertex we have chosen the operator basis $Q_1^{kl},\ldots Q_6$ to use
the results of \cite{bjlw}.
Now the LO matching at $\mu=\mu_W$ is performed from the large
logarithm $\ln x_c$ in  $G^{c,\,(0)}$. In NLO we also equate the
nonlogarithmic  part of $G^{c,\,(0)}$ to the diagrams of the
effective field theory to obtain the initial value $\cloc (\mu_W)$.
Since the initial value of the \dso coefficients $C_k$ are of
order $\alpha$ for $k=1,3,4,5,6$ and $C_k= 1+O(\alpha)$ for
$k=2,+,-$, the matching at the scale $\mu=\mu_W$ reads
\begin{eqnarray}
\lefteqn{ i G^{ct\, (0)} (\mu_W)+ O(x_c^2 \ln x_c)
    \; = \;    \langle H^{ct} (\mu_w) \rangle ^{(0)} }  \nn
& = &  \frac{\gft}{2} \lambda_c\lambda_t \lt[
      \langle \bilr{2,+}  \rangle ^{(0)}
      +   \langle \bilr{2,-}  \rangle ^{(0)}
       + \frac{\alpha(\mu_W) }{4 \pi}  \cloc^{(1)} (\mu_W)
         \langle \oloc (\mu_W)  \rangle^{(0)} \rt]   .
\label{matchctup}
\end{eqnarray}
Here we have accounted for the fact that $\cloc$ starts in order
$\alpha$ due to the inverse power in the definition in $\oloc$
giving $\langle \oloc (\mu ) \rangle ^{(0)} =
   \mc ( \mu) /( 4 \pi \alpha(\mu) ) \cdot
     \langle \oll (\mu ) \rangle ^{(0)} $ .
Hence in LO there is nothing  to adjust in \rf{matchctup} and the
coefficients of the large logarithms must be the same on both sides.
This is indeed the case.
Further for the NLO matching only the
diagrams in \fig{box} and the current--current diagrams of \fig{bi}
are needed.
For later convenience we split $H^{ct}$ up:
\begin{eqnarray}
H^{ct} (\mu )  & = & H^{ct,+} (\mu ) +  H^{ct,-} (\mu ) \nn
H^{ct,\pm } (\mu)    &=&  \frac{\gft}{2} \lambda_c \lambda_t \lt[
           C_k (\mu) C_\pm (\mu)  \bilr{k,\pm}  (\mu)
        +  \clocpm (\mu )    \oloc (\mu )   \rt]     \no
\end{eqnarray}
$H^{ct,+}$ and $H^{ct,-}$ evolve independently under the RG flow.
Also $\cloc (\mu) $ is split into
$\clocp (\mu)+\clocm (\mu) $, both of which receive different admixtures
during the RG evolution.
It is arbitrary which part of $\cloc (\mu_W) $
is put into $\cloc^{\pm } (\mu_W)$, as long as their sum
fulfils the matching condition \rf{matchctup}. We will choose
$\cloc^{\pm,(1)} (\mu_W) = 1/2 \cdot \cloc^{(1)} (\mu_W) $

The RG evolution is much more involved than in the case of $H^c$.
With the well--known initial values of the \dso coefficients
\cite{bw,bjlw} and with $\cloc (\mu_W)$ obtained from \rf{matchctup}
one knows
\begin{eqnarray}
\cvecp (\mu_W) \; = \;
\lt(
\begin{array}{c}
C_+ (\mu_W) C_1(\mu_W)  \\[1mm]
            C_+ (\mu_W) C_2(\mu_W)  \\[1mm] C_+ (\mu_W) C_3(\mu_W)  \\[1mm]
         C_+ (\mu_W) C_4(\mu_W)  \\[1mm] C_+ (\mu_W) C_5(\mu_W)  \\[1mm]
      C_+ (\mu_W) C_6(\mu_W)  \\[1mm] \clocp (\mu_W)
\end{array} \rt) ,
&\quad &
\cvecm (\mu_W) \; = \;
\lt(
\begin{array}{c}
    C_- (\mu_W) C_1(\mu_W)  \\[1mm]
            C_- (\mu_W) C_2(\mu_W)  \\[1mm] C_- (\mu_W) C_3(\mu_W)  \\[1mm]
         C_- (\mu_W) C_4(\mu_W)  \\[1mm] C_- (\mu_W) C_5(\mu_W)  \\[1mm]
      C_- (\mu_W) C_6(\mu_W)  \\[1mm] \clocm (\mu_W)
\end{array} \rt)  \label{cvecup}
\end{eqnarray}
to order $\alpha$. These are the initial conditions for the NLO
RG evolution. We will give the
relevant expressions for $C_k(\mu_W)$ and $\clocpm (\mu_W)$
in chapter \ref{calc}, where the actual
calculation is described.

In NLO all coefficients in $H^{ct}$ have a non--zero initial value.
The structure of the RG evolution has been described in sect.\
\ref{doi}. In  \rf{inhsol} the summation indices
$m,m^\prime, v$ and $v^\prime$ assume  the
values $+$ and $-$ and $t,t^\prime, n  $ and $n^\prime$ run from 1 to 6.
The two terms of the summation over $n$ correspond to
$H^{ct,+}$ and $H^{ct,-}$ and
the two
coefficient  vectors in \rf{cvecup}. The first term in \rf{inhsol}
involving the initial value $\cloc (\mu_W)$ can obviously be arbitrarily
 distributed over $H^{ct,+}$ and $H^{ct,-}$.

The structure  of the RG  evolution is probably more transparent from
\rf{rgwil2}: The anomalous dimension tensor $\g_{kn,l}$
describing the mixing of \dso coefficients into the \dst coefficient
$\cloc$
contains
12 elements $\g_{n+,7},\g_{n-,7}$, $n=1,\ldots 6$, which we
summarize into the row vectors
\begin{eqnarray}
\gabip &=& \lt( \g_{1+,7}\; , \; \g_{2+,7}\; , \; \g_{3+,7}\; ,
                \; \g_{4+,7}\; , \; \g_{5+,7}\; , \;
                 \g_{6+,7}   \rt) \nn
\gabim &=&  \lt( \g_{1-,7}\; , \; \g_{2-,7}\; ,
              \; \g_{3-,7}\; , \; \g_{4-,7}\; , \; \g_{5-,7}\; , \;
                \g_{6-,7}   \rt) \label{gabi}
\end{eqnarray}
Now \rf{rgwil2} reads in the case at hand:
\begin{eqnarray}
\dmu \cloc &=& \galoc \cloc + \g_{k+,7} \,  C_k \, C_+    \;  + \,
                              \g_{k-,7} \,  C_k \, C_-
\label{rgwilpm}
\end{eqnarray}
which is split into
\begin{eqnarray}
\dmu \clocp &=& \galoc \clocp + \g_{k+,7} \,  C_k \, C_+  \nn
\dmu \clocm &=& \galoc \clocm + \g_{k-,7} \,  C_k \, C_-  .  \no
\end{eqnarray}
Here $\galoc$ is the anomalous dimension of $\oloc$.
This yields the RG equation for $\cvecpm$ defined in
\begin{eqnarray}
\dmu \cvecpm (\mu) &=& \left(
\begin{array}{cc}
\; \g + \g_\pm \uma \; & 0   \\[2mm]
\; \gabipm     \;      & \galoc
\end{array}  \right) \cvecpm (\mu) , \label{rg7}
\end{eqnarray}
where $\g$ is the $6 \times 6 $ \dso anomalous dimension matrix
and $\g_+$ and $\g_-$  are the anomalous dimensions  of
$Q_+$ and $Q_-$. \rf{rg7} leads to a block--triangular evolution
matrix of the form \rf{blocku} and one again recognizes that
$ \cloc (\mu) = \clocp (\mu) +  \clocm(\mu)$ does not depend on how
the initial value $\cloc (\mu_W)$ is distributed over
$\clocp (\mu_W)$ and $ \clocm(\mu_W)$.

The solution of \rf{rg7} involves two
$7\times 7$ evolution matrices. They, however, encode redundant
information, because each of them contains the full $6 \times 6$ \dso
evolution matrix. Now one can do better:
Define the vector
\begin{eqnarray}
\dvec &=&
\lt(
\begin{array}{c}
\vec{C} \\[1mm]  \clocp / C_+ \\[1mm] \clocm / C_-
\end{array} \rt) , \label{defdvec}
\end{eqnarray}
where $\vec{C}=(C_1,\ldots C_6) $ is the \dso coefficient vector.
{}From \rf{rgwilpm} one finds the RG equation for the seventh and eighth
component of $\dvec$:
\begin{eqnarray}
\dmu D_{7} &=& ( \galoc - \g_+ ) D_{7} + \gabip \cdot \vec{C} , \nn
\dmu D_{8} &=& ( \galoc - \g_- ) D_{8} + \gabim \cdot \vec{C} . \nn
\end{eqnarray}
Hence $\dvec$ satisfies the following RG equation:
\begin{eqnarray}
\dmu \dvec &=&
\lt(
\begin{array}{ccc}
\g     & 0           &     0       \\[1mm]
\gabip & \galoc-\g_+ &     0       \\[1mm]
\gabim &   0         &  \galoc - \g_-
\end{array} \rt) \dvec
\label{rg8}       .
\end{eqnarray}
After obtaining  the $8 \times 8$ RG evolution matrix in the standard
way one gets $\dvec(\mu )$ which has to be multiplied with
$C_\pm (\mu )$  to obtain $\cvecpm (\mu )$ and
\begin{eqnarray}
\lefteqn{
H^{ct,\pm} (\mu) \; = \; \frac{\gft}{2} \lambda_c \lambda_t \sum_{n=+,-}
\lt(\; \bilr{1,n} (\mu) ,\ldots , \bilr{6,n} (\mu) , \oloc (\mu)
    \; \rt) \cdot \vec{C}^{(n)} (\mu) } \nn
\!&\!=\!&\!\! \frac{\gft}{2} \lambda_c \lambda_t
\lt(\, C_+ (\mu) \bilr{1,+} (\mu) + C_- (\mu) \bilr{1,-} (\mu)
       ,\ldots , C_+ (\mu) \oloc (\mu) , C_- (\mu) \oloc (\mu) \, \rt)\!\!
    \cdot \dvec (\mu)
\label{hctpm} .
\end{eqnarray}
Hence we have switched from a
$ 7 \times 7^\prime $
representation to an $ 8 \times 1 \times 1^\prime $
representation of the renormalization group.
We have programmed both the solution of \rf{rg7} and \rf{rg8} and
checked that we obtain the same result. Of course, the program
for \rf{rg8} is much faster. Further the upper left $6\times 6$
block--submatrix  of the $8\times 8$ evolution matrix is the
\dso evolution matrix of \cite{bjlw}, which provides an additional
check of the RG evolution.

Now the calculation of the elements $\g_{k\pm,7} $
of the LO and NLO anomalous dimension tensor
has required
the calculation of the divergent parts of the diagrams
of \fig{bi} and \fig{biqcd}, but now with insertions of $\bilr{ij}$ for
$j=1,2,$ rather than
$\bilo{ij}$ and of the diagrams of \fig{penglo} and
\fig{pengqcd} corresponding to
matrix elements of $\bilr{ij}$ for $j=3,\ldots 6$.
\begin{figure}[tbh]
\caption[Diagrams with penguin operators]{Diagram
       yielding the LO anomalous dimension tensor
        $\g^{(0)}_{k\pm,7}$ for $k=3\ldots 6$.
        The white cross denotes the insertion of a
        $\Delta S = 1$ current--current operator. The dark cross
        stands for one of the \dso penguin operators $Q_3,\ldots Q_6$.
        }
\label{penglo}
\end{figure}
Further there are  diagrams in which the s--quark foot of the
penguin operator is connected with the current--current vertex and
likewise diagrams with the d--quark foot involved.  An example is
shown in \fig{bigdouble}. For insertions of $Q_5$ and $Q_6$ these
diagrams vanish by chirality. For insertions of $Q_3$ and $Q_4$
the diagrams are nonvanishing, but their sum cannot give a factor of
$\mc$ in the divergent part. This follows from a subtle analysis
exploiting \rf{g1prop} and the theorems on the mixing into gauge
invariant operators.
\begin{figure}[htb]
\caption[Radiative corrections to \fig{penglo}]{
        The classes of diagrams
        contributing to $\langle \bilr{ij} \rangle$
        for $j=3,\ldots 6$, to $O(\alpha)$.
        The other members of a given class are obtained by
        left--right and up--down reflections.
        Also one--loop counterterms have to be included.
}
\label{pengqcd}
\end{figure}
\begin{figure}[ht]
\caption[Another diagram with a penguin operator]{An example of a
    diagram involving the s--quark foot of the
         penguin operators.  }
\label{bigdouble}
\end{figure}
%
%
%
\subsection{Below $\mathbf{\mu_c}$}
At the scale $\mu_c \approx m_c$ we remove the c--quark from the
theory. Hence we are left with the  single local operator $\oll$.
We have listed the low energy hamiltonian already in
\rf{s2} and \rf{leff2}:
\begin{eqnarray}
H&=& H^c + H^t +H^{ct} \nn
H^c (\mu)  &=& \frac{\gft}{16 \pi^2} \lambda_c^2 \cll{c} (\mu)
 \oll (\mu)
\; =\; \frac{\gft}{16 \pi^2} \mw \lambda_c^2 \eta_1 S(x_c) b(\mu)\oll (\mu) ,
\nn
H^t (\mu)  &=& \frac{\gft}{16 \pi^2} \lambda_t^2 \cll{t} (\mu)
 \oll (\mu)
\; = \; \frac{\gft}{16 \pi^2} \mw \lambda_t^2 \eta_2 S(x_t) b(\mu)\oll (\mu) ,
\nn
H^{ct} (\mu) &=& \frac{\gft}{16 \pi^2}
      \lambda_c \lambda_t \cll{ct} (\mu) \oll (\mu)
\; = \; \frac{\gft}{16 \pi^2} \mw 2 \lambda_c \lambda_t \eta_3
     S(x_c,x_t)  b(\mu)\oll (\mu) . \label{hbelow}
\end{eqnarray}
The only effect of the charm threshold for $\cll{t}$ is the transition
to the three--quark running $\alpha$.

The coefficient $\cll{c} (\mu_c)$
in \rf{hbelow}
is determined by the matching
of the matrix element of $H^{c} (\mu_c)$ in \rf{thehs} to those
of $H^c (\mu_c)$ in \rf{hbelow}:
\begin{eqnarray}
\langle H^c (\mu_c) \rangle &=&
\frac{G_F^2}{2} \lambda _c^2 \sum_{i,j=+,-} C_i (\mu_c) C_j (\mu _c)
     \langle {\cal O}_{ij} (\mu_c ) \rangle
\; = \; \frac{G_F^2}{16 \pi ^2 } \lambda _c^2 \; \cll{c} (\mu _c)
    \langle \oll (\mu _c) \rangle
    . \label{matchc}
\end{eqnarray}
This yields $\cll{c} (\mu_c)$ in terms of the RG improved \dso Wilson
coefficients $C_k(\mu_c)$. Here we verify the findings of sect.\
\ref{dom} that
in the NLO the finite parts of \fig{locqcd}
and of the two--loop diagrams in \fig{biqcd} are necessary.

The determination of $\cll{ct} (\mu_c)$ requires the same  procedure with
$H^{ct}$ of \rf{thehs} and \rf{hbelow}.
Consider first the LO matching: The coefficient $\cloc (\mu_c)$
is nonzero due to admixtures from $C_2 ^{(0)}(\mu_W )$ picked up
during the RG evolution from $\mu_W$ to $\mu_c$. Hence the LO matching
is done from the inverse power of $\alpha(\mu_c)$:
\begin{eqnarray}
\cll{ct}(\mu_c) &=& \frac{\mc (\mu_c)}{2} \frac{4 \pi}{\alpha (\mu_c) }
           \cloc  ( \mu_c ) . \label{cctlo}
\end{eqnarray}
In the NLO matching all coefficients participate. We define
\begin{eqnarray}
\langle \bilr{ij} (\mu)  \rangle &=&
\frac{\mc(\mu) }{16 \pi^2} \, 2 \, \rst{ij} (\mu)
\langle \oll (\mu)  \rangle^{(0)} . \label{defrij}
\end{eqnarray}
Here the definition of $\rst{ij}$ differs from the one of
$\widetilde{r}_{nm,l}$ given in \rf{defrtw} by a factor of 2.
In NLO one therefore finds:
\begin{eqnarray}
\cll{ct}(\mu_c) &=& \mc (\mu_c) \lt[
   \frac{1}{2}  \frac{4 \pi}{\alpha (\mu_c) }    \cloc  ( \mu_c )  +
    \sum_{i=+,-} \sum_{j=1}^6 \rst{ij} (\mu_c) C_i (\mu_c) C_j (\mu_c)
             \rt] .
\label{cctnlo}
\end{eqnarray}
On the side corresponding to the three--quark
theory one expects the diagrams of \fig{loc} and \fig{locqcd}
corresponding to $\langle \oll  \rangle^{(0)}$ and
$\langle \oll  \rangle^{(1)}$ to be
necessary. They, however, cancel with the terms accompanying
$\langle \oloc \rangle $ on the side of the four--quark theory.

Now one is left with the RG evolution from $\mu_c$ down to some scale
$\mu$ at which the hadronic matrix element
$\bra{ \ov{\mr{K^0}} }  \oll (\mu) \ket{ \mr{K^0} }$ is obtained.
The RG evolution of $\oll$ is well--known \cite{bw,bjw}, its
anomalous dimension is $\g_+$ given in \rf{gammapm}.
In terms of \rf{defd} and \rf{defj} the NLO solution for $\cll{j} (\mu)$,
$j=c,t,ct$,
reads:
\begin{eqnarray}
\cll{j} (\mu) &=&  \cll{j} (\mu_c)
      \lt[ \frac{\alpha (\mu_c)}{\alpha (\mu)} \rt]^{d_+^{(3)}}
      \lt( 1 - J_+^{(3)} \frac{\alpha(\mu_c) - \alpha(\mu)}{4 \pi}  \rt).
\label{cllend}
\end{eqnarray}
Finally  we can express the NLO $\eta_i$'s in \rf{hbelow} in terms of
the coefficients:
\begin{eqnarray}
\eta _1 &=& \frac{1}{\mc} \cll{c}(\mu_c)   \lt[ \alpha (\mu_c) \rt]
^{d_+^{(3)}}
     \lt( 1 - J_+^{(3)} \frac{\alpha(\mu_c) }{4 \pi}  \rt)      \nn
\eta_2  &=&  \frac{1}{\mw S(x_t)} \cll{t} (\mu_c )
       \lt[ \alpha (\mu_c) \rt] ^{d_+^{(3)}}
     \lt( 1 - J_+^{(3)} \frac{\alpha(\mu_c) }{4 \pi}  \rt)      \nn
\eta _3 &=&    \frac{1}{2 \mw S(x_c,x_t)} \cll{ct} (\mu_c)
       \lt[ \alpha (\mu_c) \rt] ^{d_+^{(3)}}
     \lt( 1 - J_+^{(3)} \frac{\alpha(\mu_c) }{4 \pi}  \rt)      .
\label{theetas}
\end{eqnarray}
The $\mu$--dependence in \rf{cllend} is absorbed into $b(\mu)$,
which equals
\begin{eqnarray}
b(\mu) &=& \lt[ \alpha (\mu) \rt] ^{ - d_+^{(3)}}
         \lt( 1 + J_+^{(3)} \frac{\alpha(\mu) }{4 \pi}  \rt)
\label{bmu}
\end{eqnarray}
in the NLO. The index $(3)$ denotes that one has to set $f=3$
in the corresponding quantities.

{}From \rf{theetas} it is evident that the $\eta_i$'s depend on the
definition of the masses. We will adopt the convention of
\cite{bjw,hn1} that the running masses in \rf{theetas}  are
evaluated at the scale at which they are integrated out,
i.e.\ $m_c=m_c(\mu_c), m_t=m_t(\mu_W)$. Whenever the $\eta_i$'s
are defined such that they multiply
$S(x_c (m_c) ) $, $S( x_t (m_t) ) $ and
$S(x_c (m_c),x_t (m_t)) $
in the effective hamiltonian in \rf{hbelow}
we mark
them with a star: $\eta _i^\ast $.

The hadronic matrix element $\langle \oll \rangle $ is parametrized
as
\begin{eqnarray}
 \langle   \ov{K^0} \mid \oll (\mu) \mid {K^0} \rangle &=&
\frac{8}{3} B_K (\mu ) f_K^2 m_K^2 .
\label{bk}
\end{eqnarray}
Here $m_K$ and $f_K$ are mass and decay constant of the neutral
K meson.
In \rf{bk} $B_K (\mu)$ measures the deviation from the vacuum insertion
approximation $B_K (\mu)=1$. It must combine with $b(\mu )$ in
\rf{bmu} to the RG invariant
\begin{eqnarray}
B_K &=& B_K (\mu ) b(\mu ), \label{bkinv}
\end{eqnarray}
because physical observables are scale independent.

\section{Why Working Beyond Leading  Logarithms?}
After all we may ask whether it is necessary to do the effort of a NLO
calculation of \kkm\/ instead  of contenting oneself with the leading log
approximation. Let us therefore close this chapter by listing the
compelling reasons for passing to the
next--to--leading order:
\begin{itemize}
\item[i)]   Only in  NLO calculations it is possible to use the NLO
            quantity $\laMSb$ defined in \rf{lbmsb}.
\item[ii)]  The dependence of $\eta_3 S(x_c,x_t)$ on the top--quark
            mass first enters in the NLO.
\item[iii)] The LO results show a huge dependence on the scales
            $\mu_W$ and $\mu_c$ at which
            particles are integrated out. Such scale dependences are
            inherent to any RG improved calculation working with
            truncated perturbation series. It is caused by the fact
            that we can always subtract an arbitrary  small logarithm
            (such as $\ln (\mu_W/M_W)$) from the summed large logarithm.
            By this one pushes small terms from the LO summation into
            the NLO. In the NLO the scale dependences therefore diminishes.
\item[iv)]  One must go to the NLO to judge whether perturbation theory
            works, i.e.\ whether the radiative corrections are small.
            Further the corrections  can be sizeable.
\end{itemize}
\cleardoublepage
\chapter{Calculation of $\mathbf{\eta_1}$ and $\mathbf{\eta_3}$}\label{calc}
The following chapter is devoted to the description of some details of
the NLO calculation of $\eta_3$ and $\eta_1$ and to the presentation of the
analytical results.
\section{The LO Analysis}
The problem of RG improved short distance QCD correction
to \kkm\/ has been first
addressed by Va\u{\i}nste\u{\i}n,  Zakharov,  Novikov and
             Shifman and by
Vysotski\u{\i} \cite{vys}. They did not use the operator
product expansion
approach, but extracted the coefficients of the leading logarithms
in a complicated way from the Standard Model amplitudes.
Pioneering work  in the development of the factorization method has been
done by Gilman and Wise \cite{gw}. Their leading log results for $\eta_1$ and
$\eta_2$ agreed with those of Vysotski\u{\i}, but they found his
result for $\eta_3$ to be incorrect.
In these early works the top--quark has been treated as light.
Later Flynn \cite{fly} has redone
the calculation for the case of a heavy top--quark. A good
approximation neglecting flavour thresholds has been derived by
Datta, Fr\"ohlich and Paschos \cite{dfp}.

Let us comment  on the LO calculation of $\eta_3$ of Gilman and Wise here:
Their approach to the RG evolution is equivalent to the one described
first in the paragraph on $H^{ct}$ in sect.\ \ref{above}, i.e.\ they
consider the mixing of $\bilr{ij}$ into $\oloc$ and work with coefficient
vectors similar to \rf{cvecup}. Yet their bilocal structures are defined
in an inconvenient way, so that they need four evolution matrices instead
of two. Further their operator basis is overcomplete resulting in
$8 \times 8$ matrices rather than $7 \times 7$ ones. Indeed, the evolution
matrices in \cite{gw} have a double eigenvalue.

With \rf{gamma1double} we can simply read off the LO anomalous
dimension tensor from the divergent parts of
$\langle \bilr{ij} \rangle^{(0)} $ depicted in \fig{bi} and
\fig{penglo}. One finds
\begin{eqnarray}
\g_{+,7}^{(0)} \; = \;
\lt(
\begin{array}{c}
-4 (N+1) \\[1mm]   -8 \\[1mm] -8 (N+1) \\[1mm] -16 \\[1mm] 8 (N+1)
 \\[1mm]  16
\end{array} \rt) ,     &\quad &
\g_{-,7}^{(0)} \; = \;
\lt(
\begin{array}{c}
4 (N-1) \\[1mm]   0 \\[1mm] 8 (N-1) \\[1mm] 0 \\[1mm] -8 (N-1)
 \\[1mm]  0
\end{array} \rt) .
\label{gatelo}
\end{eqnarray}
We have expressed the  result of \cite{gw}  in terms of \rf{gatelo}
and found agreement.

\section{NLO Strategy}
{}From the discussion of chapter \ref{operator} we know that we
need the divergent parts of two--loop diagrams  for the calculation
of the NLO anomalous dimension tensor needed for $\eta _3$ and the
finite parts of two--loop diagrams for the matching calculations
in $\eta _1$. According to sect.\ \ref{icwc} and \ref{rgo} we
can use any external state to perform the calculation and we
choose massless external quarks with zero momentum.
The QCD gauge parameter $\xi$ is kept arbitrary. Its absence
from the anomalous dimension tensor and the Wilson coefficients
provides a check of the calculation.  The same has to be true
for the IR regulators, which are small masses $m_s$ and $m_d$
for the internal quarks. The problem  therefore requires
the calculation of the diagrams of \fig{biqcd} and \fig{pengqcd}.
We remark here that the insertion of the right--handed penguin
operators $Q_5$ and $Q_6$ into the diagrams of
\fig{pengqcd} involves different integrals
than that of the left--handed ones $Q_3$ and $Q_4$.

A further check of the result will be the absence of non--local terms
involving $\ln \mu $ after the inclusion of subloop counterterm
graphs.

Yet an incorrect treatment of the evanescent operators
would pass all the checks mentioned above. In the calculation of
$\eta_3$ we will
therefore keep
$\wt{a}_{1}$ and $\wt{b}_{1}$
parametrizing the order--$\e$  term in the definition
 \rf{DefEvan1} and
\rf{defe2} of $E_1[\oloc]$ and  $E_2[\oloc]$
arbitrary. While the individual $Z$--factors in
\rf{twoloop} depend on  $\wt{b}_{1}$'s in \rf{defe2} this
dependence cancels in $\g^{(1)}$. We will insert the evanescent counterterms
with a factor of $\lambda$ and with $\lambda=1/2$ the dependence
on  $\wt{b}_{1}$ correctly vanishes. Conversely the non--local
terms in the two--loop renormalization constants $Z^{(2)}_{k,E_{1k}}$
responsible for the mixing of physical operators into evanescents
vanish for $\lambda=1$.

Clearly the task at hand is simplified by an algorithm for the calculation
of the two--loop diagrams. For the Dirac algebra the computer package
{\sc Tracer} \cite{jl} has been used. Further
two--loop tensor vacuum bubble integrals up to rank six  are
involved. An algorithmic way to calculate them is described
in the following section.

\section{Master Formula for the Two--loop Integrals}\label{integrals}
We define the general two--loop vacuum bubble tensor integral as
\begin{eqnarray}
\lefteqn{T^{\nu _1 \ldots \nu _s , \mu _1 \ldots \mu _r }_{LNM}
   \lt( \{ m_j \} , \{ M_{1j} \} , \{ M_{2j} \} \rt) \; =} \nn
&&\frac{\mu ^{4 \e } e^{2 \euga \e }  }{\lt[ i \pi^{D/2}  \rt]^2 }
  \int \!\!\! \int \! d^D k d^D p
  \frac{ p^{\nu _1} \cdots  p^{\nu _s} \cdot
         k^{\mu _1} \cdots  k^{\mu _r}   }{
      \lt( m_1^2-k^2 \rt) \cdots  \lt( m_L^2-k^2 \rt) \cdot
         \lt( M_{11}^2-p^2 \rt) \cdots  \lt( M_{1N}^2-p^2 \rt)}
         \cdot \nn
&& \quad \quad \quad \quad
    \frac{1}{\lt( M_{21}^2-(p+k)^2 \rt) \cdots
       \lt( M_{2M}^2-(p+k)^2 \rt)  }
\label{deft} .
\end{eqnarray}
Here the $M_{1j}$'s and  $M_{2j}$'s  are arbitrary masses, while
the $m_j$'s are small masses to regulate IR singularities. This
is the most complicated two--loop integral
which can appear in the calculation
of QCD corrections in light hadron systems. In our case of \kkm\/
all cases with $0\leq L \leq 4$ show up in the two--loop diagrams.
Let us  set $T$ small masses exactly to zero while keeping $L-T$
of them nonzero. In practice  one chooses  $T$  just as large
that all the remaining $L-T$ masses suffice to regulate the IR
singularities. In our calculation we needed at most two nonzero
light masses, which were quark masses $m_s$ and $m_d$. But also
the case of on--shell dimensional
IR regulators will be contained in the result,
it simply corresponds to  $L=T$. The heavy masses $M_{1j}$ and
$M_{2j}$ in \rf{deft} correspond to the remaining masses
in the problem, i.e.\ $M_W$, $m_t$, $m_c$ and $m_u$ in the
Standard Model diagrams and $m_c$ and $m_u$ in the effective theory
diagrams. We have only needed the latter, but for different combinations
for the internal masses, so that we also keep them arbitrary.

For the scalar case $r=s=0$ the result for \rf{deft} is known for
all masses  being arbitrary \cite{bv},
it involves a $F_4$ function and is not
very useful for the situation at hand.

Finally the prefactor in \rf{deft} has been introduced for later convenience.

Let us now sketch the solution of \rf{deft}: The first step consists
of a decomposition into partial fractions of the propagators involving
heavy masses:
\begin{eqnarray}
\lefteqn{ T^{\nu _1 \ldots \nu _s , \mu _1 \ldots \mu _r }_{LNM}
   \lt( \{ m_j \} , \{ M_{1j}\}  , \{ M_{2j}\}  \rt) \; = } \nn
&& \sum _{l=1}^M \sum _{i=1}^N
\lt(\prod^M_{
\begin{array}{c}
\scriptstyle k=1 \\[-1mm] \scriptstyle  k \neq l
\end{array} }
\frac{1}{M_{2k}^2 - M_{2l}^2  }  \rt)
\lt(\prod^N_{
\begin{array}{c}
\scriptstyle k=1 \\[-1mm] \scriptstyle k \neq i
\end{array} }
\frac{1}{M_{1k}^2 - M_{1i}^2 }  \rt)
T^{\nu _1 \ldots \nu _s , \mu _1 \ldots \mu _r }_{L11}
   \lt( \{ m_j \} ,  M_{1i}  ,  M_{2l}  \rt)
\no
\end{eqnarray}
For equal heavy masses one finds derivatives of
$T^{\nu _1 \ldots \nu _s , \mu _1 \ldots \mu _r }_{L11}  $
with respect
to $M_{1i}$ or $M_{2l}$   instead.
$T^{\nu _1 \ldots \nu _s , \mu _1 \ldots \mu _r }_{L11}  $
is depicted in \fig{vacuum}. We refrain from using partial fractions
also in the line flown through by the loop--momentum $k$, because this
would yield inverse powers of the infrared regulators $m_j$.
\begin{figure}[btp]
\caption[Vacuum bubble diagram]{The vacuum bubble diagram with two
         light masses $m_1$ and
         $m_2$. The tensor structure  is arbitrary.}
\label{vacuum}
\end{figure}

The first step to solve
$T^{\nu _1 \ldots \nu _s , \mu _1 \ldots \mu _r }_{L11}  $
is the expression of the sub--loop integral over $p$ in
\rf{deft} in terms of  a Feynman parameter integral.
The corresponding expression for a general one--loop tensor integral
has been derived first by Davydychev \cite{dav}.
Clearly only one Feynman parameter $x$ is involved.
Now one is faced with a sum of tensor integrals corresponding to
all possibilities to build a rank--s tensor out of $\lambda$
copies of the metric tensors and $s-2 \lambda$
copies of the other momentum $k$.
Each of the remaining  integrals over
$k$ is proportional to the totally symmetric tensor
\begin{eqnarray}
\{  [ g ] ^{\frac{r+s}{2} - \lambda  }
    \}^{\nu _{\rho_{2 \lambda +1 } } \ldots \nu_{\rho_{s}} ,
                     \mu _1 \ldots \mu_r  } , \no
\end{eqnarray}
where $\{ \rho_{ 2 \lambda +1} \ldots \rho_{s}  \}$ is a subset of
$\{1,\ldots s   \} $. The result reads
\begin{eqnarray}
T^{\nu _1 \ldots \nu _s , \mu _1 \ldots \mu _r }_{L11}\!\!\!\! &=&
\!\! \mu ^{4 \e}
\sum_{\lambda =0}^{\lt[\frac{s}{2} \rt] } \;
\sum_{ \{ \rho _1,\ldots \rho _{2 \lambda }   \} \cup
       \{ \rho _{2 \lambda+1},\ldots \rho _{s }   \}
     =  \{ 1,\ldots s \} } \;
\{  [ g ] ^{\lambda  }
    \}^{\nu _{\rho_{1 } } \ldots \nu_{\rho_{2 \lambda }} }
\{  [ g ] ^{ \frac{s+r}{2}  -\lambda  }
    \}^{\nu _{\rho_{2 \lambda +1 } } \ldots \nu_{\rho_{ s }}
       \mu_1 \ldots \mu_r  } \nn
&& (-1)^{\frac{3}{2} s + \frac{r}{2}}
    \lt(\frac{1}{2} \rt)^{\frac{s+r}{2}}
     \frac{\Gamma (-\lambda +\e ) e^{2 \e \euga}  }{
           \Gamma \lt(\frac{s+r}{2}+2 -\lambda -\e \rt)  } \cdot
    \int_0^1 \! dx \, x^{s-2 \lambda } \lt[ x (1-x)  \rt]^{\lambda -\e}
\cdot \nn
&& \quad \quad \quad \quad \quad  \quad \quad \quad
j_{1+\frac{s+r}{2} -\lambda-T , \lambda  }
    ( M (x) , m_{T+1},\ldots m_L )
\no    .
\end{eqnarray}
 Here the second sum runs over all
$\lt( \begin{array}{c}  s \\[-1mm]  \rho_{ 2 \lambda}
\end{array} \rt)$
possibilities to select
$\rho_{2 \lambda } $ numbers from the set $\{ 1,\ldots s  \}$.
Further
$M^2(x)= M_{1j}^2/x-M_{2l}^2/(1-x)$ and the integration over
$z=\sqrt{-k^2}$ is contained in
\begin{eqnarray}
j_{p,q} (M,m_{T+1},\ldots m_L) &=&
\int_0^\infty \!\! d z \,
   \frac{ z^{p-\e }
       \lt( M^2 +z  \rt)^{q -\e} }{(m_{T+1}^2+z) \cdots (m_L^2+z) }
\label{defjpq}
\end{eqnarray}
Here the original IR singularities show up at the lower end $z=0$ of the
integration in the limit $m_i\rightarrow 0$. We have put $T$ masses exactly to
zero with $ T \leq 1 + [ (r+1)/2 ] $, so that the remaining masses regulate
the IR singularity at $z=0$.
We can  disentangle the IR regulators from the integral by noticing
that the integration path in \rf{defjpq} connects two branch points of
the integrand.  Hence we can express the integral in \rf{defjpq}
by a contour integral
\begin{eqnarray}
\int_0^\infty \!\! dz \, z^{p-\e} f(z)
   &=& \frac{(-1)^p}{2 i \sin (\pi \e )}  \int_{C} dz \,  (-z)^{p-\e} f(z) ,
\no
\end{eqnarray}
where the contour $C$ starts below the real axis at $z=+\infty$, encircles
the point $z=0$ clockwise and returns to $z=+\infty$ above the real axis.
Here $C$ is arbitrary as long as the poles $z=-m_{T+1}^2,\ldots z=-m_L^2$
and the third branch point $z=-M^2$
are to the left of $C$. Next one deforms $C$ such that it encircles the other
branch cut in \rf{defjpq} connecting
$z=-M^2$ with $z=\infty$ along the negative real axis.
When doing this one has to pull $C$ over the
poles thereby picking up $ 2  \pi i $ times the sum of the residues
of the integrand at $z=-m_{T+1}^2,\ldots z=-m_L^2 $.
One then obtains:
\begin{eqnarray}
j_{p,q} (M,m_{T+1},\ldots m_L) &=& (-1)^{p+q}
\int_0^\infty \!\! d z \,
   \frac{ z^{q-\e }
    \lt( M^2 +z  \rt)^{p -\e} }{(-m_{T+1}^2+M^2 +z) \cdots (-m_L^2+M^2 +z) }
\nn
&&    + (-1)^{p} \frac{\pi}{\sin (\pi \e)}  \sum_{l=T+1}^{L}
          \frac{(m_l^2)^{p-\e} (M^2-m_l^2)^{q-\e} }{
   \prod_{
\begin{array}{@{}l}
\scriptstyle k= T+1   \\[-1.5mm] \scriptstyle  k \neq l
\end{array}
}^L \lt( m_k^2 - m_l^2 \rt)  } . \label{jcont}
\end{eqnarray}
In the first integral we can safely set the small masses to zero and
express the integral in terms of Euler's beta--function.
This integral represents the result  for the case of
dimensional IR regularization.
All IR regulators $m_j$  are contained in an additive term. This must be
so, because we know that IR terms in the full and effective theory
factorize with the LO Wilson coefficient.
If one  kept
small external momenta as IR regulators, they would appear in the same
way in an additive term to the first integral (with $m_i=0$) in
\rf{jcont}.
Now we can do the final integration  over the Feynman parameter $x$
yielding:
\begin{eqnarray}
\lefteqn{ T^{\nu _1 \ldots \nu _s , \mu _1 \ldots \mu _r }_{L11}
   \lt( \{ m_j \} ,  M_{1j}  ,  M_{2j}  \rt) \; =} \nn
&& \sum_{\lambda =0}^{\lt[\frac{s}{2} \rt] } \;
\sum_{ \{ \rho _1,\ldots \rho _{2 \lambda }   \} \cup
       \{ \rho _{2 \lambda+1},\ldots \rho _{s }   \}
     =  \{ 1,\ldots s \} } \;
\{  [ g ] ^{\lambda  }
    \}^{\nu _{\rho_{1 } } \ldots \nu_{\rho_{2 \lambda }} }
\{  [ g ] ^{ \frac{s+r}{2}  -\lambda  }
    \}^{\nu _{\rho_{2 \lambda +1 } } \ldots \nu_{\rho_{ s }}
       \mu_1 \ldots \mu_r  } \cdot  \nn
&&
    \lt(\frac{1}{2} \rt)^{\frac{s+r}{2}}
   \lt[
     (-1)^{ \frac{r-s}{2} }
    \frac{\Gamma (-2 +L - \frac{s+r}{2} +2 \e ) e^{2 \e \euga}  }{
           \lt( 2 - L +\frac{s+r}{2} -\lambda -\e \rt)_L  }
        h^{2-L +\frac{s+r}{2}  }_{\frac{s-r}{2} +L-2-\lambda , \:
               -\frac{s+r}{2}+L-2 +\lambda   }
          \lt( M_{2l}, M_{1j} \rt) \rt.
 \nn
&& \quad \quad \quad \quad + \lt.
   T^{s,r,\lambda, T}_{L11, \mr{IR}}
   \lt( \{ m_j \} ,  M_{1j}  ,  M_{2j}  \rt) \rt] .
\label{tires}
\end{eqnarray}
Here
$T^{s,r,\lambda, T}_{L11,\mr{IR}}  $
is the part comprising the IR regulators. In \rf{tires}
$(k)_l=k\cdot (k+1) \cdots (k+l-1)$ is the Pochhammer symbol.
The function $h$ in \rf{tires} reads:
\begin{eqnarray}
h^{p}_{k,n}(M_{i},M_{j})&:=&\mu^{4\varepsilon}
  \int_{0}^{1} \!\! dx x^{k+\varepsilon}
    (1-x)^{n+\varepsilon} \left[x (M_{i}^{2}-M_{j}^{2})+M_{j}^{2} \right]
    ^{p-2 \varepsilon} \; = \; h^{p}_{n,k}(M_{j},M_{i}) \nn
&=& (M_{j}^{2})^{p} \left( \frac{M_{j}^{2}}{\mu ^{2}}\right)^{-2\e}
\mbox{B}(1+k+\e,1+n+\e)  \nn
 &&     {}_{2}\mbox{F}_{1}(-p+2\e,1+k+\e,2+k+n+2\e,
          1- \frac{M_{i}^{2}}{M_{j}^{2}} )  .
\label{defh}
\end{eqnarray}
The second expression in terms of Gau\ss' hypergeometric function
is useful for the case of one small mass $M_i$, e.g.\
$M_i=m_u$, in which case one can use Kummer's relations to transform
to ${}_2F_1 (.,.,.; M_i^2/M_j^2)$.
In general the integral representation is more useful, because it can be more
easily  expanded in terms of $\e$. It further exhibits more symmetries
and functional relations such as
\begin{eqnarray}
h^p_{k,n} (M_i,M_j) \; = \; h^p_{n,k} (M_j,M_i) , \quad
h^p_{k,n} (M_i,M_j) \; = \;
h^p_{k+1,n} (M_i,M_j)  +
h^p_{k,n+1} (M_i,M_j) && \nn
\frac{\partial }{\partial M_j^2} h^p_{k,n} (M_i,M_j) \; = \;
(p-2 \e) h^{p-1} _{k,n+1} (M_i,M_j) \label{func}.
\end{eqnarray}
For equal masses  one simply has
\begin{eqnarray}
h^{p}_{k,n} (M,M) &=&   \mu ^{4 \e} \lt(  M^2 \rt)^{p-2 \e}
  B(1+k+\e, 1+n +\e) \no .
\end{eqnarray}
For different masses  the expansion of the integral in \rf{defh}
in terms of $\e$ turns out to be the only cumbersome task
to be done. One finds the familiar dilogarithm in the finite part and
higher polylogarithms in the higher order terms in $\e$.
Yet one only has to expand a few of the $h$'s and one can obtain
the others by the use of the functional equations \rf{func}.
The expanded $h$--functions have been stored in a {\sc Mathematica}
program  to build  a database.

Now for a dimensional regularization of the IR  singularities one
finds
$ T^{s,r,\lambda, T}_{L11, \mr{IR}} =0$.
For the use of small masses as done by us one obtains:
\begin{eqnarray}
\lefteqn{   T^{s,r,\lambda, T }_{L11, \mr{IR}}
   \lt( \{ m_j \} ,  M_{1j}  ,  M_{2j}  \rt) \; = } \nn
&&
(-1)^{r+1-T-\lambda } \frac{\pi}{\sin (\pi \e) }
 \frac{\Gamma ( - \lambda + \e ) e^{2 \e \euga} \mu ^{4 \e}  }{
       \Gamma (\frac{s+r}{2}+2 - \lambda -\e  )  }
   \sum_{j=1}^{L-T} \sum_{l=0}^{L-2 -\frac{s+r}{2}+\lambda  }
    \frac{ \lt( m_j^2 \rt)^{1+\frac{s+r}{2}-T-\lambda +l-\e }  }{
         \prod_{
\begin{array}{@{}c} \scriptstyle n=1   \\[-1.5mm] \scriptstyle  n \neq j
\end{array}
       }^{L-T} \lt( m_n^2 - m_j^2  \rt)   } \cdot \nn
&& \sum_{k=0}^l
\lt( \begin{array}{c}  l \\[-1.5mm]  k
\end{array} \rt)
(-1)^{k+l} \frac{1}{(\lambda +1 -\e)_{s+k+1-2 \lambda } } \cdot \nn
&& \quad \quad \quad   \frac{\partial ^{s+k+l-2 \lambda  } }{
        \partial \lt( M_{2l}^2  \rt)^{s+l+k-2 \lambda } }
  \frac{\lt( M_{2l}^2  \rt)^{s+k+1- \lambda -\e  } -
      \lt( M_{1j}^2  \rt)^{s+k+1- \lambda -\e  }  }{M_{2l}^2- M_{1j}^2  }.
\label{ir}
\end{eqnarray}
In practice the summation indices only run over a small range, in the
calculation of $\eta _1$ and $\eta _3$ only situations with
$L-T \leq 2$  and $k=l=0$ appeared.

We remark that the master formula \rf{tires} is a good investment into the
future of short distance QCD corrections: First the formula
holds in $D$ dimensions. It can therefore be used for the matching
calculations in a NNLO calculation and also for the two--loop
counterterm diagrams in a $n$--loop calculation, which requires
two--loop integrals up to the order $n-3$ in $\e$, if the
divergent part is needed.
Probably all  calculations beyond NLO involve so many diagrams
that a complete computerization is unavoidable. Then formulas
of the type \rf{tires} also for higher loop integrals are
indispensable.
Further \rf{tires} can
simply be extended to more powers in the small masses. Such integrals are
needed in NLO calculations for the coefficients of
higher  dimension operators.

\section{Result for the NLO Anomalous Dimension Tensor}
In this section we list the result for the NLO anomalous dimension
tensor $\mathbf{\gabipm}$.

{}From \rf{thehs}, \rf{bilocs} and \rf{eff2} one readily finds the
relation between the two--loop renormalization tensor
$\lt[ Z^{-1,(2)} \rt]_{ij,7}$ and the bare $O(\alpha)$ matrix
elements of $\bilr{ij}$:
\begin{eqnarray}
\lefteqn{
- \lt[ Z^{-1,(2)} \rt]_{ij,7} \langle \oloc \rangle ^{(0)}
\; = } \nn
&&
\langle \bilr{ij}   \rangle^{(1) ,\ba} +
     \frac{1}{\e}
  \lt[  2 Z^{(1)}_{\psi ,1} \delta_{ii^\prime}  \delta_{jj^\prime}
      - \lt[ Z_1^{(1)}  \rt]_{ii^\prime} \delta_{jj^\prime}  -
        \lt[ Z_1^{(1)}  \rt]_{jj^\prime} \delta_{ii^\prime} \rt]
       \langle \bilr{i^\prime j^\prime}   \rangle^{(0) ,\ba}      \nn
&& + 2 Z_{\psi,1}^{(1)} \frac{1}{\e^2} \lt[ Z_1^{-1,(1)}  \rt]_{ij,7}
     \langle \oloc \rangle ^{(0)} + \frac{1}{\e^2}
   \lt[ Z_1^{-1,(1)}  \rt]_{ij,7} \langle \oloc \rangle^{(1), \ba}
\label{findz}   ,
\end{eqnarray}
where we have used the notation \rf{matexr} and \rf{exp} for the
expanded matrix elements and $Z$--factors.
Now one has to take into consideration that
\begin{eqnarray}
\langle \oloc \rangle^{\ba} &=&
    \langle \frac{ m_{c,\,\ba }^{2} }{g_{\ba }^{ 2}} \oll
      \rangle^{\ba} \; = \; Z_m^2 Z_g^{-2}
     \frac{ m_c^{2} }{g^{ 2} \ov{\mu}^{2 \e}  } \langle \oll
      \rangle^{\ba}  .
\end{eqnarray}
Hence
\begin{eqnarray}
\langle \oloc \rangle^{(1),\ba} &=&
   \frac{ m_c^{2} }{g^{ 2} \ov{\mu}^{2 \e}  }
      \lt( 2 Z^{(1)}_{m,1} +\beta_0      \rt) \frac{1}{\e}
     \langle \oll \rangle ^{(0)} +
     \frac{ m_c^{2} }{g^{ 2} \ov{\mu}^{2 \e}  }
    \frac{1}{\e} \langle \oll \rangle ^{(1), \ba} . \label{expsev}
\end{eqnarray}
Now the divergent parts of the diagrams of
\fig{pengqcd} and \fig{biqcd} and the corresponding subloop diagrams
yield  the terms in the first line of \rf{findz} and the last
term in \rf{expsev}. The remaining divergences therefore correspond
to
\begin{eqnarray}
\lt[ Z^{-1,(2)}_2 +
  \lt( 2 Z_{m,1}^{(1)}  +\beta_0 +2 Z_{\psi,1}^{(1)}  \rt)
  Z^{-1,(1)}_1      \rt]_{ij,7} \frac{1}{\e^2} +
    \lt[ Z^{-1,(2)}_1 \rt]_{ij,7} \frac{1}{\e}  \label{remdiv}
\end{eqnarray}
Hence we can simply read off $Z^{-1,(1)}_1$ from the
$1/\e$--divergences of the diagrams of \fig{biqcd} and \fig{pengqcd}
after the inclusion of subloop counterterms. We have also found that
the dependence on $\xi$ vanished from the $1/\e^2$--terms in
\rf{remdiv} after subtracting the term involving $Z_{\psi,1}^{(1)}$.
Further it has been checked that  $Z^{-1,(2)}_2 $ fulfills the relation
to the $1/\e$ divergences of the one--loop renormalization constants.

Now $\g_{k\pm,7}^{(1)}$ is related to $\lt[ Z^{-1,(1)}_1 \rt]_{k\pm,7}$
via \rf{gamma1double}, i.e.\ the evanescents must be inserted with a
factor of $1/2$.

To show the scheme dependence found in chapter \ref{eva} we keep
the definitions of some
evanescents arbitrary:
\begin{eqnarray}
E_1[Q_{1/2}] &=& \g_ \mu \g_ \nu \g_ \vartheta L \otimes
               \g^\vartheta   \g^ \nu \g^\mu L - (4 - 8  \e)
      \g_ \mu L \otimes   \g^ \mu L \nn
E_1[Q_{3/4}] &=& \g_ \mu \g_ \nu \g_ \vartheta L \otimes
               \g^\vartheta   \g^ \nu \g^\mu L - (4 + a_1 \e)
      \g_ \mu L \otimes   \g^ \mu L \nn
E_1[Q_{5/6}]   &=& \g_ \mu \g_ \nu \g_ \vartheta R \otimes
               \g^\vartheta  \g^ \nu \g^\mu L - (16 + a_2 \e)
               \g_ \mu R \otimes   \g^ \mu L  . \nn
E_1[\wt{Q}_{7}] &=& \g_ \mu \g_ \nu \g_ \vartheta L \otimes
               \g^\vartheta   \g^ \nu \g^\mu L - (4 + \wt{a}_1 \e)
      \g_ \mu L \otimes   \g^ \mu L   \label{ev1to7}
\end{eqnarray}
Insertions of $Q_1$ and $Q_2$ into the diagrams of \fig{biqcd}
also involve
\begin{eqnarray}
E_2[\wt{Q}_7] &=&
    \g_ \mu \g_ \nu \g_ \vartheta \g_\sigma \g_\rho L \otimes
       \g^\rho \g^\sigma  \g^\vartheta   \g^ \nu \g^\mu  L
    - \lt[ (4 + \wt{a}_1 \e)^2 + \wt{b}_1 \e \rt]
      \g_ \mu L \otimes   \g^ \mu L  \label{ev27}.
\end{eqnarray}

Now one finds for $N=3$:
\begin{eqnarray}
\g_{+,7}^{(1)} \; = \;
\lt(
\begin{array}{@{}c}
\displaystyle
\frac{404}{3}  + \frac{130 \wt{a}_1}{3}
\\[3mm]
\displaystyle
\frac{172}{3} + \frac{32 \wt{a}_1}{3}
\\[3mm]
\displaystyle
- \frac{1720}{3} - \frac{88 a_1}{3}  + \frac{32 \wt{a}_1}{3}
\\[3mm]
\displaystyle
-  \frac{584}{3} - \frac{80 a_1}{3} + \frac{28 \wt{a}_1}{3}
\\[3mm]
\displaystyle
\frac{1576}{3} + \frac{80 a_1}{3} - \frac{32 \wt{a}_1}{3} + \frac{8 a_2}{3}
 \\[3mm]
\displaystyle
\frac{1664}{3} + \frac{28 a_1}{3} - \frac{28 \wt{a}_1 }{3} + \frac{52 a_2}{3}
\end{array} \rt) , \nn
\g_{-,7}^{(1)} \; = \;
\lt(
\begin{array}{@{}c}
\displaystyle
- \frac{52}{3} - \frac{110 \wt{a}_1}{3}
\\[3mm]
\displaystyle
-60 + 4 \wt{a}_1
\\[3mm]
\displaystyle
\frac{1592}{3}  + \frac{8 a_1}{3} - \frac{16 \wt{a}_1}{3}
\\[3mm]
\displaystyle
-88 + 8 a_1 +4 \wt{a}_1
\\[3mm]
\displaystyle
- \frac{1160}{3} - \frac{16 a_1}{3} + \frac{16 \wt{a}_1}{3} +
 \frac{8 a_2}{3}
\\[3mm]
\displaystyle
-128 - 4 a_1 - 4 \wt{a}_1 - 4 a_2
\end{array} \rt) .
\label{gatenlo}
\end{eqnarray}
The dependence on $\wt{b}_1$ cancels from \rf{gatenlo}, although it
is present in the individual contributions to
the first two components of $\g_{\pm ,7}^{(1)} $.
Yet we discover a dependence on the $a_i$'s and $\wt{a}_1$ as stated
in \rf{SchemeAnomBi1} and \rf{SchemeAnomBi2}.

For the numerical analysis we will use
\begin{eqnarray}
a_1 &=& -8, \quad \quad a_2 \; = \; -16, \quad\quad \wt{a}_1  \; =\;
-8 \label{standard}  .
\end{eqnarray}
The first two choices have to made to comply with \cite{bjlw}, from
which we have taken the \dso anomalous dimension matrix and the
initial condition for the \dso coefficient vector.

\section[$\eta_3$]{$\mathbf{\eta_3}$}\label{6eta3}
We describe the initial condition and the RG evolution for the
formalism of sect.\ \ref{above} using $\vec{D}$ according to
\rf{defdvec} and \rf{rg8}. All formulas refer to the NLO.
The LO expressions can be simply obtained from them by only keeping
the leading terms and  using the LO running  $\alpha$.

\subsection{Initial Condition}
The initial condition for the \dso coefficient vector can be taken
from \cite{bjlw}:
\begin{eqnarray}
\vec{C}(\mu_W) &=&
\lt(
\begin{array}{c}
0
\\[1mm]
1
\\[1mm]
0
\\[1mm]
0
\\[1mm]
0
\\[1mm]
0
\end{array} \rt)
+ \frac{\alpha( \mu_W )}{4 \pi}  \lt[
\lt(
\begin{array}{c}
\g_{21}^{(0)}
\\[1mm]
\g_{22}^{(0)}
\\[1mm]
\g_{23}^{(0)}
\\[1mm]
\g_{24}^{(0)}
\\[1mm]
\g_{25}^{(0)}
\\[1mm]
\g_{26}^{(0)}
\end{array} \rt)  \ln \frac{\mu_W}{M_W} +
\lt(
\begin{array}{c}
B_1
\\[1mm]
B_2
\\[1mm]
- \wt{E} ( x_t(\mu_W) ) /( 2 N)
\\[1mm]
 \wt{E} ( x_t(\mu_W) )  /2
\\[1mm]
- \wt{E} ( x_t(\mu_W) ) /(2 N)
\\[1mm]
\wt{E} ( x_t(\mu_W) )   /2
\end{array} \rt)
\rt] \label{inic},
\end{eqnarray}
with
\begin{eqnarray}
B_1 &=& \frac{11}{2} \quad\quad B_2 \; = \; - \frac{11}{2 N} \nn
\wt{E} (x) &=& -\frac{2}{3} \ln x +
\frac{ x^2 ( 15- 16 x + 4 x^2) }{6 (1-x)^4 }  \ln x +
\frac{x (18 -11 x -x^2 ) }{12 (1-x)^3} - \frac{2}{3} \no
\end{eqnarray}
In \rf{inic} we have introduced a term involving the second row of
the LO \dso anomalous dimension matrix $\g^{(0)}$ to allow for the
choice $\mu_W \neq M_W$.

Further $\vec{D}$ in \rf{defdvec} involves the initial coefficients
of the operators $Q_+$ and $Q_-$:
\begin{eqnarray}
C_+  (\mu_W) &=&  B_1+B_2 + \g_+^{(0)} \ln \frac{\mu_W}{M_W}, \quad \quad
C_-  (\mu_W) \;=\;  -B_1+B_2 + \g_-^{(0)} \ln \frac{\mu_W}{M_W} .
\end{eqnarray}

For the initial coefficient of $\cloc$ we have to perform the matching
calculation with the diagrams of \fig{box} and \fig{bi}.
If the  evanescent operator of $\oloc$ is defined as in \rf{ev1to7},
one easily finds:
\begin{eqnarray}
\cloc (\mu_W) &=& \lt[ - 8 \ln \frac{\mu_W}{M_W}
                       + 4 F( x_t (\mu_W) )  - (6 +\wt{a}_1) \rt] \nn
 &=&  \clocp (\mu_W) + \clocm (\mu_W) ,
\end{eqnarray}
where $F(x_t)$ is the top--dependent part of $S(x_c,x_t)$:
\begin{eqnarray}
S(x_c,x_t) &=& -x_c \ln x_c  \, + x_c F(x_t) . \no
\end{eqnarray}

\subsection[RG Evolution between $\mu_W$ and $\mu_c$]{RG Evolution
         between $\mathbf{\mu_W}$ and $\mathbf{\mu_c}$}
The $8 \times 8$ anomalous dimension matrix $\ov{\g}$ has been introduced
in \rf{rg8}:
\begin{eqnarray}
\ov{\g} &=& \lt(
\begin{array}{ccc}
\g     & 0           &     0       \\[1mm]
\gabip & \galoc-\g_+ &     0       \\[1mm]
\gabim &   0         &  \galoc - \g_-
\end{array} \rt)
\label{gabar}       .
\end{eqnarray}
While the \dso mixing matrix $\g$ can be taken from \cite{bjlw} and
$\g_\pm$ and $\gabipm$ have been given in \rf{gammapm}, \rf{gatelo}
and \rf{gatenlo}, we still have to list the anomalous dimension
$\galoc$ of $\oloc$. Here one has to take into account the running of
the mass and coupling in its definition:
\begin{eqnarray}
\galoc &=& \g_+ + 2 \g_m -2 \beta_0 .
\end{eqnarray}
The anomalous dimension $\g_m$
of the mass has been given in \rf{gammam}.

Now one can build the matrices $\ov{U}^{(0)}$ and $\ov{J}^{(f)}$ as
described in
\rf{los} and \rf{defj}.
The NLO evolution matrix \rf{nlos} from $\mu_W$ down to $\mu_c$ reads:
\begin{eqnarray}
\dvec( \mu_c) &=&
\lt[ \uma + \frac{\alpha(\mu_c)}{4 \pi} \ov{J}^{(4)} \rt]
\ov{U}^{(0)} (\mu_c, \mu_b )
\lt[ \uma + \frac{\alpha(\mu_b)}{4 \pi} \lt( \ov{J}^{(5)} -
           \ov{J}^{(4)} + \ov{r}^{T\, (5)} - \ov{r}^{T\, (4)} \rt) \rt]
\nn
&& \ov{U}^{(0)} (\mu_b, \mu_W )
\lt[ \uma - \frac{\alpha(\mu_W)}{4 \pi} \ov{J}^{(5)} \rt]  \dvec (\mu_W) .
\label{evo8}
\end{eqnarray}
Here the matching correction at the b--threshold has been taken into
account, $\ov{r}^{(f)}$ coincides in the upper left $6 \times 6$ submatrix
with the result of \cite{bjlw}. The 7th and 8th row and column are
zero,  because there is no b--quark involved in the diagrams of
\fig{biqcd} and \fig{pengqcd}.

Now after obtaining $D_{7/8}(\mu_c)$ in \rf{evo8}
we can find the coefficient $\cloc(\mu_c)$:
\begin{eqnarray}
\cloc(\mu_c) &=& C_+ (\mu_c) D_7 (\mu_c) + C_- (\mu_c) D_8 (\mu_c).
\end{eqnarray}

\subsection[Matching at $\mu_c$]{Matching at $\mathbf{\mu_c}$}
The matching condition at $\mu_c$  has been given in \rf{cctnlo}.
Now both the diagrams of \fig{bi} and  those of \fig{penglo} have to
be calculated.
One finds for the quantities $\rst{ij}$ defined in \rf{defrij}:
\begin{eqnarray}
\rst{ij} (\mu_c) &=&
\lt\{
\begin{array}{cl}
\lt[ -4 \ln (m_c/\mu_c) + (3+\wt{a}_1/2) \rt] \tau _{ij}
   \quad & \mbox{for } j\leq2 \\
\lt[ -8 \ln (m_c/\mu_c) - 4  \rt] \tau_{ij}
   \quad & \mbox{for } 3\leq j \leq 4 \\
\lt[ 8 \ln (m_c/\mu_c) + 4  \rt] \tau_{ij}
    \quad & \mbox{for } 5 \leq j \leq 6
\end{array} \label{rsc}\rt.
\end{eqnarray}
with the colour factors
\begin{eqnarray}
\tau_{\pm1} =\tau_{\pm3} =\tau_{\pm5} = \frac{1}{2} (1 \pm N), \nn
\tau_{+j} = 1 \quad \mbox{and} \quad
\tau_{-j} = 0 \quad \mbox{for $j$ even}. \nn
\end{eqnarray}

Finally with \rf{theetas} we write down the NLO result for $\eta_3$:
\begin{eqnarray}
\eta_3 &=& \frac{ x_c ( \mu_c) }{ S \lt( x_c (\mu_c) , x_t(\mu_W) \rt) }
 \lt[ \alpha ( \mu_c) \rt]^{d_+^{(3)}} \lt[
 \frac{1}{4} \frac{4 \pi }{\alpha(\mu_c)}
  \lt( 1- \frac{\alpha(\mu_c)}{4 \pi} J_+^{(3)}  \rt) \cloc (\mu_c)
 \rt. \nn
&& \quad \quad \quad \quad
+ \lt. \frac{1}{2} \sum_{i=+,-} \sum_{j=1}^6 \rst{ij}
 C_i(\mu_c) C_j (\mu_c) \rt]
\cdot \lt( 1 + O(\alpha^2) \rt)
\label{eta3res}.
\end{eqnarray}
Order $\alpha$ terms should be consistently removed here.

In most cases it is more useful to define $\eta_3$ such that it
multiplies $S(x_c^\ast , x_t^\ast ) $ in the hamiltonian, where
$x_c^\ast= x_c (m_c)$ and $ x_t^\ast = x_t (m_t) $.
We then mark $\eta_3$ with a star:
\begin{eqnarray}
\eta_3^\ast &=& \frac{ x_c( \mu_c) }{ S \lt( x_c^\ast , x_t^\ast \rt) }
 \lt[ \alpha ( \mu_c) \rt]^{d_+^{(3)}} \lt[
 \frac{1}{4} \frac{4 \pi }{\alpha(\mu_c)}
  \lt( 1- \frac{\alpha(\mu_c)}{4 \pi} J_+^{(3)}  \rt) \cloc (\mu_c)
\rt. \nn
&& \quad \quad \quad \quad
+ \lt. \frac{1}{2} \sum_{i=+,-} \sum_{j=1}^6 \rst{ij}
 C_i(\mu_c) C_j (\mu_c) \rt]
\cdot \lt( 1 + O(\alpha^2) \rt)
\label{eta3resst}.
\end{eqnarray}
I.e.\ one has
\begin{eqnarray}
\eta_3 S \lt( x_c(\mu_c), x_t (\mu_W) \rt)  &=& \eta_3^\ast
S \lt( x_c^\ast, x_t^\ast \rt) . \no
\end{eqnarray}

\rf{eta3res}  has been checked by expanding all evolution matrices
to the first order in $\alpha$ to verify that the leading logarithm
in $S(x_c,x_t)$  is correctly reproduced. Indeed it has been found
that $\eta_3 b(\mu) =1 +O(\alpha)$, when the RG improvement is
switched off in this way.
Further $\eta_3^\ast$ has been expanded around $\mu_c = m_c$
to check whether the logarithm in \rf{rsc} cancels
the dependence of the LO term  on $\mu_c$ in the calculated order.
This is also the case.

\section[$\eta_1$]{$\mathbf{\eta_1}$}
As discussed in sect.\ \ref{above} the RG factors in $\eta_1$
only involve the known RG evolution of $C_+$ and $C_-$.
The only complicated task is the matching at the scale $\mu_c$
shown in \rf{matchc}, which
requires the finite parts of the diagrams of
\fig{biqcd} and \fig{locqcd}. We will be sketchy here, for more
details on the calculation we refer to \cite{hn1} and \cite{stefdiss}.

The LO matrix element of $\bilo{ij}$ reads
\begin{eqnarray}
\langle {\cal O}_{ij} (\mu) \rangle^{(0)}&=&
      \tau_{ij} \frac{ \mc (\mu ) }{8 \pi^ 2}
                   \langle \oll \rangle^{(0)}  \label{tau}
\end{eqnarray}
with the colour factors
\begin{eqnarray}
\tau_{++} &=& \frac{N+3}{4}, \quad
\tau_{+-} \; = \; \tau_{-+} \; = \; \frac{-N+1}{4}, \quad
\tau_{--} \; = \;  \frac{N-1}{4}.  \no
\end{eqnarray}

In the NLO one obtains:
\begin{eqnarray}
\langle {\cal O}_{ij} \lt( \mu \rt) \rangle^{(1)}   &=&
                \frac{\mc (\mu)}{8 \pi^2} \lt[
                \langle \oll  \rangle ^{(0)}   a_{7}^{(ij)} (\mu)
             +  \langle \hat{T}  \rangle ^{(0)}   \tau_{ij} h_T
              +  \langle \hat{U}  \rangle ^{(0)}   \tau_{ij} h_U \rt]
              .
\label{bime}
\end{eqnarray}
The coefficients multiplying the unphysical matrix elements are of
course the same as those of the Standard Model amplitude given
in \rf{ltu}.
We split $a_{7}^{(ij)}$ in (\ref{bime})
into its physical part $c_{7}^{(ij)}$ and those parts which depend
on the infrared regulators or involve the gauge parameter:
\begin{eqnarray}
a_{7}^{(ij)}(\mu) &=& c_{7}^{(ij)}(\mu) + \tau _{ij} \lt\{
       \frac{N-1}{2 N} 3 \log \frac{\md \ms}{\mu ^4}  \rt.     \nn
&&   \quad \quad  \quad \quad + \,
   \xi \lt[  \lt(C_F + \frac{N-1}{2 N}  \rt)
    \lt( 2 - 2 \frac{\ms \log (\ms/\mu^2) -
                     \md \log (\md/\mu^2) }{\ms-\md}     \rt) \rt. \nn
&& \quad \quad \quad \quad \quad \quad \quad \lt. \lt. +
     \frac{N-1}{2 N} \log \frac{\md \ms}{\mu^4}  \rt]   \rt\} . \label{all}
\end{eqnarray}
The desired physical parts in (\ref{all}) are found as
\begin{eqnarray}
c_{7}^{(++)} (\mu) &=& \tau_{++} \: 3 (1-N) \log \frac{\mc (\mu)}{\mu^2}   \nn
&&                   + \frac{102-73 N -32 N^2 +3 N^3}{8 N}  +
               \pi^2 \frac{-6+6 N+N^2 -N^3}{12 N} , \no \\[2mm]
c_{7}^{(+-)}(\mu) \; = \; c_{7}^{(-+)} ( \mu)
&=& \tau_{+-}  \: 3 (-1-N) \log \frac{\mc (\mu)}{\mu^2}    \nn
&&                   + \frac{34-39 N +8 N^2 -3 N^3}{8 N}  +
               \pi^2 \frac{-2+4 N-3 N^2 +N^3}{12 N} , \no  \\[2mm]
c_{7}^{(--)}(\mu) &=& \tau_{--} \: 3 (-3-N) \log \frac{\mc
      (\mu)}{\mu^2}  \nn
&&                   + \frac{-34+19 N +12 N^2 +3 N^3}{8 N}  +
               \pi^2 \frac{2-6 N+5 N^2 -N^3}{12 N} . \label{physll}
\end{eqnarray}

The matrix element of $\oll$ can be found in \cite{bjw,bw}.
Writing
\begin{eqnarray}
\langle \oll \lt( \mu \rt) \rangle   &=&
\langle \oll \lt( \mu \rt) \rangle^{(0)}  + \frac{\alpha (\mu) }{4 \pi}
          \lt[
                \langle \oll  \rangle ^{(0)}   a (\mu)
             +  \langle \hat{T}  \rangle ^{(0)}    h_T
              +  \langle \hat{U}  \rangle ^{(0)}   h_U \rt]  \label{lome}
\end{eqnarray}
the coefficient $a(\mu)$ reads
\begin{eqnarray}
a (\mu) &=& c  +  \lt\{
       \frac{N-1}{2 N} 3 \log \frac{\md \ms}{\mu ^4}  \rt.     \nn
&&   \quad \quad  \quad \quad + \,
   \xi \lt[  \lt(C_F + \frac{N-1}{2 N}  \rt)
    \lt( 2 - 2 \frac{\ms \log (\ms/\mu^2) -
                     \md \log (\md/\mu^2) }{\ms-\md}     \rt) \rt. \nn
&& \quad \quad \quad \quad \quad \quad \quad \lt. \lt. +
     \frac{N-1}{2 N} \log \frac{\md \ms}{\mu^4}  \rt]   \rt\} , \nn
c  &=&
 -3 \, C_F -5 \, \frac{N-1}{2 N} .  \label{acloc}
\end{eqnarray}

Now we can solve the matching condition \rf{matchc} for $\cll{c}(\mu_c)$:
\begin{eqnarray}
\cll{c} (\mu_c) &=&
 \! \! \mc(\mu_c ) \sum_{i,j=+,-}
C_i(\mu_c) C_j (\mu_c)
\lt[  \tau_{ij}
      + \frac{\alpha(\mu_c)}{4 \pi}   r_{ij} (\mu_c) \rt] . \label{wccloc}
\end{eqnarray}
with
\begin{eqnarray}
r_{ij} (\mu_c) &=& c_{7}^{ij} (\mu_c)- \tau_{ij} c . \label{rij}
\end{eqnarray}
Again the unphysical terms have canceled from the Wilson coefficient.

Hence \rf{theetas} gives for $\eta_1$:
\begin{eqnarray}
\eta_1 \!  &=&
 \!  \lt( \alpha (\mu_c) \rt)^{d_+^{(3)}}
    \! \! \sum_{i,j=+,-} \!
    C_i(\mu_c) C_j (\mu_c)
    \lt[ \tau_{ij}
    + \frac{\alpha(\mu_c)}{4 \pi}
      \lt( r_{ij}  (\mu_c) - \tau_{ij} J_+^{(3)} \rt)     \rt] .
\label{4eta1}
\end{eqnarray}
$\eta_1^\ast$ is simply given by
\begin{eqnarray}
\eta_1^{\ast} &=&
\frac{\mc(\mu_c)}{m_c^{\ast 2}} \;
\eta_1  .
\label{4eta1s}
\end{eqnarray}

We close the section on $\eta_1$ by looking back to the matching at
the scale $\mu_W$ as displayed in \rf{matchcup} to analyze the
findings of \cite{wit} and sect.\ \ref{above}:

Consider first  the diagrams $D_1$ and $D_3$. They can only be divided
into their \dso substructures by a three-particle cut. This cut
separates two \dso subdiagrams with four external (anti--) quarks
and one external gluon. The cut is flown through by two integration
momenta, but both refer to loop integrals which are  finite
after using the GIM mechanism, which reduces the
UV behaviour of the integral involving the internal up--type quark
propagators.
Hence these diagrams match their counterparts
in \fig{boxqcd} with  the tree--level Wilson coefficients
$C_i^{(0)} C_j^{(0)} $. The same reasoning applies to the self--energy
insertion in diagram $D_5$. Now the remaining diagrams can be divided
by a two--particle cut into their \dso substructures. Again the
divergences of the loop
integration associated with the momentum flowing through the cut
are canceled by the GIM mechanism. But now the cut separates a
tree--level \dso vertex from a dressed one, which involves an
UV--divergent loop. These UV divergences are responsible for the
$O(\alpha)$ corrections to the \dso coefficients
$C_\pm (\mu_W)$. In fact the colour singlet
diagram $D_6$ does not contribute to the \dso Wilson coefficient.
We have verified that only $D_2$, $D_4$ and $D_7$ contain terms
requiring $C^{(1)}(\mu_W)$ to match their Standard Model counterparts.
Finally only $D_7$ gives a term involving $\ln \mu_W/M_W$, which
appears in the NLO matching to compensate the LO scale ambiguity.
This can be traced back to the fact that $D_7$ involves
diagram (3) of \fig{ds1ccnlo} as a subdiagram,
which is the only source of the
LO anomalous dimension of the \dso operators.
\cleardoublepage
\chapter{Numerical Results and Phenomenology}\label{phen}
\section[$\eta_3$]{$\mathbf{\eta_3}$}
At first we will investigate the dependence of $\eta_3$ and
$\eta_3^\ast$  on the various physical parameters and on the scales
at which particles are integrated out.

Recall the relevant part of the effective low--energy
hamiltonian from \rf{hbelow}:
\begin{eqnarray}
H^{ct} (\mu) &=&
     \frac{\gft}{16 \pi^2} \mw 2 \lambda_c \lambda_t \eta_3
     S(x_c (\mu_c) ,x_t (\mu_W ))  b(\mu)\oll (\mu) \nn
&=&   \frac{\gft}{16 \pi^2} \mw 2 \lambda_c \lambda_t \eta_3^\ast
     S(x_c^\ast,x_t^\ast)  b(\mu)\oll (\mu) \label{rebelow} .
\end{eqnarray}
We will start with the discussion of the scale dependence:
$\eta_3$ and $\eta_3^\ast$ depend on the initial scale $\mu_W$ and on
$\mu_c\approx m_c$,
at which the c--quark is removed from the theory. There is
also a dependence on the scale $\mu_b\approx m_b$, at which we pass from a
five--quark theory to a four--quark theory. But $\mu_b$ only appears
in the running $\alpha$ and the dependence is so small that one could
even set $\mu_b=\mu_c$.
Now $\eta_3$ in \rf{rebelow} multiplies $S(x_c (\mu_c) ,x_t (\mu_W ))$,
which also
depends on the scales, and ideally the product is scale independent.
Clearly for the discussion of this issue the notion of $\eta_3^\ast$
is more useful, because it multiplies a scale independent function and
should therefore also be independent of $\mu_W$ and $\mu_c$.
In sect. \ref{6eta3} we have mentioned that the LO scale dependences
of $\eta_3^\ast$ are analytically
canceled by the NLO terms within the calculated order. Hence we can
use the remaining scale dependence as a measure of the accuracy of the
calculation.  Since W--boson and top--quark are simultaneously
integrated out at $\mu=\mu_W$, the range for $\mu_W$ is
\begin{eqnarray}
M_W \sims \mu_W \sims m_t .  \no
\end{eqnarray}
We expect the dependence of $\eta_3^\ast$ on $\mu_c\approx m_c$
to be larger than the one on $\mu_W$ because of the larger coupling
constant involved.

Next consider the dependence on the QCD scale parameter $\laqcd$:
In the following $\laqcd$ is understood to be defined
with respect to four active flavours,  the corresponding quantities
in the three-- and five--flavour theory are obtained by imposing
continuity on the coupling at $\mu_b$ and $\mu_c$. The world average
for $\alpha(M_Z)=0.117$ \cite{pdg}
correspond to $\laMSb=310 \mr{MeV}$ for
$\mu_b= 4.6 \gev$ and $\laMSb=315 \mr{MeV}$ for $\mu_b=5.0 \gev$.
The leading order $\laqcd$, however,  differs from $\laMSb$ by an
overall $\mu$--dependent factor, see \rf{lbmsb}.
If one equates the LO coupling and the NLO $\ov{\mr{MS}}$--coupling
constant at the scale $\mu=M_Z=91 \gev$,
one finds that $\laMSb=315 \mr{MeV}$ corresponds
to $\laqcd^\mr{LO}=110 \mr{MeV} $. If the relation is imposed at the low scale
$\mu = 1.3 \gev$, one finds $\laqcd^\mr{LO}= 180 \mr{MeV} $.  This shows one
contribution to the error bar in LO calculations. It is, however, more
pronounced in $\eta_1$ than in $\eta_3$.

Let us therefore pick the following set of parameters:
\begin{eqnarray}
\begin{array}{lll}
\laMSb = 0.31 \gev , & \laqcd^\mr{LO} = 0.15 \gev , &  \\
m_c = \mu_c = 1.3 \gev ,\;\;\; & \mu_b = 4.8 \gev , &  \\
M_W = 80 \gev ,& m_t (m_t) = 170 \gev ,& \mu_W = 130 \gev .
\end{array} \label{stapa}
\end{eqnarray}
The value for $\eta_3^\ast$ corresponding to this set reads:
\begin{eqnarray}
\eta_{3} ^{\ast \mr{LO}} &=& 0.365  ,  \quad
\eta_{3} ^{\ast \mr{NLO}} \; = \;  0.468  \label{cent} .
\end{eqnarray}
Hence the NLO calculation has enhanced $\eta_3^\ast$
by $27\% $. From the difference of $0.103$ between the two values in
\rf{cent} 0.019 originates in the change from the LO to the NLO running
$\alpha$. The smallness of this contribution is of course caused
by the adjustment of $\laqcd^\mr{LO}$ to fit the
NLO running coupling as described in the previous paragraph.
The explicit $O(\alpha)$--corrections  from the NLO mixing and matching
contribute 0.084. Let us list the dominant sources of
the enhancement: At the initial scale $\mu_W$ the coefficients
$C_2$, $C_+$ and $C_-$  have a size  of the order 1, while
all other coefficients are almost negligible.
The RG evolution from  $\mu_W$ to
$\mu_c$ enhances the coefficient $C_-$ by roughly 75 \%
because of the negative sign of the anomalous dimension of $Q_-$, while
the coefficient $C_+$  is damped by 25\%.
Now the penguin coefficients $C_3$ to $C_6$ are still negligible
at $\mu=\mu_c$, only $C_1(\mu_c), C_2(\mu_c)$ and $\wt{C}_7 (\mu_c)$
are important. In the matching at $\mu_c$ the contribution
of $C_+ (\mu_c) C_1(\mu_c)$ numerically cancels the one of
$C_- (\mu_c) C_1(\mu_c)$, because the RG damping of the former
is compensated by a larger colour factor $\tau_{+1}=2$ having the opposite
sign of $\tau_{-1}=-1$. Finally $\wt{C}_7 (\mu_c)\approx 0.7$ has
become large due to the RG admixtures from $C_2$. Hence only
$C_- (\mu_c) C_2 (\mu_c) $,
$C_+ (\mu_c) C_2 (\mu_c) $  and $\wt{C}_7(\mu_c)$ are important.
In LO only $\wt{C}_7(\mu_c)$ enters $\eta_3$.
Keeping only $C_7 (\mu_c) $ in the NLO expression, however,
overestimates the  NLO enhancement by a factor of roughly 1.5,
because
$C_2(\mu_C)$
contributes with a negative sign to $\eta_3$ (for the standard
definition of the evanescents: $\wt{a}_1=-8$), see \rf{eta3res}.

Next we will display the dependence of $\eta_3$ and $\eta_3^\ast$
on the various parameters.
In all plots we
take the fixed quantities from the standard data set \rf{stapa}.
First consider the dependence of $\eta_3^\ast$
on the initial scale $\mu_W$ as depicted
in \fig{p1x}. We find a sizeable scale dependence of $14 \%$ in the LO
result. It is almost totally removed in the NLO, where we find
less than $3 \%$ change in $\eta_3^\ast$, when $\mu_W$ is varied
between $70 \gev$ and $190 \gev$.
\begin{figure}[hbt]
\caption[Dependence of $\eta_3^\ast$ on the scale $\mu_W$]{Dependence of
$\eta_3^\ast$ on the scale $\mu_W$, at which
the initial condition is defined.  }
\label{p1x}
\end{figure}
The situation is not so nice in the case of the dependence on $\mu_c$.
To show the large effects related to the running c--quark mass
in \rf{rebelow} we first display the running of $\eta_3$ with
$\mu_c$. In \rf{rebelow} it multiplies $S(x_c(\mu_c), x_t(\mu_W)) $,
which grows with increasing $x_c$ and therefore falls off with
increasing $\mu_c$. Hence $\eta_3$ has to grow with increasing
$\mu_c$.
\begin{figure}[hbt]
\caption[Dependence of $\eta_3$ on the scale $\mu_c$]{Dependence
of $\eta_3 $ on the scale $\mu_c$, at which
the charm quark is integrated out. $\eta_3$ has to compensate
the large running of the charm quark mass. See also \fig{p2x}.}
\label{p4x}
\end{figure}
{}From  \fig{p4x} one realizes that the LO result for $\eta_3$ worsens
the effect of the running mass in $S(x_c(\mu_c), x_t(\mu_W)) $, while
the NLO $\eta_3$ correctly grows with increasing $\mu_c$.

The corresponding plot for $\eta_3^\ast$ is shown in \fig{p2x}.
\begin{figure}[hbt]
\caption[Dependence of $\eta_3^\ast$ on the scale $\mu_c$]{Dependence
of $\eta_3^\ast$ on the scale $\mu_c$. For the scales below $1 \gev$
one recognizes the breakdown of perturbation theory. For a realistic
estimate of the scale ambiguity take
$1.1\gev \sims \mu_c \sims 1.6 \gev$.
 }
\label{p2x}
\end{figure}
We have intentionally extended the range for $\mu_c$ to
the unphysically  low value  of $\mu_c \approx 0.7 \gev$ to visualize
the breakdown of perturbation theory. Varying $\mu_c$ within the range
$1.1\gev \leq \mu_c \leq  1.6 \gev$ yields the following estimate
of the scale uncertainty:
\begin{eqnarray}
0.33 \leq  \eta_3^{\ast \,\mr{LO}} \leq 0.40 ,
&\quad& 0.43 \leq  \eta_3^{\ast \, \mr{NLO}}  \leq  0.50 . \label{interv}
\end{eqnarray}
This corresponds to a reduction of the scale uncertainty from $19 \%$
to $15 \%$. One reason for the poor improvement is the fact that the
NLO running of the mass is stronger than the LO one.
One realizes from \rf{interv} that the scale ambiguity alone is not
always a good measure of the accuracy of the calculation, because
the central value for $\eta_3^{\ast \, \mr{NLO}}$ is not in the
range quoted for $\eta_3^{\ast \,\mr{LO}}$.
Yet in the NLO order one may also judge the contribution of the
uncalculated $O(\alpha^2)$--terms  by
squaring the calculated
$O(\alpha)$ corrections.
By this one is lead to the same interval as in
\rf{interv}, so that we may consider \rf{interv} as a realistic
estimate for   $\eta_3^{\ast \, \mr{NLO}}$.

Let us now look at the dependence of $\eta_3^\ast$ on the physical
parameters.  From the smallness of the initial coefficient
$\wt{C}_7$ we expect the QCD correction to be almost independent of
$m_t^\ast=m_t(m_t)$. The relevant quantity to be discussed is the product
$\eta_3^\ast S(x_c^\ast,x_t^\ast)$ which enters \rf{rebelow} and
physical observables. It is shown in \fig{p10x}.
\begin{figure}[hbt]
\caption[Dependence of $\eta_3^\ast S(x_c^\ast,x_t^\ast)$
on $m_t^\ast$]{Dependence
of $\eta_3^\ast S(x_c^\ast,x_t^\ast) $ on $m_t^\ast=m_t(m_t)$. There is no
dependence on $m_t$ in the leading log approximation.
}
\label{p10x}
\end{figure}
Clearly the LO curve is flat. We also display the dependence of
$\eta_3^\ast$ alone on $m_t$, see \fig{p3x}. One cannot get much
physical insight from this plot, it only serves to verify that we
can treat $\eta_3^\ast$ as a constant function
of $m_t$ in phenomenological analyses.
\begin{figure}[hbt]
\caption[Dependence of $\eta_3^\ast $ on $m_t^\ast$]{Dependence
of $\eta_3^\ast $ on $m_t^\ast$.}
\label{p3x}
\end{figure}

On the other hand there is a sizeable dependence of
$\eta_3^\ast S(x_c^\ast,x_t^\ast)$
on $m_c^\ast=m_c(m_c)$ as
shown in \fig{p9x}. In $\eta_3^\ast$ alone, however, the LO dependence
on $m_c^\ast$ happens to be washed out in the NLO as can be seen in
\fig{p7x}.
\begin{figure}[btp]
\caption[Dependence of $\eta_3^\ast S(x_c^\ast,x_t^\ast) $ on
         $m_c^\ast$]{Dependence
of $\eta_3^\ast S(x_c^\ast,x_t^\ast)$ on $m_c^\ast$.}
\label{p9x}
\end{figure}
\begin{figure}[btp]
\caption[Dependence of $\eta_3^\ast$ on $m_c\ast$]{Dependence
of $\eta_3^\ast$ on $m_c^\ast$. }
\label{p7x}
\end{figure}
We close this section with a plot  of the dependence of
$\eta_3^\ast$ on $\laqcd$. \fig{p8x} reveals a very moderate
dependence on the QCD scale parameter. Recall that the actual
measurements of
$\alpha(M_Z)$ corresponds to LO values  for $\laqcd$ close to the left edge
of the plot, while the NLO scale parameter $\laMSb$ is close to the
highest values displayed.

Now we can easily summarize the result: The largest uncertainty of our
estimate of $\eta_3$ is due to the the scale ambiguity stemming from
$\mu_c$ and we obtain:
\begin{eqnarray}
\eta_3^\mr{\ast\,NLO}  &=& 0.47 \begin{array}{l}
\scriptstyle +0.03 \\[-.7mm]
\scriptstyle -0.04
\end{array} \label{3res}
\end{eqnarray}
\clearpage
\begin{figure}[hbt]
\caption[Dependence of $\eta_3^\ast$ on $\laqcd$]{Dependence
of $\eta_3^\ast$ on $\laqcd$. The NLO curve corresponds to $\laqcd=\laMSb$.
Realistic values are $300 \mr{MeV} \sims \laMSb \sims 330 \mr{MeV} $
and $100 \mr{MeV} \sims \laqcd^\mr{LO} \sims 200 \mr{MeV}$.
}
\label{p8x}
\end{figure}

\section[$\eta_1$]{$\mathbf{\eta_1}$}
Again we only sketch the NLO result for $\eta_1$ and refer to
\cite{hn1,stefdiss} for more details.
The part of the effective hamiltonian involving $\eta_1$ has been
found to be
\begin{eqnarray}
H^c (\mu)  &=&
 \frac{\gft}{16 \pi^2} \lambda_c^2 \eta_1  \mc (\mu_c) b(\mu)\oll
(\mu) \nn
&=& \frac{\gft}{16 \pi^2} \lambda_c^2 \eta_1^\ast  m_c^{\ast\, 2}
       b(\mu)\oll (\mu) \label{rebelowc} .
\end{eqnarray}

For the set of parameters given in \rf{stapa} one finds:
\begin{eqnarray}
\eta_1^{\ast \,\mr{LO} } &=& 0.80
\begin{array}{l}
\scriptstyle +0.20 \\[-.7mm]
\scriptstyle -0.16                  
\end{array} \, ,
\quad
\eta_1^{\ast \, \mr{NLO}} \;=\; 1.32
\begin{array}{l}
\scriptstyle +0.21 \\[-.7mm]        
\scriptstyle -0.23
\end{array} \, .
\label{1res}
\end{eqnarray}
For $m_c^\ast=1.4 \gev$ one likewise finds
\begin{eqnarray}
\eta_1^{\ast \,\mr{LO} } &=& 0.78
\begin{array}{l}
\scriptstyle +0.17 \\[-.7mm]
\scriptstyle -0.14                  
\end{array} \, ,
\quad
\eta_1^{\ast \, \mr{NLO}} \;=\; 1.22
\begin{array}{l}
\scriptstyle +0.16\\[-.7mm]        
\scriptstyle -0.18
\end{array} \,  .
\label{1res14}
\end{eqnarray}

Here the the dominant source of the error is again the
dependence on $\mu_c$, which has also been varied
between $1.1 \gev$ and $1.6 \gev$ in \rf{1res} and
between $1.2 \gev$ and $1.7 \gev$ in \rf{1res14}.
Now the dependence on $\laqcd$ is more
pronounced than in the case of $\eta_3^{\ast }$.
Therefore  we have included the variation with $\laqcd^\mr{LO}$
for $110 \mr{MeV} \leq  \laqcd^\mr{LO} \leq 180 \mr{MeV}$ in the error bar
for the LO result. The dependence on $\laqcd$ is displayed in
\fig{figeta1laMSb}.
The NLO calculation has therefore yielded a reduction of the
theoretical error from $45 \%$ to $33 \%$ for $m_c^\ast=1.3 \gev$
and from $40 \%$ to $26 \%$ for $m_c^\ast =1.4 \gev$.

\begin{figure}[htb]
\caption[Dependence of $\eta_1^{\ast}$ on $\laqcd $]{The dependence of
$\eta_1^{\ast}$ on $\laqcd^\mr{LO}, \laMSb$.
The choice $m_c^\ast=1.4 \gev$ has been made.
All other fixed parameters are as in
\rf{stapa}.}
\label{figeta1laMSb}
\end{figure}

\rf{1res} reveals a much larger scale dependence for $\eta_1^\ast$
than we have found for $\eta_3^\ast $in \rf{3res}. Let us therefore
investigate the origin of this uncertainty:
There are three different  contributions to $\eta_1^\ast$ in \rf{4eta1}
stemming from the products $C_+(\mu_c) C_+(\mu_c)$,   $C_+(\mu_c) C_-(\mu_c)$
and $C_-(\mu_c) C_-(\mu_c)$.  The last one is clearly the largest,
because of the negative anomalous dimension of $Q_-$. These different
sources are displayed in \fig{figeta1indivmuc}:
One can easily see, that the NLO curves related
to $C_+C_+$ and
$C_+C_-$ are almost flat as they should be, but
they contribute with opposite sign to $\eta_1^{\ast}$ and therefore
almost cancel in the sum.
On the other hand the part stemming from $C_-C_-$
contains a large residual scale
dependence and is identified as the source of the large theoretical error
in \rf{1res}.
\begin{figure}[htb]
\caption[Dependence of $\eta_1^{\ast}$ on $\mu_c$]{The dependence of the
individual  contributions
to $\eta_1^{\ast}$ on $\mu_c$ for $\laqcd^\mr{LO}, \laMSb = 300 \mr{MeV}$.
$m_c^\ast=1.4 \gev$ has been chosen.
The line labeled by \emph{(all)}\/ refers to $\eta_1^\ast$.
}
\label{figeta1indivmuc}
\end{figure}

We have already noticed in \rf{1res} and \rf{1res14} that $\eta_1^\ast$
exhibits a sizeable dependence on $m_c^\ast$. We therefore plot
$\eta_1^\ast$ as a function of $m_c^\ast$ in \fig{e1x}. Physical
quantities involve $m_c^{\ast\, 2} \eta_1^\ast$, which is shown in
\fig{e2x}.
\begin{figure}[htb]
\caption[Dependence of $\eta_1^{\ast}$ on $m_c^\ast$.]{The
 dependence of $\eta_1^{\ast}$ on $m_c^\ast$.
For very low values
$m_c^\ast \approx 1.1 \gev$ as favoured by some sum rule analyses
our perturbative calculation of $\eta_1^\ast$ is not very
reliable.    }
\label{e1x}
\end{figure}
\begin{figure}[htb]
\caption[Dependence of $\eta_1^{\ast} m_c^{\ast \, 2}$
on $m_c^\ast$.]{The dependence of $\eta_1^{\ast} m_c^{\ast \, 2}$,
which enters the hamiltonian \rf{rebelowc},
on $m_c^\ast$. The (dominant) contribution of $\eta_1^\ast$ to the
short distance contribution to the \kkmd\/ is obtained by multiplying
the displayed value for $\eta_1^{\ast} m_c^{\ast \, 2}$ by
$1.371 \cdot 10^{-15} \gev^{-1} \cdot B_K$, where $B_K$ is defined in
\rf{bkinv} }
\label{e2x}
\end{figure}
\section[Indirect CP--Violation in
$\mr{K^0\!-\!\ov{K^0}}\,$--Mixing]{Indirect CP--Violation in
$\mathbf{K^0\!-\!\ov{K^0}}\,$--Mixing}
Here we briefly describe the indirect CP--violation in the neutral
Kaon system (cf.\ e.g.\ \cite{tqs}).

The immediate consequence of \kkm\/ is that the weak interaction eigenstates
$\ket{K^0}$ and $\ket{\ov{K^0}}$ do not coincide with the mass eigenstates.
Yet if the CP symmetry were conserved, the latter would be identical
to the CP eigenstates $\ket{K_1}$ and $\ket{K_2}$:
\begin{eqnarray}
\ket{K_1} &=&   \frac{1}{\sqrt{2}} \lt( \ket{K^0} + \ket{\ov{K^0}} \rt)
\quad \quad \mbox{(CP=+1)}, \nn
\ket{K_2} &=&   \frac{1}{\sqrt{2}} \lt( \ket{K^0} - \ket{\ov{K^0}} \rt)  \no
\quad \quad \mbox{(CP=--1)}.
\end{eqnarray}
Now the  mass eigenstates $\ket{K_L}$ and $\ket{K_S}$ differ from
$\ket{K_1}$ and $\ket{K_2}$ by small admixtures of the other CP eigenstate:
\begin{eqnarray}
\ket{K_L} &=& \frac{\ket{K_2}+\ov{\e} \ket{K_1}}{ \sqrt{1+|\ov{\e}|^2} }, \quad
\ket{K_S} \; =\; \frac{\ket{K_1}+\ov{\e} \ket{K_2}}{
                                \sqrt{1+|\ov{\e}|^2} } . \no
\end{eqnarray}
The decay of $K_L$ into the CP--even two--pion state may proceed
via a CP--conserving decay of the small admixture of $\ket{K_1}$ or via
a CP--violating  decay of the dominant contribution of $\ket{K_2}$.
These possibilities are termed \emph{indirect}\/ and \emph{direct}\/
CP--violation and are compactly explained by the following
picture:
\begin{displaymath}
\begin{array}{rclccl}
\ket{K_L} & = & \displaystyle
  \frac{1}{\sqrt{2 (1+ |\ov{\e}|^2) }} & \left[
   \underbrace{   \ket{K^0} - \ket{\ov{K^0}} } +\right. & \ov{\varepsilon}
  \left.    \left( \underbrace{ \ket{K^0} + \ket{\ov{K^0}} }  \right)
   \right] &
 \label{cp} \\[5mm]
&&&\!\!\! \overbrace{\mbox{CP=--1 }} &\, \overbrace{\mbox{CP=+1 }}&
\nonumber \\
& \multicolumn{2}{c}{\mbox{Decay:}} & \multicolumn{2}{c}{
\parbox[t]{70mm}{
\setlength{\unitlength}{1mm}
\begin{picture}(70,10)
\put(10,0){\line(0,1){10}}
\put(10,0){\vector(1,0){55}}
\put(50,3){\line(0,1){7}}
\put(50,3){\vector(1,0){15}}
\put(52,5){\mbox{indirect}}
\put(25,2){\mbox{direct}}
\end{picture}}} & \mbox{ \Large$\pi^0 \pi^0$} \, \mbox{(CP=+1)}
\end{array}
\end{displaymath}
\clearpage
A measure of the indirect CP--violation which is independent of
phase conventions is given by the ratio of the amplitudes
\begin{eqnarray}
\e_K &=& \frac{ {\cal A} \lt( K_L \rightarrow (\pi \pi)_{I=0} \rt)  }{
         {\cal A} \lt( K_S \rightarrow (\pi \pi)_{I=0} \rt)   } .
\label{defek}
\end{eqnarray}
In terms of the off--diagonal element of the mass matrix
\begin{eqnarray}
M_{12} &=& \bra{K^0} H^\mr{\Delta S = 2} \ket{\ov{K^0}}
\end{eqnarray}
$\e_K$ reads
\begin{eqnarray}
\e_K &=& \frac{e^{i \pi/4}}{\sqrt{2}\Delta m } \lt( \imag M_{12} +
        2 \xi \real M_{12} \rt) ,   \no
\end{eqnarray}
where $\xi$ is a small quantity involving the  amplitude of
$\Delta I=1/2$ transitions and $\Delta m$ is the \kkmd.
Experimentally $\e_K$ is very well measured \cite{pdg}:
\begin{eqnarray}
\e_K &=& (2.266 \pm 0.023 ) \cdot 10^{-3} e^{i \pi/4 } .
\label{expek}
\end{eqnarray}
\subsection{Parametrization of the CKM--Matrix}
Since one may redefine the quark mass eigenstates by an arbitrary phase
factor, there are many physically equivalent forms of the
CKM--matrix. We adopt the standard  convention, in which
$V_{ud},V_{us},V_{cb}$ and $V_{tb}$ are real and positive.
It is further
advantageous to have a simple parametrization of the CKM--matrix.
Here we use the improved Wolfenstein parametrization, which is
used in \cite{blo}. It reads
\begin{eqnarray}
V_{CKM} &=& \lt(
\begin{array}{ccc}
1-\frac{\lambda^2 }{2} &\lambda & A \lambda^3 (\rho - i \eta ) \\
- \lambda -i A^2 \lambda^5 \eta
& 1 -  \frac{\lambda^2 }{2} & A \lambda ^2 \\
A \lambda^3 (1 - \ov{\rho} - i \ov{\eta} )
& -A \lambda^2 -i A \lambda ^4 \eta & 1
\end{array} \rt) \label{wolf}
\end{eqnarray}
with
\begin{eqnarray}
\ov{\rho } \; = \; \rho \lt( 1- \frac{\lambda ^2}{2} \rt) ,
&\quad &
\ov{\eta } \; = \; \eta \lt( 1- \frac{\lambda ^2}{2} \rt).\no
\end{eqnarray}
\rf{wolf}
is an expansion in the parameter
$\lambda = |V_{us}|+O(\lambda^6)$. \rf{wolf} is exact to order
$\lambda ^3$ and contains the phenomenologically important terms
up  to the order $\lambda^5$.
One of the relations provided by the  unitarity of $V_{CKM}$ is
\begin{eqnarray}
V_{ud} V_{ub}^* + V_{cd} V_{cb}^* + V_{td} V_{tb}^* &=& 0 . \no
\end{eqnarray}
When expressed in terms of the parameters introduced above, this
relation corresponds to a \emph{unitarity triangle}\/ in the
$(\ov{\rho},\ov{\eta})$--plane with edges
$(0,0),(1,0)$ and $(\ov{\rho},\ov{\eta})$.

One finds to an accuracy of $0.2\%$
\cite{blo} for $\lambda _j =V_{jd} V_{js}^*$:
\begin{eqnarray}
\imag \lambda_t	&=& - \imag \lambda_c \; = \; \eta A^2 \lambda	^5.
\label{imags}
\end{eqnarray}
CP--violation occurs for $\eta \neq 0$. Because of \rf{imags} we
may  use $\imag \lambda _t$ to 	parametrize
all CP--violating quantities in the Standard Model.

\mathversion{bold}
\subsection[$\mathnormal
\e_K$--Phenomenology]{$\mathbf{\e_K}$--Phenomenology\footnote{for more
details see \cite{hn3}}}
\mathversion{normal}
Inserting \rf{hbelow}, \rf{bk} and \rf{expek} into \rf{defek}
yields the following relation between $\ov{\rho}$ and $\ov{\eta} $
\cite{blo}:
\begin{eqnarray}
\ov{\eta } \lt[ (1-\ov{\rho}) A^2 \eta_2 S(x_t) + P_0 \rt]
A^2 B_K &=& 0.223 \label{hyp}  .
\end{eqnarray}
with
\begin{eqnarray}
P_0 &=& \lt[ \eta _3 S(x_c,x_t) - \eta _1 x_c \rt] \frac{1}{\lambda^4 }
\no      .
\end{eqnarray}
When $A$ is fixed, \rf{hyp} defines a hyperbola in the
$(\ov{\rho},\ov{\eta})$--plane.

Now from \rf{wolf} we notice that $A$ is simply related to $V_{cb}$
via $A=|V_{cb}|/\lambda^2 $. We remark here that much progress has been
made in the recent past to determine $|V_{cb}|$.  Exclusive tree--level
$b\rightarrow c$ decays allow the extraction from the endpoint
of the spectrum, where heavy quark  symmetry reduces the problem
to the determination of a single form factor. Here  typically values
in the range $0.036 \leq |V_{cb}| \leq 0.050$ are stated \cite{suv}.
The theoretical predictions for inclusive analyses still suffer from
large scheme and scale dependences, because they are proportional to the
fifth power of the bottom quark mass \cite{bn}. The  latter can be determined
from $\Upsilon$ spectroscopy \cite{suv}  or  by the
use of heavy quark symmetry
from  inclusive D--decays \cite{ls,bn}. The different lifetimes of the charmed
hadrons, however, make the use of heavy quark symmetry troublesome here.
Inclusive analyses give  lower central values for $|V_{cb}|$ than the
exclusive method.

To fix the third edge $(\ov{\rho}, \ov{\eta})$ of the unitarity triangle,
we need a further input. It is straightforward to   verify that
the knowledge of $|V_{ub}/V_{cb}|$ fixes a circle in the
$(\ov{\rho},\ov{\eta})$--plane:
\begin{eqnarray}
\ov{\rho}^2 + \ov{\eta}^2 &=& \lt( 1-\frac{\lambda^2}{2}\rt)^2
            \frac{1}{\lambda^2 } \frac{|V_{ub}|^2}{|V_{cb}|^2}
 . \label{circ}
\end{eqnarray}
If the phase $\delta$ in the CKM--matrix is the only source
of indirect CP--violation in the $K^0$--system,
the equations \rf{hyp} and \rf{circ} must have a common solution
$(\ov{\rho}, \ov{\eta})$. In general the hyperbola will intersect
the circle in two points to give two solutions.
Yet there is a critical set
$\{ (m_t, V_{cb}, |V_{ub}/V_{cb}|$,\linebreak $B_K) \}$,
for which the hyperbola only touches the circle.
This set encodes a lower bound on any of the four quantities as
a function of the other parameters \cite{blb,blo}.

Fig.\ \ref{fighyp} shows the hyperbola for the LO and the NLO result
for $\eta_3$. One can see that the hyperbola has moved downwards.
Therefore  the NLO value for $\eta _3$ allows for a wider
range of
$ m_t, V_{cb}, |V_{ub}/V_{cb}|$ and $ B_K $ than the older imprecise
LO result. In this sense the Standard Model mechanism
for CP--violation
has been vindicated.  For example a future determination
$( m_t, V_{cb}, \linebreak |V_{ub}/V_{cb}|, B_K) =
(170 \gev, 0.040,  0.08 , 0.7 ) $ would have been interpreted as a signal
of new physics, if the LO values $\eta_1=0.78$, $\eta_3=0.37$ had been
used. With the NLO results  $\eta_1=1.32 $, $\eta_3= 0.47$ , however,
the hyperbola \rf{hyp} intersects the circle \rf{circ} and there
is no conflict with the CKM mechanism of CP--violation.
\begin{figure}[htb]
\caption[$\e_K$: The hyperbola in the
$(\ov{\rho},\ov{\eta})$--plane.]{$\e_K$:
The hyperbolas in the $(\ov{\rho},\ov{\eta})$--plane for
$\eta_3^{\ast\,LO}=0.37$ and $\eta_3^{\ast\, NLO}=0.47$.
The other parameters are $\eta_1^\ast=1.32$, $\eta_2^\ast=0.57$,
$m_t=165 \gev$, $|V_{cb}|=0.041$, $m_c=1.3 \gev$
and $B_K=0.7$. The circle corresponds
to $|V_{ub}/V_{cb}|=0.08$. The NLO value for $\eta_3$ permits
two solutions for $(\ov{\rho},\ov{\eta})$ in this example.  }
\label{fighyp}
\end{figure}

The effect of the NLO calculation for $\eta_3^\ast$
on the  bounds on $m_t$ and $B_K$ is shown in \fig{newmt} and
\fig{newbk}.
\begin{figure}[phtb]
\caption[Lower bound on $m_t$]{Lower bound on $m_t$ from $\e_K$
for $|V_{ub}/V_{cb}|=0.08$, $B_K=0.5,\,0.7$ and $0.9$,
$\eta_1^*=1.32$, $\eta_2^*=0.57$, $\eta_3^\mr{*\,NLO}=0.47$ and
$\eta_3^\mr{*\, LO}=0.37$.}
\label{newmt}
\end{figure}
\begin{figure}[phtb]
\caption[Lower bound on $B_K$.]{Lower bound on $B_K$ from $\e_K$
for $m_t=165 \gev$, $|V_{ub}/V_{cb}|=0.06,\, 0.08$ and $0.10$,
$\eta_1^*=1.32$, $\eta_2^*=0.57$, $\eta_3^\mr{*\,NLO}=0.47$ and
$\eta_3^\mr{*\, LO}=0.37$.}
\label{newbk}
\end{figure}
\mathversion{bold}
\subsubsection[Table of $\mr{Im}\, \lambda_t $]{Table of
 $\mathbf{Im\, \lambda_t} $}
\mathversion{normal}
Since $\imag \lambda_t$ enters all CP--violating quantities, we
now give a table of the two solutions for
this quantity for various
values of  $ m_t, V_{cb}, |V_{ub}/V_{cb}|$ and $ B_K $.
The fixed quantities are as in \rf{stapa}. The corresponding
$\eta_j$'s are
\begin{eqnarray}
\eta_1 &=& 1.32, \quad \quad \quad \eta_2 \;=\; 0.57 \quad
\mbox{and} \quad \quad
\eta_3 \; = \; 0.47 \no   .
\end{eqnarray}
Both solutions are listed, the entries have to be multiplied by
$10^{-4} $. \emph{n.s.}\/ means that there is no solution.

\paragraph{$B_K=0.5$ and $m_t=160 \, \gev $:}
\begin{displaymath}
\begin{array}{||r||c|c|c|c|c|cc||}
\hline \hline
|V_{cb}|= &0.034 & 0.036 & 0.038 & 0.04 & 0.042 & 0.044 &
 \\ \hline \hline
\frac{|V_{ub}|}{|V_{cb}|}=0.06 &
 n.s. &  n.s. &  n.s. &  n.s. &  n.s. &  n.s. & \\ \hline
\frac{|V_{ub}|}{|V_{cb}|}=0.08 &
 n.s. &  n.s. &  n.s. &  n.s. &  n.s. &  n.s. & \\ \hline
\frac{|V_{ub}|}{|V_{cb}|}=0.1 &
 n.s. &  n.s. &  n.s. &  n.s. & \begin{array}{c} 1.61\\1.75\end{array} &
\begin{array}{c} 1.37\\1.87\end{array} &
\\ \hline
\end{array}\\
\end{displaymath}

\paragraph{$B_K=0.5$ and $m_t=170 \, \gev $:}
\begin{displaymath}
\begin{array}{||r||c|c|c|c|c|cc||}
\hline \hline
|V_{cb}|= &0.034 & 0.036 & 0.038 & 0.04 & 0.042 & 0.044 &
 \\ \hline \hline
\frac{|V_{ub}|}{|V_{cb}|}=0.06 &
 n.s. &  n.s. &  n.s. &  n.s. &  n.s. &  n.s. & \\ \hline
\frac{|V_{ub}|}{|V_{cb}|}=0.08 &
 n.s. &  n.s. &  n.s. &  n.s. &  n.s. & \begin{array}{c} 1.43\\1.54\end{array}
&
\\ \hline
\frac{|V_{ub}|}{|V_{cb}|}=0.1 &
 n.s. &  n.s. &  n.s. &  n.s. & \begin{array}{c} 1.41\\1.76\end{array} &
\begin{array}{c} 1.25\\1.81\end{array} &
\\ \hline
\end{array}\\
\end{displaymath}

\paragraph{$B_K=0.5$ and $m_t=190 \, \gev $:}
\begin{displaymath}
\begin{array}{||r||c|c|c|c|c|cc||}
\hline \hline
|V_{cb}|= &0.034 & 0.036 & 0.038 & 0.04 & 0.042 & 0.044 &
 \\ \hline \hline
\frac{|V_{ub}|}{|V_{cb}|}=0.06 &
 n.s. &  n.s. &  n.s. &  n.s. &  n.s. &  n.s. & \\ \hline
\frac{|V_{ub}|}{|V_{cb}|}=0.08 &
 n.s. &  n.s. &  n.s. &  n.s. &  n.s. & \begin{array}{c} 1.15\\1.49\end{array}
&
\\ \hline
\frac{|V_{ub}|}{|V_{cb}|}=0.1 &
 n.s. &  n.s. &  n.s. & \begin{array}{c} 1.35\\1.6\end{array} &
\begin{array}{c} 1.17\\1.67\end{array} &
\begin{array}{c} 1.06\\1.68\end{array} &
\\ \hline
\end{array}\\
\end{displaymath}

\paragraph{$B_K=0.7$ and $m_t=160 \, \gev $:}
\begin{displaymath}
\begin{array}{||r||c|c|c|c|c|cc||}
\hline \hline
|V_{cb}|= &0.034 & 0.036 & 0.038 & 0.04 & 0.042 & 0.044 &
 \\ \hline \hline
\frac{|V_{ub}|}{|V_{cb}|}=0.06 &
 n.s. &  n.s. &  n.s. &  n.s. &  n.s. &  n.s. & \\ \hline
\frac{|V_{ub}|}{|V_{cb}|}=0.08 &
 n.s. &  n.s. &  n.s. &  n.s. & \begin{array}{c} 1.13\\1.38\end{array} &
\begin{array}{c} 1.01\\1.39\end{array} &
\\ \hline
\frac{|V_{ub}|}{|V_{cb}|}=0.1 &
 n.s. &  n.s. &  n.s. & \begin{array}{c} 1.14\\1.54\end{array} &
\begin{array}{c} 1.03\\1.55\end{array} &
\begin{array}{c} 0.94\\1.54\end{array} &
\\ \hline
\end{array}\\
\end{displaymath}

\paragraph{$B_K=0.7$ and $m_t=170 \, \gev $:}
\begin{displaymath}
\begin{array}{||r||c|c|c|c|c|cc||}
\hline \hline
|V_{cb}|= &0.034 & 0.036 & 0.038 & 0.04 & 0.042 & 0.044 &
 \\ \hline \hline
\frac{|V_{ub}|}{|V_{cb}|}=0.06 &
 n.s. &  n.s. &  n.s. &  n.s. &  n.s. & \begin{array}{c} 1.03\\1.16\end{array}
&
\\ \hline
\frac{|V_{ub}|}{|V_{cb}|}=0.08 &
 n.s. &  n.s. &  n.s. & \begin{array}{c} 1.19\\1.27\end{array} &
\begin{array}{c} 1.02\\1.33\end{array} &
\begin{array}{c} 0.92\\1.32\end{array} &
\\ \hline
\frac{|V_{ub}|}{|V_{cb}|}=0.1 &
 n.s. &  n.s. & \begin{array}{c} 1.19\\1.44\end{array} &
\begin{array}{c} 1.04\\1.49\end{array} &
\begin{array}{c} 0.94\\1.49\end{array} &
\begin{array}{c} 0.86\\1.46\end{array} &
\\ \hline
\end{array}\\
\end{displaymath}

\paragraph{$B_K=0.7$ and $m_t=190 \, \gev $:}
\begin{displaymath}
\begin{array}{||r||c|c|c|c|c|cc||}
\hline \hline
|V_{cb}|= &0.034 & 0.036 & 0.038 & 0.04 & 0.042 & 0.044 &
 \\ \hline \hline
\frac{|V_{ub}|}{|V_{cb}|}=0.06 &
 n.s. &  n.s. &  n.s. &  n.s. & \begin{array}{c} 0.98\\1.06\end{array} &
\begin{array}{c} 0.86\\1.07\end{array} &
\\ \hline
\frac{|V_{ub}|}{|V_{cb}|}=0.08 &
 n.s. &  n.s. &  n.s. & \begin{array}{c} 0.97\\1.23\end{array} &
\begin{array}{c} 0.87\\1.23\end{array} &
\begin{array}{c} 0.79\\1.2\end{array} &
\\ \hline
\frac{|V_{ub}|}{|V_{cb}|}=0.1 &
 n.s. & \begin{array}{c} 1.18\\1.28\end{array} &
\begin{array}{c} 0.99\\1.38\end{array} &
\begin{array}{c} 0.89\\1.38\end{array} &
\begin{array}{c} 0.81\\1.36\end{array} &
\begin{array}{c} 0.74\\1.33\end{array} &
\\ \hline
\end{array}\\
\end{displaymath}

\paragraph{$B_K=0.9$ and $m_t=160 \, \gev $:}
\begin{displaymath}
\begin{array}{||r||c|c|c|c|c|cc||}
\hline \hline
|V_{cb}|= &0.034 & 0.036 & 0.038 & 0.04 & 0.042 & 0.044 &
 \\ \hline \hline
\frac{|V_{ub}|}{|V_{cb}|}=0.06 &
 n.s. &  n.s. &  n.s. &  n.s. & \begin{array}{c} 0.93\\1.05\end{array} &
\begin{array}{c} 0.83\\1.04\end{array} &
\\ \hline
\frac{|V_{ub}|}{|V_{cb}|}=0.08 &
 n.s. &  n.s. & \begin{array}{c} 1.08\\1.15\end{array} &
\begin{array}{c} 0.93\\1.2\end{array} &
\begin{array}{c} 0.84\\1.19\end{array} &
\begin{array}{c} 0.76\\1.16\end{array} &
\\ \hline
\frac{|V_{ub}|}{|V_{cb}|}=0.1 &
 n.s. & \begin{array}{c} 1.09\\1.29\end{array} &
\begin{array}{c} 0.95\\1.34\end{array} &
\begin{array}{c} 0.86\\1.33\end{array} &
\begin{array}{c} 0.78\\1.31\end{array} &
\begin{array}{c} 0.72\\1.28\end{array} &
\\ \hline
\end{array}\\
\end{displaymath}

\paragraph{$B_K=0.9$ and $m_t=170 \, \gev $:}
\begin{displaymath}
\begin{array}{||r||c|c|c|c|c|cc||}
\hline \hline
|V_{cb}|= &0.034 & 0.036 & 0.038 & 0.04 & 0.042 & 0.044 &
 \\ \hline \hline
\frac{|V_{ub}|}{|V_{cb}|}=0.06 &
 n.s. &  n.s. &  n.s. &  n.s. & \begin{array}{c} 0.84\\1.01\end{array} &
\begin{array}{c} 0.76\\0.99\end{array} &
\\ \hline
\frac{|V_{ub}|}{|V_{cb}|}=0.08 &
 n.s. &  n.s. & \begin{array}{c} 0.96\\1.14\end{array} &
\begin{array}{c} 0.85\\1.15\end{array} &
\begin{array}{c} 0.77\\1.13\end{array} &
\begin{array}{c} 0.7\\1.09\end{array} &
\\ \hline
\frac{|V_{ub}|}{|V_{cb}|}=0.1 &
 n.s. & \begin{array}{c} 0.99\\1.27\end{array} &
\begin{array}{c} 0.87\\1.29\end{array} &
\begin{array}{c} 0.79\\1.27\end{array} &
\begin{array}{c} 0.72\\1.25\end{array} &
\begin{array}{c} 0.66\\1.21\end{array} &
\\ \hline
\end{array}\\
\end{displaymath}

\paragraph{$B_K=0.9$ and $m_t=190 \, \gev $:}
\begin{displaymath}
\begin{array}{||r||c|c|c|c|c|cc||}
\hline \hline
|V_{cb}|= &0.034 & 0.036 & 0.038 & 0.04 & 0.042 & 0.044 &
 \\ \hline \hline
\frac{|V_{ub}|}{|V_{cb}|}=0.06 &
 n.s. &  n.s. &  n.s. & \begin{array}{c} 0.8\\0.93\end{array} &
\begin{array}{c} 0.71\\0.92\end{array} &
\begin{array}{c} 0.65\\0.89\end{array} &
\\ \hline
\frac{|V_{ub}|}{|V_{cb}|}=0.08 &
 n.s. & \begin{array}{c} 0.94\\1.04\end{array} &
\begin{array}{c} 0.81\\1.07\end{array} &
\begin{array}{c} 0.72\\1.05\end{array} &
\begin{array}{c} 0.66\\1.02\end{array} &
\begin{array}{c} 0.6\\0.98\end{array} &
\\ \hline
\frac{|V_{ub}|}{|V_{cb}|}=0.1 &
\begin{array}{c} 0.97\\1.15\end{array} &
\begin{array}{c} 0.83\\1.2\end{array} &
\begin{array}{c} 0.75\\1.19\end{array} &
\begin{array}{c} 0.68\\1.16\end{array} &
\begin{array}{c} 0.62\\1.13\end{array} &
\begin{array}{c} 0.57\\1.09\end{array} &
\\ \hline
\end{array}\\
\end{displaymath}

\section[\kkmd]{$\mathbf{K_L\!-\!K_S}\,$--mass difference}
We will now discuss the short distance contribution to the
\kkmd : Neglecting the small
imaginary parts of the CKM elements, one finds
\begin{eqnarray}
(\Delta m)_{SD} &=& \frac{\gft}{6 \pi^2} m_K f_K^2 B_K
 \lt[ (\real \lambda _c)^2 m_c^{\ast\, 2} \eta_1^{\ast } +
      2 (\real \lambda _t)  (\real \lambda _c)
        \mw S(x_c^\ast,x_t^\ast) \eta _3^{\ast} \rt. \nn
&& \quad\quad\quad \lt.
        + (\real \lambda _t )^2 \mw S(x_t^\ast) \eta_2^\ast . \rt]
\label{kkmd}
\end{eqnarray}
Here the relative size of the three terms in the brackets is
roughly $100:10:1$. Hence the \kkmd\/ is dominated by $\eta_1$.
This contribution has already been tabulated in \cite{hn1,stefdiss},
so that we  comment  on the corrections stemming from the
second term involving $\eta_3$ here: Since
\begin{eqnarray}
\real \lambda_t &=& - \lt( 1- \frac{\lambda^2}{2} \rt) A^2
\lambda^5 (1- \ov{\rho} )
\end{eqnarray}
to an accuracy of $2\% $, the term with $\eta_3$ depends on
$|V_{cb}|$ and $\ov{\rho}$. Its average contribution to
$(\Delta m)_{SD}/((\Delta m)_{exp})$ is about 0.06. Hence one
can obtain the full $(\Delta m)_{SD}$ from the tabulated
values for the first term in \rf{kkmd} in \cite{hn1,stefdiss}
by multiplying with 1.1.
With the central value for $\eta_1^\mr{NLO}$ listed in
\rf{1res14}
our short distance calculation  therefore reproduces
$68 \%$ of the observed mass difference for $m_c=1.3$ and
$73 \%$ of $(\Delta m)_{exp})$  for $m_c=1.4$,
if the $1/N$--result $B_K=0.7$ is used. Taking into
account the large theoretical error  in  \rf{1res14}
the short distance contribution from \rf{kkmd} is between
$52 \%$ and $95 \%$.

In most textbooks the \kkmd\/ is termed to be dominated by
poorly calculable
long--distance effects. Indeed, with the old LO result for
$\eta_1$ and with the old smaller values for $\laqcd^\mr{LO}$
the short distance part of $\Delta m$ is less than $50 \%$.
Long distance effects come from the operators with only
light quarks in \rf{thehs}, which are also present
below the charm threshold and may generate meson poles
in the low--energy matrix elements.
But a long distance  dominance is  clearly a puzzle,
because by power counting these contributions are suppressed
with $\laqcd^2/\mc$ with respect to the short distance
part, because the coefficient of the leading dimension--six
operator in \rf{hbelow} is proportional to $\mc$
(for a discussion see e.g.\ \cite{sh88}).
This fact has even stimulated speculations about new
physics contributing to the \kkmd\/  (e.g.\ in \cite{cee}).
With the new NLO result for
$\eta_1$ the short distance part is well above potential long distance
contributions. Keeping in mind the large uncertainties
in the problem
it appears to be hopeless to find  evidence
of new physics in the \kkmd.

\cleardoublepage
\newcommand{\pr}{Phys.\ Rev.\ }
\newcommand{\np}{Nucl.\ Phys.\ }
\newcommand{\prl}{Phys.\ Rev.\ Lett.\ }
\newcommand{\pl}{Phys.\ Lett.\ }
\newcommand{\cmp}{Comm.\ Math.\ Phys.\ }
\newcommand{\zp}{Z.\ Phys.\ }
\newcommand{\prp}{preprint }
\newcommand{\ajb}{A.~J.~Buras}

\cleardoublepage
\chapter*{Acknowledgements}
In the first place I would like to thank my advisor, Andrzej
Buras, for guiding my way into the exciting field of QCD corrections
to weak processes. I am grateful for the suggestion of the topic and
the permanent encouragement during my work. In my future places   I will
certainly miss the  stimulating and
creative atmosphere of his \emph{Munich next--to--leading order club}.

Next I would like to thank Stefan Herrlich for exploring the field
of \kkm\/ and the associated field theoretic subtleties
together with me. The collaboration with him has been very
pleasant and certainly very fruitful.
I also owe most of my knowledge on computers to him.

I do not want to miss the numerous stimulating talks with
Patricia Ball, Gerhard Bu\-chal\-la, Mat\-thi\-as Jamin, Markus Lautenbacher,
Miko\l aj Misiak and  Manfred M\"unz. I would like to thank Patricia
for her patience in our collaboration on inclusive semileptonic decays.
I owe a  large part of my knowledge about tree--level decays of heavy
hadrons to her.
I am grateful for Gaby Ostermaier's help with figs.\ \ref{newmt}
and \ref{newbk} and for Stefan's preparation of some of the figures,
which have been used in our common publications.
Hubert Simma's thorough explanation of the role of the unphysical
operators described in sect.\ \ref{unphy} is gratefully acknowledged.

I have enjoyed many discussions on other fields of particle physics
with many members of the TUM Elementary Particle Physics Division,
especially with Manfred Lindner, Kurt Rie\ss elmann and Erhard
Schnapka. I have also learned a lot from friends outside
the department, in particular from Dirk Kreimer and David Broadhurst.
This thesis would not have been completed in time, if
Stephan Stieberger had not provided me with many lifts from Garching
to Munich after the departure of the last bus. I am grateful to
Andrzej Buras, Stefan Herrlich and Manfred M\"unz for proofreading
the manuscript.

Finally I would like to thank everybody in the group for creating an
extraordinary pleasant social environment.

\vfill

Financial support of \emph{Studienstiftung des deutschen Volkes},
\emph{Bundesministerium f\"ur Forschung und Technologie} and
\emph{Freistaat Bayern}\/ is gratefully acknowledged.
\end{document}